\documentclass[12pt,oneside]{report}
\usepackage{amssymb,amsthm,amsmath}
\usepackage{color}
\usepackage{graphicx}
\usepackage[all]{xy}
\usepackage{suthesis-2e}

\DeclareMathOperator{\tr}{Tr}

\DeclareMathOperator{\E}{\mathbb{E}}

\submitdate{March, 2005}
\begin{document}

\def\qedsymbol{\rule{7pt}{7pt}}
\def\supp{ {\rm{supp \,}}}
\def\dist{ {\rm{dist }}}
\def\dim{ {\rm{dim \,}}}
\def\oti{{\otimes}}
\def\bra#1{{\langle #1 |  }}
\def\lb{ \left[ }
\def\rb{ \right]  }
\def\tilde{\widetilde}
\def\bar{\overline}
\def\*{\star}
\def\({\left(}		\def\BL{\Bigr(}
\def\){\right)}		\def\BR{\Bigr)}
	\def\BBL{\lb}
	\def\BBR{\rb}
%
%%%%%%%%%%%%%%%%%%%%%%%%%%%%%%%%%%%%%%%%%%%%%%%%%%%%%%%%%%%%%%%
%\newcommand{\E}{{\mathbb{E}}}
%\newcommand{\1}{{\openone}}

%\def\E{{\mathbb{E} }}
\def\1{{\mathbf{1} }}

\def\bb{{\bar{b} }}
\def\ab{{\bar{a} }}
\def\zb{{\bar{z} }}
\def\zbar{{\bar{z} }}
\def\inv#1{{1 \over #1}}
\def\half{{1 \over 2}}
\def\d{\partial}
\def\der#1{{\partial \over \partial #1}}
\def\dd#1#2{{\partial #1 \over \partial #2}}
\def\vev#1{\langle #1 \rangle}
\def\ket#1{ | #1 \rangle}
\def\rvac{\hbox{$\vert 0\rangle$}}
\def\lvac{\hbox{$\langle 0 \vert $}}
\def\2pi{\hbox{$2\pi i$}}
\def\e#1{{\rm e}^{^{\textstyle #1}}}
\def\grad#1{\,\nabla\!_{{#1}}\,}
\def\dsl{\raise.15ex\hbox{/}\kern-.57em\partial}
\def\Dsl{\,\raise.15ex\hbox{/}\mkern-.13.5mu D}
\def\b#1{\mathbf{#1}}
\newcommand{\proj}[1]{\ket{#1}\bra{#1}}
\def\braket#1#2{\langle #1 | #2 \rangle}
%
%%%%%%%%%%%%%%%%%%%%GREEK LETTERS%%%%%%%%%%%%%%%%%%%%%%%%%%%%%%
%
\def\th{\theta}		\def\Th{\Theta}
\def\ga{\gamma}		\def\Ga{\Gamma}
\def\be{\beta}
\def\al{\alpha}
\def\ep{\epsilon}
\def\vep{\varepsilon}
\def\la{\lambda}	\def\La{\Lambda}
\def\de{\delta}		\def\De{\Delta}
\def\om{\omega}		\def\Om{\Omega}
\def\sig{\sigma}	\def\Sig{\Sigma}
\def\vphi{\varphi}
%
%%%%%%%%%%%%%%%%%%%CALIGRAPHIC LETTERS%%%%%%%%%%%%%%%%%%%%%%%%%
%
\def\CA{{\cal A}}	\def\CB{{\cal B}}	\def\CC{{\cal C}}
\def\CD{{\cal D}}	\def\CE{{\cal E}}	\def\CF{{\cal F}}
\def\CG{{\cal G}}	\def\CH{{\cal H}}	\def\CI{{\cal J}}
\def\CJ{{\cal J}}	\def\CK{{\cal K}}	\def\CL{{\cal L}}

\def\CM{{\cal M}}	\def\CN{{\cal N}}	\def\CO{{\cal O}}
\def\CP{{\cal P}}	\def\CQ{{\cal Q}}	\def\CR{{\cal R}}
\def\CS{{\cal S}}	\def\CT{{\cal T}}	\def\CU{{\cal U}}
\def\CV{{\cal V}}	\def\CW{{\cal W}}	\def\CX{{\cal X}}
\def\CY{{\cal Y}}	\def\CZ{{\cal Z}}

\def\rvac{\hbox{$\vert 0\rangle$}}
\def\lvac{\hbox{$\langle 0 \vert $}}
\def\comm#1#2{ \BBL\ #1\ ,\ #2 \BBR }
\def\2pi{\hbox{$2\pi i$}}
\def\e#1{{\rm e}^{^{\textstyle #1}}}
\def\grad#1{\,\nabla\!_{{#1}}\,}
\def\dsl{\raise.15ex\hbox{/}\kern-.57em\partial}
\def\Dsl{\,\raise.15ex\hbox{/}\mkern-.13.5mu D}
\def\beq{\begin {equation}}
\def\eeq{\end {equation}}
\def\to{\rightarrow}
\newtheorem{lem}{Lemma}
\newtheorem{prop}{Proposition}
\newtheorem{theo}{Theorem}
\newtheorem{dfn}{Definition}
\newtheorem{cor}{Corollary}
\newtheorem*{cor*}{Corollary}

\newtheorem*{property}{Property}
\newtheorem*{theo1'}{Theorem 1'}
\newtheorem*{theo2'}{Theorem 2'}
\newtheorem*{prop*}{Proposition}

\theoremstyle{definition}
\newtheorem*{ex}{Example}

\theoremstyle{definition}
\newtheorem*{rem}{Remark}

\def\diag{\mbox{diag}}

\def\argmax{\mbox{argmax}}

\definecolor{gray}{gray}{.9}
\def\com#1{\vspace{.1in}\fcolorbox{black}{gray}{\begin{minipage}{5.75in}#1\end{minipage}}\vspace{.1in}}

\def\bC{\mathbb{C}}
\def\bR{\mathbb{R}}
\def\v#1{\vec{#1}}
\def\h#1{\widehat{#1}}

\def\bpm{\begin{pmatrix}}
\def\epm{\end{pmatrix}}
\def\1s2{\frac{1}{\sqrt{2}}}
\def\12{\frac{1}{2}}

\def\nn{\nonumber}

% Title Page
\title{Simultaneous Classical-Quantum Capacities of Quantum Multiple Access Channels}
\author{Jon Thomas Yard}
\principaladviser{Thomas M. Cover}
\firstreader{Yoshihisa Yamamoto}
\secondreader{Abbas El Gamal}
\dept{Electrical Engineering}
%\twoside

\figurespagefalse
\tablespagefalse

\beforepreface

\prefacesection{}
\vspace{1.75in}
\begin{center}
\emph{To my father} \\  Carl A. Yard Jr.
\end{center}

\prefacesection{Acknowledgments}
It is hard to express the gratitude I have for my advisor, Tom Cover.  I thank him for not only giving me the freedom to learn about and work on any topic of my choosing, but also for his unending encouragement and belief that I would eventually figure out how to do what I needed to do in order to graduate.   
 
I am particularly indebted to my collaborators Igor Devetak and Patrick Hayden.  
The assistance and guidance that I received from them has helped me to learn an extraordinary amount about quantum information theory, a task which would have been insurmountable without their help.  In fact, this dissertation is based on my collaboration \cite{qmac} with them.  
I would also like to give thanks to my reading committee, Yoshi Yamamoto, Abbas El Gamal and Alexander Fetter.

I thank the current members of the information theory group, Young-Han, Charles, Styrmir, George and Navid, as well as the past members Josh, David, Michael, Arak and Assaf, for years of stimulating conversations around the office and at our weekly group meetings.
 
Much thanks to my family for their love and support, especially Mom, who has been anticipating my graduation for quite some time.  
I also thank my many friends for being there for me over the years, most notably Glenn and Jeremy, who have helped me out of quite a few mathematical jams in which I have found myself over the years.
I am gracious for the roommates I have had the pleasure of living with during graduate school, most notably Andy and Sophia, who lived with me during the writing of this dissertation.  Thanks for dealing with me and my endless piles of paper.

I also thank Martin Morf, whose support and encouragement throughout my stay at Stanford has certainly help to keep me going.  I am appreciative of Denise Murphy for rescuing me numerous times from various administrative messes in which I found myself over the years.
I am also grateful to John Preskill for hosting me for a week at the Caltech Institute for Quantum Information where I worked on completing this dissertation.
Finally, a million thanks to Meagen for her love and support.

\afterpreface

\chapter{Introduction}
Information is embodied in physical objects.  The paint on the ceiling of the Sistine chapel, the groove of a record, a single strand of DNA, and the spin of an individual electron each reflect a configuration of a particular physical system which governs the way in which it interacts with the rest of the world.  Michaelangelo's painting interacts with ambient light, emitting a spectrum of colors viewed by churchgoers and tourists alike.  The spinning record induces vibrations in a needle which are converted to a flow of electrons which, when amplified, cause pressure waves to travel through the air.  The structure of DNA encodes instructions for both self-replication and the construction of living beings.  Interactions between individual electrons are mediated by photons and can be modeled with great precision using the tools of quantum electrodynamics.  The physics of simplified models can be calculated using the rules of quantum mechanics.
Quantum particles embody \emph{quantum information}.

% There are a number of mantras in the culture of physics which cater to the relevance of information to the physical world.  Wheeler coined the phrase ``it from bit" which, according to him [CITE], 
% \begin{quote}
% symbolises the idea that every item of the physical world has at bottom --- at a very deep bottom, in most instances --- an immaterial source and explanation; that which we call reality arises in the last analysis from the posing of yes-no questions and the registering of equipment-evoked responses; in short, that things physical are information-theoretic in origin.
% \end{quote}

Claude Shannon, motivated by engineering problems in communication theory,
initiated the study of information theory as an abstract discipline.
On the first page of his seminal paper \cite{shannon}, he writes  
\begin{quote}
\begin{singlespace}
The fundamental problem of communication is that of reproducing at one point either exactly or approximately a message selected at another point. \ldots the actual message is one \emph{selected from} a set of possible messages.
\end{singlespace}
\end{quote}
He later states ``we wish to consider certain general problems involving communication systems."  These general problems, the outgrowth from which is referred to nowadays as ``Shannon theory," concern characterizing the possibilities of reliably transmitting certain ``information sources" over  ``information-bearing channels." 
By making probabilistic assumptions regarding the behaviors of the sources and channels, a rich mathematical theory emerges which, in many cases, reasonably approximates the underlying physics.  Specifically, Shannon showed that if a channel is modeled by a probability transition matrix $p(y|x)$, its capacity for the transmission of classical information is given by 
\[C = \max_{p(x)} I(X;Y).\]
This formula will be discussed in Section~\ref{section:mutualinformation}, where we introduce the mutual information $I(X;Y)$, and in  Section~\ref{section:classicalcap} where we discuss the proof of Shannon's theorem.   Shannon theory applies to network communication as well.  A probability transition matrix $p(z|x,y)$ models a situation where two senders transmit to a single receiver, subject to noise and interference.  The rates at which the senders can transmit independent information were determined by Ahlswede \cite{ahlswede} and Liao \cite{liao} to admit a  single-letter characterization, given by the convex hull of the closure of the set of pairs of nonnegative rates $(R_X,R_Y)$ satisfying
\begin{eqnarray*}
R_X &<& I(X;Z|Y) \\
R_Y &<& I(Y;Z|X) \\
R_X+R_Y &<&I(XY;Z)
\end{eqnarray*}
for some $p(x)p(y).$
Further analysis by Cover, El Gamal and Salehi
\cite{ces} gives single-letter characterizations of a set of correlated sources which can be reliably transmitted over a
multiple access channel,
generalizing the above, as well as
Slepian-Wolf source coding and cooperative multiple access channel capacity.  They also give a multi-letter expression for the capacity region, showing that an i.i.d.\ source $(U,V)$ can be reliably transmitted if and only if
\begin{eqnarray*}
H(U|V) &<& \frac{1}{n} I(X^n;Z^n|U^nY^n) \\
H(V|U) &<& \frac{1}{n} I(Y^n;Z^n|V^nX^n) \\
H(U,V) &<& \frac{1}{n} I(X^nY^n;Z^n)
\end{eqnarray*}
for some $n$ and $p(x^n|u^n),p(y^n|v^n)$, where by $x^n$ we mean the sequence of symbols $(x_1,\dotsc,x_n)$.  Here, $H(U,V)$ and $H(U|V)$ respectively denote the \emph{entropy} and \emph{conditional entropy} of the pair of random variables $(U,V)$ which model the source.  Such a characterization is of limited practical use, however, as it does not apparently lead to a finite computation for deciding whether or not a source can be transmitted.

The concept of the entropy of a physical system initially arose out of attempts to characterize the optimal efficiency of physical machines such as steam engines, as well as to rule out the possibility of such constructions as perpetual motion machines.  The extensivity of entropy demands that it be additive for independent physical systems.  Boltzmann defined the entropy of a physical system to be proportional to the logarithm of the number of microstates, or indistinguishable configurations of its constituents, a definition he was likely led to because $\log(W_1 W_2) = \log(W_1) + \log(W_2)$, making additivity of entropy evident.  In order to circumvent mathematical subtleties which arise due to course grainings of the system's configuration space, a probabilistic approach can be taken, allowing rigorous mathematical statements to be made about related systems which are essentially hidden Markov models \cite{coverjulian}.  A crucial philosophical step was taken by Boltzmann in his work; he assumed that things were made of atoms.  In his framework, heat was not a fluid that flowed from warm to cold bodies; rather, vibrational energy of the constituents of a physical system induces similar behavior in neighboring systems.  Without direct physical evidence to support the existence of atoms, Boltzmann provided a mechanism for the flow of heat which assumed such ingredients did in fact exist.  The existence of atoms was experimentally verified soon after Boltzmann's untimely death.

In the years that followed, the structure of atoms was intensely investigated.  The assumption that atoms obey the laws of Newtonian mechanics quickly resulted in various logical inconsistencies in the form of predictions which did not agree with experimental results.  Quantum mechanics was reluctantly developed as a collection of fundamental assumptions about the nature of the physics of atoms and their constituents.  A collection of mathematical rules was thus constructed which allowed theoretical calculations of certain aspects of experimental results.  A caveat was that the new theory introduced randomness as a fundamental assumption of the theory, a feature which was quite unsettling to even the creators of quantum mechanics, most notably Einstein, who thought that ``God does not play dice with the universe."

Today, we live in a quantum world.  Progress in applied physics and engineering has begun to make manipulation of matter on the quantum scale a reality.  It is a strange world, at least when viewed with a classical mind.  From the other side of the fence, however, classical physics can be seen to be part of a quantum world.  The emergence of classicality due to phase transitions in systems of many particles is one way this occurs.  Mathematically, as we will see in Sections~\ref{section:classical} and  \ref{section:cqsystem}, the tools and language of quantum theory enable the expression of concepts from classical probability theory.

This opens up the possibility of analyzing communication scenarios in which the senders and receivers process quantum information.  In this case, the medium quite literally is the message, whereas rather than sending information by selecting a message from a set, physical systems are suitably prepared for transmission to a receiver.  The possible types of quantum communication range from transmitting particles from sender to receiver to generating entanglement between the users of a channel. Quite remarkably, certain basic components from the classical theory find a place in the quantum extension.  The techniques used to separate possible quantum information processing tasks from the impossible are directly analogous to those used Shannon's in original program.  Possibility questions of this nature have much in common with the original motivations of thermodynamics. The ways in which entropy arises in characterizing the answers further deepens this connection.  

In this sense, one aspect of quantum information theory involves generalizing existing classical results to include quantum resources.  While network Shannon theory is already a rich active area of research, 
its quantum extension has new aspects which do not fit into the former framework.  This leads to a theory which, while including the old one as a special case, asks new questions leading to a deeper understanding of the physical nature of information.  Apparently, quantum information is something new which cannot be properly analyzed with classical tools alone.

In this dissertation, we will analyze quantum channels with many senders and a single receiver, used in a variety of ways for the simultaneous transmission of classical and quantum information, 
representing an expanded version of the manuscript \cite{qmac}.  
At a high level, the results and approaches contained within mirror those of classical Shannon theory.  Yet, the mathematical tools utilized are distinctly quantum.
Let us end this introduction by giving a quote from Asher Peres and Daniel Terno's paper \cite{peresinforel} on quantum information and relativity, where it is written that ``the goals of quantum information theory are the intersection of those of quantum mechanics and information theory, while its tools are the union of these two theories."  Well said.

\chapter{Background} \label{section:background}
\section{The basics}
\subsection{Quantum mechanics} \label{section:quantum}
Let us briefly review some elements of quantum mechanics. 
The physical state of an isolated system with $d$ quantum degrees of freedom is described by a complex unit vector $\ket{\psi}\in \bC^d$.  The notation $\ket{\psi}$, known as a \emph{ket} or \emph{ket vector}, refers to a normalized column vector with $d$ complex components:
\[\ket{\psi} = \begin{pmatrix} \alpha_1 \\ \alpha_2 \\ \vdots \\ \alpha_d \end{pmatrix}, \,\, \text{where} \,\, \sum_a|\alpha_a|^2 = 1.\]
The conjugate transpose of $\ket{\psi}$ is a row vector 
\[
 \bra{\psi} = \begin{pmatrix} \alpha_1^* & \alpha_2^* & \hdots & \alpha_d^* \end{pmatrix}.
\]
$\bra{\psi}$ is called a \emph{bra} or \emph{bra vector}.  This notation (and nomenclature) was introduced by Dirac, partly to emphasize the inner product structure of $\bC^d$.  Indeed, the inner product between two state vectors $\ket{\phi}$ and $\ket{\psi}$ is written as a bra times a ket, or bra-ket
\[\braket{\phi}{\psi} = \braket{\psi}{\phi}^*.\]
It is often useful to write a basis for $\bC^d$ by defining a collection of kets as 
\[\ket{1}= \begin{pmatrix} 1 \\ 0 \\ \vdots \\ 0 \end{pmatrix} \,\,,\,\,
 \ket{2} = \begin{pmatrix} 0 \\ 1 \\ \vdots \\ 0 \end{pmatrix} \,\,,\,\,
 \dotsc\,\,,\,\,
 \ket{d}= \begin{pmatrix} 0 \\ 0 \\ \vdots \\ 1 \end{pmatrix}.
\]
Then, a state such as $\ket{\psi}$ can be expanded in terms of this basis as  
\[\ket{\psi} = \al_1\ket{1} + \cdots + \al_d \ket{d}.\]
A measurement can be performed on the quantum system, obtaining classical information regarding the system's current quantum state.  Quantum mechanics is only able to predict the probabilities of occurrence for each outcome of the measurement.  Further, the state of the system will generally be disturbed by the act of obtaining this classical inforamation.  The simplest measurement to describe is a \emph{pure state measurement}, which is completely described in terms of some orthogonal basis for $\bC^d$.  Such a basis will be called a \emph{measurement basis}.  Suppose that a pure state measurement in the measurement basis $\{\ket{1},\dotsc,\ket{d}\}$ is made on the state $\ket{\psi}$.  Then, 
\begin{itemize}
\item The measurement will return $y$ with probability 
\[p(y) \equiv \Pr\{\text{measure }\, \ket{y}\} = |\braket{y}{\psi}|^2.\]
\item If the measurement returns $y$, the post-measurement state is then $\ket{y}$.
\end{itemize}
In other words, the measurement result is modeled by a $\CY = \{1,\dotsc,d\}$-valued random variable $Y$, distributed as 
$p(y) = |\braket{y}{\psi}|^2 = |\al_y|^2$, and the post-measurement state is a random vector $\ket{Y}$.  If the same measurement is performed again, the same result $Y$ is obtained with certainty, leaving the system in the same state $\ket{Y}$ after the measurement. 

\subsection{Pure state ensembles} \label{section:ensembles}
Here, let us fix a basis $\{\ket{y}\}_{y=1}^d$ for $\bC^d$. 
Imagine now a game with two parties, Alice and Bob.  Assume that Alice has the ability to prepare any pure state from the finite collection of states $\{\ket{\psi_x}\}_{x\in\CX}$.  
Then, the probability that Bob obtains measurement result $y$ given that Alice prepares state $\ket{\psi_x}$ is given by 
\begin{equation}
p(y|x) = |\braket{y}{\psi_x}|^2. \label{pureprob1}
\end{equation}
Notice that this can be rewritten as 
\begin{eqnarray}
|\braket{y}{\psi_x}|^2 
&=& \braket{y}{\psi_x}(\braket{y}{\psi_x})^* \nn\\
&=& \braket{y}{\psi_x}\!\braket{\psi_x}{y} \nn\\
&\equiv& \bra{y}\big(\proj{\psi_x}\big)\ket{y}.\label{pureprob2}
\end{eqnarray}
We may interpret this as saying that if the 1-dimensional projection matrix $\proj{\psi_x}$ is written in the $\{\ket{y}\}$ basis, then $p(y|x)$ is equal to the diagonal matrix element corresponding to $\ket{y}$.  

% Recall the cyclicity of the trace: for arbitrary matrices \[A\in \bC^{m\times n} \,\text{ and } \,B\in\bC^{n\times m},\] 
% it is always true that 
% \[\tr AB = \tr BA.\]
% The case where $m = 1$ and $n = d$ allows the previous expression to be rewritten as 
% \begin{eqnarray*}
% \braket{y}{\psi_x}\!\braket{\psi_x}{y}
% &=& \tr \braket{y}{\psi_x}\!\braket{\psi_x}{y} \\
% &=& \tr \proj{y}\proj{\psi_x}.
% \end{eqnarray*} 
% We have thus written $p(y|x)$ as the \emph{trace inner product} between the projection onto the original state $\proj{\psi_x}$ and the projection onto the measurement outcome $\proj{y}$. 

Now, suppose that Alice gives Bob a random state, choosing $\ket{\psi_x}$ with probability $p(x)$.  In this case, we say that Alice is preparing an \emph{ensemble} $\{p(x),\ket{\psi_x}\}$ of pure states.  Together with elementary probability, (\ref{pureprob2}) can be used to write the probability that Bob measures $y$ as 
\begin{eqnarray}
p(y) &=& \sum_x p(x)p(y|x) \nn\\
&=& \sum_x p(x) \bra{y}\,\proj{\psi_x}\,\ket{y} \nn\\
&=& \bra{y}\left(\sum_xp(x)\proj{\psi_x}\right)\ket{y}\nn\\
&\equiv& \bra{y}\rho\ket{y}. \nn%\label{pureprob2}
\end{eqnarray}
where the third line is by linearity.  The fourth line defines the \emph{density matrix}
\[\rho = \sum_x p(x)\proj{\psi_x}\]
of the ensemble $\{p(x),\ket{\psi_x}\}$, 
which contains all the data required to compute all probabilities associated with any possible measurement on the ensemble, under the assumption that Bob doesn't know the identities of the individual states.  
Note that $\rho$ is Hermitian
\[\rho^\dag = \sum_x p(x) \big(\proj{\psi_x}\big)^\dagger  = 
 \sum_x p(x) \proj{\psi_x} = \rho\]
and satisfies 
\[\tr \rho = \sum_x p(x) \tr \proj{\psi_x} = \sum_x p(x) = 1.\] 
$\rho$, as we've constructed it, is also nonnegative definite.  This is because for any $\ket{\phi}$, we have 
\[\bra{\phi} \rho \ket{\phi} = 
\sum_x p(x) \bra{\phi} \proj{\psi_x} \ket{\phi}
= \sum_x p(x) |\braket{\phi}{\psi_x}|^2 \geq 0\]
where the last inequality is because each term in the sum is nonnegative.

\subsection{Density matrices} \label{section:density}
We have now seen that if a quantum system is prepared in a random pure state, one can write down its density matrix.  This contains all of the data necessary to compute the probabilities of the outcomes of any measurement that can be made on that system, provided that the identities of the random pure states are unknown to the measurer.  
For a system in a pure state $\ket{\psi}$, we will use the abbreviation 
$\psi \equiv \proj{\psi}$ for the density matrix corresponding to that pure state (this is just the matrix which projects onto the subspace spanned by $\ket{\psi}$.  Let us define here the collection of all density matrices of a $d$-level quantum system as 
\[\CD^d = \{\rho \in \bC^{d\times d} \colon \rho = \rho^\dagger, \rho\geq 0, \tr\rho = 1\}.\]
In other words, a density matrix $\rho\in \CD^d$ is a Hermitian, nonnegative definite normalized matrix.   
We give the following facts about $\CD^d$ without proof, as they are proven in detail in many texts on quantum mechanics \cite{preskill,cn,peres}:  
\begin{property}
$\CD^d$ is convex.
\end{property}
\begin{property}
The extremal points of $\CD^d$ are the projections onto rank 1 subspaces of $\bC^d$, corresponding to equivalence classes of pure states which are identified up to a global phase factor $e^{i\theta}$.
\end{property}
\begin{property} 
$\CD^d$ is compact. 
\end{property}
We may interpret the first fact as saying that if with probability $p$, one chooses to prepare a quantum system so that its density matrix is $\rho$, while with probability $1-p$, it is instead prepared so that its density matrix is $\sigma$, someone who measures the resulting system (and is also ignorant about which preparation was made) computes measurement probabilities with the state 
$p\rho + (1-p)\sigma.$

The second fact illustrates the fact that every density matrix can arise from some pure state ensemble.  This can be seen more directly, since the Hermiticity of $\rho$ implies that it is diagonalizable as 
\[\rho = \sum_i \lambda_i \proj{i}\]
for some orthogonal basis $\{\ket{i}\}$ for $\bC^d$.  The positivity of $\rho$ implies that $\lambda_i\geq 0$, and the fact that $\rho$ is normalized implies that the $\lambda_i$ may be interpreted as probabilities, implying the existence of the required pure state ensemble.  Note that there is in fact an uncountable number of ways in which a density matrix can arise by probabilistically preparing pure states. 

More importantly, the fact that the extremal points of $\CD^d$ are pure states implies that pure states are special, in that they cannot arise as nontrivial probabilistic preparations of other states.  A quantum system in a pure state is in a \emph{definite} state.  

\subsection{Trace norm} \label{section:tr}
For an arbitrary $M \in \bC^{d\times d},$ its \emph{trace norm}
$|M|_1$ is defined as
\[|M|_1 = \tr\sqrt{MM^\dag}.\]
This is easily seen to be equal to the 
sum of the singular values of $M$.  Indeed, writing a singular value decomposition $M=U\Lambda V^\dag$, it follows that
\[|M|_1 = \tr\sqrt{U\Lambda V^\dag V\Lambda U^\dag} = 
\tr U\sqrt{\Lambda^2} U^\dag = \sum_{i}^d |\lambda_i|,\]
where $\Lambda = \diag(\lambda_1,\dotsc,\lambda_d)$.
As $|\cdot|_1$ is a norm (or rather, a \emph{unitarily invariant matrix norm}), it satisfies the following properties: 
\begin{property}[Positivity]
$|M|_1 \geq 0$, while $|M|_1=0$ if and only if $M=0$.
\end{property}
\begin{property}[Homogeneity]
for any $c\in \bC$, $|cM|_1 = |c||M|_1$
\end{property}
\begin{property}[Unitary invariance]
$|M|_1 = |UMU^\dag|_1$ for any unitary $U$
\end{property}
\begin{property}[Triangle inequality]
$|M+N|_1 \leq |M|_1 + |N|_1$
\end{property}
\begin{property}[Submultiplicativity]
$|MN|_1 \leq  |M|_1|N_1|$
\end{property}
Positivity follows because the singular values of any matrix $M$ are always nonnegative, and are all equal to zero if and only if $M=0$.  Homogeneity is true because the singular values of $cM$ equal $|c|$ times those of $M$, and unitary invariance holds because $UMU^\dagger$ and $M$ have the same singular values.  For proofs of the triangle inequality and submultiplicativity, the reader is referred to \cite{hj}.

The trace norm gives a natural metric space structure to $\bC^{d\times d}$ which we will exploit considerably throughout this dissertation.  
Given two matrices 
$M,N \in \bC^{d\times d}$ their \emph{trace distance} is thus defined as the trace norm of their difference $|M-N|_1$. 
For two density matrices $\rho$ and $\sigma$ of a $d$-level quantum system, their trace distance satisfies 
\[0 \leq |\rho-\sig|_1 \leq 2,\]
where the lower bound is saturated if and only $\rho=\sig$, while the upper bound is saturated if and only if $\rho$ and $\sig$ are supported on orthogonal subspaces.  Let us mention here the following alternative characterization of the trace distance between two density matrices \cite{cn}
\[|\rho - \sigma|_1 = 2\max_{0\leq\Lambda\leq 1} \tr\Lambda(\rho - \sigma).\]
The maximization above is over all nonnegative definite matrices $\Lambda$ with spectrum bounded above by 1.

\subsection{Fidelity}
Given two density matrices $\rho$ and $\sigma$ of a $d$-level system, their \emph{fidelity} 
is defined \footnote{Note that many authors (such as \cite{cn}) define this quantity as the square root of our definition.}
as 
\[F(\rho,\sigma) = \left(\tr\sqrt{\sqrt{\rho}\sig\sqrt{\rho}}\right)^2.\]
Fidelity can be expressed in terms of the trace norm as 
\[F(\rho,\sigma) = \left|\sqrt{\rho}\sqrt{\sig}\right|_1^2,\]
a form which makes apparent the symmetry of fidelity in its two arguments.
The following bounds are always satisfied whenever the arguments are density matrices
\[0 \leq F(\rho,\sigma)\leq 1.\]
The lower bound is saturated if and only if $\rho$ and $\sigma$ have orthogonal support, while the upper bound is saturated if and only if $\rho = \sigma$.  Contrary to the situation with the trace norm, a large value of the fidelity between two states signifies that they are close. Fidelity is not a norm, but it can be related to the trace norm in various ways which are summarized in Section~\ref{section:distance}.

If one of the arguments of the fidelity is a pure state, (say $\rho = \phi),$ then 
\begin{eqnarray*}
F(\ket{\phi},\sigma)&=&\left(\tr\sqrt{\proj{\phi}\sig\proj{\phi}}\right)^2 \\
&=& \left(\tr\proj{\phi}\sqrt{\bra{\phi}\sig\ket{\phi}}\right)^2 \\
&=& \bra{\phi}\sig\ket{\phi}.
\end{eqnarray*}
So $F(\ket{\phi},\sigma)$ is just the diagonal matrix element of $\sig$ corresponding to $\ket{\phi}$, when $\sig$ is written in a basis including $\ket{\phi}$.  Note that this is the success probability for a pure state measurement which tests a system prepared in the state $\sig$ for the presence of the state $\ket{\phi}$. 
When both arguments are pure states, we obtain 
\[F(\ket{\phi},\ket{\psi}) = |\braket{\phi}{\psi}|^2.\]
Finally observe the following easily verifiable property.
\begin{property}[Linearity of fidelity]
Fidelity is linear in each argument, i.e. 
\[F(c\rho,\sig) = c F(\rho,\sig) = F(\rho,c\sig).\]
\end{property}

\subsection{POVMs} \label{section:povm}
We describe here a certain general type of measurement which can be performed on a $d$-level quantum system, called a positive operator valued measurement (POVM).  A POVM is specified in terms of a finite collection of matrices 
$\{\Lambda_x\in \bC^{d\times d}\}_{x \in \CX}$
which are positive ($\Lambda_x \geq 0$) and sum to the $d\times d$ identity matrix $1_d$
\[\sum_y \Lambda_y = 1_d.\]
It is often said that the matrices $\{\Lambda_x\}_{x \in \CX}$ form a \emph{partition of unity}.
If the quantum system is in the state $\rho$, the probability of obtaining a measurement result $y$ is given by 
\[p(y) = \Pr\{\text{measure }\,\Lambda_y \} =  \tr \Lambda_y \rho.\]
Conditioned on having received the measurement result $y$, the post-measurement state after such a measurement is computed as 
\[\rho \mapsto \rho_y = \frac{\sqrt{\Lambda_y}\rho\sqrt{\Lambda_y}}{p(y)}.\]
Here, $\sqrt{\Lambda}$ is defined as the unique, positive operator which satisfies $\sqrt{\Lambda}\sqrt{\Lambda} = \Lambda$.  \footnote{Note that some authors use a more general kind of measurement, described by matrices $\{M_y\}$ satisfying $\sum_y M_y^\dagger M_y = 1_d$.  This amounts to choosing a different square root of each $\{\Lambda_y\}$, giving post-measurement states which are unitarily equivalent to those of the convention above, conditioned on the measurement result.  Such measurements can be modeled using the tools introduced in Section~\ref{section:instruments}.}
The measurement results in an ensemble of density matrices $\{p(y),\rho_y\}$.
A pure state measurement in the basis $\{\ket{x}\}$ can be expressed as the POVM 
$\{\proj{x}\}$ consisting of 1-dimensional projection matrices. 

\subsection{Classical systems} \label{section:classical}
Let $\CX$ be a finite set and let $X$ be an $\CX$-valued random variable, distributed according to $p(x)$.  We can define a vector space $\bC^{|\CX|}$ with a fixed orthonormal basis $\{\ket{x}^X\}_{x\in\CX}$, labeled by elements of the set $\CX$.  This sets up an identification
$\ket{\,\cdot}^X\colon\CX \rightarrow\bC^{|\CX|}$ between the elements of $\CX$ and that particular basis.  By this correspondence, the probability mass function $p(x)$ can be mapped to a density matrix 
\begin{equation}
\rho = \sum_{x\in\CX}p(x)\proj{x} \label{classicaldensity}
\end{equation}
which is diagonal in the basis $\{\ket{x}\}_{x\in\CX}.$  Further, to every subset $S\subseteq \CX$ corresponds a projection matrix $\Pi_S = \sum_{x\in S}\proj{x}$ which commutes with $\rho$.  In addition, the projections $\Pi_S$ and $\Pi_T$ corresponding to any two subsets $S,T\subseteq\CX$ commute. 
This way, we can express concepts from classical probability theory in the language of quantum probability.  Consider the following translations from classical to quantum language:  
\begin{eqnarray*}
\Pr\{X\in S\} &=& \tr \rho \Pi_S \\
\Pr\{X\notin S\} &=& 1 - \tr \rho \Pi_S = \tr \rho (1^X - \Pi_S) \\
&\equiv& \tr \rho \Pi_{S^c} \\
\Pr\{X\in S \,\text{and}\, X\in T\} &=& \tr\rho\Pi_S\Pi_T \\&\equiv& \tr\rho\Pi_{S\cap T}\\
\Pr\{X\in S \,\text{or}\, X\in T\} &=& 1 - \Pr\{X\notin S \,\text{and}\, X\notin T\} \\
&=& \tr\rho\big(1^X-(1^X-\Pi_S)(1^X-\Pi_T)\big)\\ &\equiv& \tr\rho\Pi_{S\cup T}.
\end{eqnarray*}
From the early development of quantum mechanics, noncommutativity has been seen to be the hallmark of quantum behavior.  It is to be expected that classical probability, embedded in quantum theory's framework, is described entirely with commuting matrices.

\section{Composite quantum systems} \label{section:composite}
%Two quantum systems can interact, evolving together as if they were a single quantum system.  At this point we introduce a formalism whereby we assign labels to the constituent quantum systems.  
%\paragraph{Composite quantum systems}
Let us begin by introducing a number of conventions which will be used when dealing with multiple quantum systems. We will use capital letters from the beginning of the alphabet $A,B,C,\dotsc$ as labels for quantum systems. 
If $A$ is a quantum system, we will abbreviate its level as $|A|$ (which will \emph{always} be finite), so that its pure states are unit vectors in $\bC^{|A|}$.  A generic pure state of $A$ will then be written as $\ket{\psi}^A$, while a generic density matrix of $A$ will be written $\rho^A$, to remind the reader to which system the state refers.  Whenever we initially introduce a state, the superscript will identify the system it is describing, although later references to that state will not always include the superscript.  This convention will not be cause for confusion, as different symbols will refer to different states.  We will also write the $|A|\times |A|$ identity matrix on $\bC^{|A|}$ as $1^A$.

If $B$ is another quantum systems, then $A$ and $B$ may be combined to form a \emph{composite quantum system} $AB$.  This new system has $|A|\cdot|B| \equiv |AB|$ levels.  The pure states of the new system are instead unit vectors in the \emph{tensor product} $\bC^{|A|}\otimes\bC^{|B|}$ vector space of the individual vector spaces.
The simplest way to define $\bC^{|A|}\otimes\bC^{|B|}$ is as follows.  First, fix arbitrary bases $\{\ket{a}^A\}_{a=1}^{|A|}$ and $\{\ket{b}^B\}_{b=1}^{|B|}$ for $\bC^{|A|}$ and $\bC^{|B|}$.  
Then, $\bC^{|A|}\otimes\bC^{|B|}$ can be formally defined as the linear span of the basis vectors formed by the product of the two individual bases
\[\Big\{\ket{a}^A\otimes \ket{b}^B\Big\}_{a,b = 1}^{|A|,|B|}.\]
A convenient shorthand for the tensor product of pure states is to write  \[\ket{a}^A\ket{b}^B\equiv\ket{a}^A\otimes\ket{b}^B.\]
Then, any pure state of the quantum system can be written as 
\begin{equation}
\ket{\Psi}^{AB} = \sum_{a=1}^{|A|}\sum_{b=1}^{|B|} c_{ab} \ket{a}^A\ket{b}^B.\label{genericentangled}
\end{equation}
Observe that this new vector space we have constructed has dimension $|A|\cdot |B|$.  It is not difficult to show that this construction is universal, meaning that it is independent of the particular bases chosen for $A$ and for $B$.

It will be useful here to describe a certain convention which can be used to write down the tensor product of two column vectors as a single column vector.  This will amount to fixing a way to enumerate the components of the tensor. Suppose that $\vec{v} \in \bC^{|A|}$ and $\vec{w}\in \bC^{|B|}$ are arbitrary column vectors 
\[\vec{v} = \begin{pmatrix}
v_1 \\ v_2 \\ \vdots \\ v_{|A|}
            \end{pmatrix}
\,\,\,\text{ and }\,\,\,
\vec{w} = \begin{pmatrix}
w_1 \\ w_2 \\ \vdots \\ v_{|B|}
            \end{pmatrix}.
\]
As $\bC^{|A|}\otimes\bC^{|B|} \simeq \bC^{|A|\cdot |B|}$, we can ``flatten" $\vec{v}\otimes\vec{w}$ into a single column vector, organizing its components according to the following convention 
\[\text{flatten}\colon\vec{v}\otimes\vec{w}  \mapsto 
\begin{pmatrix}
v_1 \vec{w}\\ v_2 \vec{w}\\ \vdots \\ v_{|A|}\vec{w}
\end{pmatrix}.
\]
In this way, the earlier generic state (\ref{genericentangled}) can be expressed as
\[\text{flatten}\colon \ket{\Psi}^{AB} \mapsto 
\begin{pmatrix}
c_{11} \\ \vdots \\ c_{1|B|} \\ c_{21} \\ \vdots \\ c_{2|B|} \\ \vdots \\ c_{|A||B|}
\end{pmatrix}
\]
It is often the case that a pure state such as $\ket{\Psi}^{AB}$ cannot be written as a tensor product of pure states of its constituent systems, i.e. 
\[\ket{\Psi}^{AB} \neq \ket{\psi}^A\ket{\phi}^B\]
for any pure states $\ket{\psi}^A$ and $\ket{\psi}^B$.
If this is the case, then $\ket{\Psi}^{AB}$ is said to be \emph{entangled}.  Nevertheless, for any pure state of the composite quantum system, there exists a pair of orthonormal bases $\{\ket{i}^A\}$ and $\{\ket{i}^B\}$ such that 
\[\ket{\Psi}^{AB} = \sum_i d_i \ket{i}^A\ket{i}^B.\]
This form is called the \emph{Schmidt decomposition} of $\ket{\Psi}^{AB}$.  Together, the combination of the orthonormal bases $\{\ket{i}^A\ket{i}^B\}$ is called the \emph{Schmidt basis}, while the $\{d_i\}$ are called the \emph{Schmidt coefficients}.  These are easily calculated from the singular value decomposition of the matrix $[c_{i,j}]$ of coefficients in (\ref{genericentangled}), where the Schmidt basis consists of the left and right eigenvectors, while the Schmidt coefficients are the singular values themselves.

Just as the tensor product builds larger vector spaces out of pairs of smaller ones, it also builds larger matrices from pairs of smaller ones.  Fix two matrices 
$M\in \bC^{|C|\times |A|}$ and $N\in \bC^{|D|\times |B|}$. Recall that these are linear operators 
\[M\colon \bC^{|A|}\rightarrow \bC^{|C|} \,\,\text{and}\,\,N\colon \bC^{|B|}\rightarrow \bC^{|D|}.\]
Their tensor product $M\otimes N$ is another linear operator 
\[(M\otimes N) \colon \bC^{|A|}\otimes \bC^{|B|}\rightarrow \bC^{|C|}\otimes \bC^{|D|}.\]
We will abbreviate this by writing
\[M\colon A\rightarrow C, \,\,\,\,\, N\colon B\rightarrow D\,\,\text{ and }\,\,(M\otimes N)\colon AB\rightarrow CD.\]
This new object acts on the tensor product of vectors as 
\[(M\otimes N)(\ket{\psi}^A\otimes\ket{\phi}^B) = (M\ket{\psi}^A)\otimes(N\ket{\phi}^B)\]
and linearity defines the action of $M\otimes N$ on all of $\bC^{|A|}\otimes \bC^{|B|}$.
The tensor product is also \emph{bilinear}, i.e. for any $c\in \bC$, 
\[c(M\otimes N) = (cM)\otimes N = M\otimes (cN).\]
In the same vein as the ``flattened" representation $\bC^{|A|}\otimes\bC^{|B|} \simeq \bC^{|A|\cdot |B|}$ for the tensor product of vectors, there is more general mapping $\bC^{|A|\times |C|}\otimes \bC^{|B|\otimes |D|}\simeq \bC^{|A|\cdot |B|\times |C|\cdot |D|}$ given by
\[\text{flatten}\colon M\otimes N \mapsto
\begin{pmatrix}
m_{11} N  & m_{12} N & \hdots & m_{1|C|}N   \\
m_{21} N  & m_{22} N &        &             \\
\vdots    &          & \ddots &             \\
m_{|A|1}N &          &        & m_{|A||C|}N
\end{pmatrix}.
\]
Note our convention, where the blocks are labelled by elements of the left-most component of the tensor product.  We will use that convention throughout this dissertation.
It is easy to see that calculations can be made in this representation, namely that 
\[\text{flatten}\{M\otimes N\}\text{flatten}\{\ket{\psi}\ket{\phi}\}
= \text{flatten} \{(M\otimes N)\ket{\psi}\ket{\phi}\}.
\]

As the composite system $AB$ is a quantum system itself, it includes a (strictly) larger collection of von Neumann measurements and unitary evolutions.
Indeed, given any two bases $\{\ket{i}^{AB}\}$ and $\{\ket{i'}^{AB}\}$ for $\bC^{|A|}\otimes\bC^{|B|}$, they are related by a particular unitary matrix $U$, defined as
\[U = \sum_{i'i} \ket{i'}\bra{i}.\]
It is not hard to see that any joint von Neumann measurement on the combined system $AB$ can be performed using separate product measurements on $A$ and $B$, provided that the unitary which takes intended measurement basis to the required product basis (and its inverse) are implementable.

Of particular interest is the subject of \emph{local measurements} on a composite quantum system.  Suppose that a measurement $\{\Lambda_x\}_{x\in \CX}$ is made on the $A$ part of the bipartite state $\rho^{AB}$.  New measurement operators $\{\Lambda_x\otimes 1^B\}_{x\in \CX}$ can be constructed, so that 
\[p(x) = \tr \rho (\Lambda_x\otimes 1^B).\]
The post-measurement states are given as before 
\[\rho\mapsto \rho_x = \frac{\big(\sqrt{\Lambda_x}\otimes 1^B\big)\rho\big(\sqrt{\Lambda_x}\otimes 1^B\big)}{p(x)}.\]
It is instructive to see what happens if a local pure state measurement is made on part of a bipartite pure state $\ket{\Psi}^{AB}$.  Here, $\Lambda_x = \proj{x}$, and we obtain 
\[p(x) = \tr \rho (\proj{x}\otimes 1^B).\]  
As a first step, express $\ket{\Psi}^{AB}$ in terms of the new basis for $A$ as
\begin{equation}
\ket{\Psi}^{AB} = \sum_{xb} d_{xb} \ket{x}^A\ket{b}^b.
\label{genericentangled2} 
\end{equation}
Note that the new coefficients $d_{xb}$ are related to the old ones via 
\[\sum_aU_{xa}c_{ab} = d_{xb},\]
where $U\colon\{\ket{a}\} \mapsto \{\ket{x}\}$ is the unitary change of basis matrix.
Then, 
\begin{eqnarray*}
\ket{\Psi}^{AB} &=& \sum_{xb} d_{xb} \ket{x}^A\ket{b}^B \\
&=&\sum_x \ket{x}^A\Big(\sum_b d_{xb}\ket{b}^B\Big) \\
&\equiv&\sum_x \ket{x}^A\ket{\tilde{\psi}_x}^B \\
&\equiv&\sum_x \beta_x \ket{x}^A\ket{\psi_x}^B.
\end{eqnarray*}
The third step above defines the unnormalized vector $\ket{\tilde{\psi}_x}^B$, where in the last, the normalization constant $\beta_x \equiv \sqrt{\braket{\tilde{\psi}_x}{\tilde{\psi}_x}}= \sqrt{\sum_b|d_{xb}|^2}$ and normalized state $\ket{\psi_x} \equiv \beta_x^{-1}\ket{\tilde{\psi}_x}$ are defined.
Now, it is a simple task to compute 
\begin{eqnarray*}
p(x) &=& \tr \big(\proj{x}\otimes 1^B\big)\Psi^{AB} \\
&=& \bra{\Psi}^{AB} \big(\proj{x}\otimes 1^B\big)\ket{\Psi}^{AB} \\
&=& \Big(\sum_{x''} \beta_{x''}^* \bra{x''}^A\bra{\psi_{x''}}^B\Big)
\big(\proj{x}\otimes 1^B\big)
\Big(\sum_{x'} \beta_{x'} \ket{x'}^A\ket{\psi_{x'}}^B\Big) \\
&=& \sum_{x''x'}\beta_{x''}^*\beta_{x'} \braket{x''}{x}\braket{x}{x'}\braket{\psi_{x''}}{\psi_{x'}}\\
&=& |\beta_x|^2.
\end{eqnarray*}
Conditioned on having received the measurement result $x$, the post-measurement state is 
\begin{eqnarray*}
\Psi_x^{AB} &=& \frac{\big(\proj{x}\otimes 1^B\big)\Psi^{AB}\big(\proj{x}\otimes 1^B\big)}{p(x)} \\
&=& \frac{\big(\proj{x}\otimes 1^B\big)
\Big(\sum_{x''x'}\beta_{x''}\beta_{x'}^* \ket{x''}\bra{x'}\otimes\ket{\psi_{x''}}\bra{\psi_{x'}}\Big)
\big(\proj{x}\otimes 1^B\big)}{|\beta_x|^2} \\
&=& \proj{x}\otimes\proj{\psi_x}.
\end{eqnarray*}
Or rather, 
\[\ket{\Psi_x}^{AB} = \ket{x}^A\ket{\psi_x}^B.\]
So, we see that a measurement on $A$ causes the state of $B$ to ``collapse" as well.  
Rather, we see that the measurement on $A$ creates a pure state ensemble $\{p(x),\ket{\psi_x}^B\}$ on $B$.  If an arbitrary POVM is performed on $A$, an ensemble of density matrices on $B$ will generally result.  To see this, we need to introduce the partial trace.

\subsection{Partial trace} \label{section:ptrace}
If we are instead concerned only with the measurement probabilities, and not with the post-measurement states, it is convenient to work with a density matrix on $A$ to compute the measurement probabilities.  This density matrix is defined in terms of the \emph{partial trace} over $B$.  Fixing a bipartite density matrix $\Omega^{AB}$, the partial trace over $B$ of $\Om^{AB}$ can be defined as the unique density matrix $\tr_B \Om$ on $A$ such that for every $M\in \bC^{|A|\times |A|}$, 
\[\tr M (\tr_B\Om)\equiv \tr(M\otimes 1^B)\Om.\]
An equivalent way to define $\tr_B \Om$ is as follows.  If we write $\Om_{a'ab'b} \equiv \bra{a'}\bra{b'}\Om\ket{b}\ket{a}$ and $(\tr_B\Om)_{a'a} \equiv \bra{a'}(\tr_B\Om)\ket{a}$, then 
\[(\tr_B\Om)_{a'a} = \sum_b\Om_{aa'bb}.\]
With this in hand, we can express 
\begin{eqnarray*}
\tr \big(\Lambda_x\otimes 1^B\big)\Om
= \tr\Lambda_x(\tr_B\Om).
\end{eqnarray*} 
% &=& \sum_{xy} \bra{x}\bra{y}\rho\big(\Lambda_x\otimes 1^B\big)\ket{y}\ket{x} \\
% &=& \sum_{xy} \tr \big(\proj{x}\otimes\proj{y}\big) \rho\big(\Lambda_x\otimes 1^B\big)\\
% &=& \sum_{xy} \tr \big(\proj{x}\otimes 1^B\big)\big(1^A\otimes\proj{y}\big)\rho\big(\Lambda_x\otimes 1^B\big)\\
% &=& \sum_{x}\big(\proj{x} \otimes 1^B\big) 
%      \Big(\sum_y\big(1^A\otimes\proj{y}\big)\rho\big(\Lambda_x\otimes 1^B\big)\Big) \\
% &=& \sum_s
% \end{eqnarray*}
% It is instructive to see what happens if a local pure state measurement is made on part of a bipartite pure state $\ket{\Psi}^{AB}$.  Here, $\Lambda_x = \proj{x}$, and we obtain 
% \[p(x) = \tr \rho (\proj{x}\otimes 1^B)\]
For any square matrix $M$ on $AB$, the partial traces over $A$ and $B$ satisfy the following easily verifyable properties: 
\begin{eqnarray*}
\tr M = \tr_{AB} M = \tr_A\tr_B M = \tr_B \tr_A M.
\end{eqnarray*}
A perhaps more concrete definition of the partial trace is obtained by writing a bipartite density matrix in the flattened representation
\[\Om^{AB} = 
\begin{pmatrix}
\om_{11} & \hdots & \om_{1|A|} \\
\vdots    & \ddots &             \\
\om_{|A|1}&        & \om_{|A||A|}
\end{pmatrix}
\]
where each $\om_{aa'}\in \bC^{|B|\times |B|}$. 
Then, $\tr_A \Om^{AB}$ is obtained by summing the blocks on the diagonal
\[\tr_A \Om^{AB} = \sum_a \om_{aa}\]
and $\tr_B \Om^{AB}$ by taking the trace of each block separately
\[\tr_B \Om^{AB} = 
\begin{pmatrix}
\tr\om_{11} & \hdots & \tr\om_{1|A|} \\
\vdots    & \ddots &             \\
\tr\om_{|A|1}&        & \tr\om_{|A||A|}
\end{pmatrix}.
\]
In fact, this representation will allow us to define the following partial product 
\[\bra{a'}\Om\ket{a} \equiv \om_{a'a},\]
allowing the partial trace over $A$ to be expressed in the same was as with the usual trace 
\[\tr_A \Om = \sum_a\bra{a}\Om\ket{a} = \sum_a\om_{aa}.\]
For the generic state $\Psi^{AB}$ written in the form (\ref{genericentangled2}), let us compute 
\begin{eqnarray*}
\tr_A \Psi^{AB} &=& 
\sum_x \bra{x}^A\Big(\sum_{x''x'}\beta_{x''}^*\beta_{x'} \ket{x''}\bra{x'}\otimes\ket{\psi_{x''}}\bra{\psi_{x'}} 
\Big)\ket{x}^A \\
&=& \sum_x |\beta_x|^2\psi_x^B
\end{eqnarray*}
\subsection{Purifications and extensions}
Given an arbitrary density matrix $\rho^B$, it is easy to construct a pure state 
$\ket{\Psi}^{AB}$ such that $\tr_A\Psi = \rho$.  The state $\ket{\Psi}^{AB}$ is called a \emph{purification} of $\rho$.  The construction is as follows.  First, choose any pure state ensemble $\{p(x),\ket{\psi_x}^B\}$ giving rise to $\rho^B$, in the sense that 
\[\sum_x p(x) \psi_x = \rho.\]
Then, the state 
\[\ket{\Psi}^{AB} = \sum_x \sqrt{p(x)}\ket{x}^A\ket{\psi_x}^B\]
is a purification of $\rho^B$.  This is easy to see by computing the partial trace over 
$A$, which was done for a pure state of the same form in the last subsection.

More generally we will speak of an \emph{extension} $\Omega^{AB}$ of a density matrix $\rho^A$, which is just any density matrix (not necessarily a pure state) for which $\tr_B\Om=\rho$.  It is easy to see that any purification $\ket{\Psi}^{ABC}$ of $\Om^{AB}$ is a purification of $\rho^A$ as well, since 
\[\tr_{BC}\Psi = \tr_B(\tr_C\Psi) = \tr_B\Om = \rho.\]

to do: purifications!  relate ensembles, purifications and measurements

\subsection{Classical-quantum (cq) systems} \label{section:cqsystem}
Consider now a collection of density matrices $\big\{\sigma^A_x\big\}_{x\in\CX},$ indexed by the finite set $\CX$.  If those states occur according to the probability mass function $p(x)$, we may speak of an \emph{ensemble} $\big\{p(x),\sigma^A_x\big\}$ of quantum states.  In order to treat classical and quantum probabilities in the same framework, a joint density matrix can be constructed
\[\sigma^{XA} = \sum_{x\in\CX}p(x)\proj{x}^X\otimes\sigma^A_x.\]
This is known as a \emph{cq state}, and describes the classical and quantum aspects of the ensemble on the \emph{extended Hilbert space} $\bC^{|\CX|}\otimes\bC^{|A|}$ \cite{dcr}.  The semiclassical nature of the ensemble is reflected in the embedding of a direct sum of Hilbert spaces $\bigoplus_{x\in\CX}\bC^{|A|}$ into $\bC^{|\CX|}\otimes\bC^{|A|}$.  This should be compared with what was done in Section~\ref{section:classical}, where a direct sum of one-dimensional vector spaces $\bigoplus_{x\in\CX}\mathbb{C}$ was embedded into $\bC^{|\CX|}$.
Just as the classical density matrix $\rho$ from (\ref{classicaldensity}) was diagonal in a basis corresponding to elements of $\CX$, the cq density matrix $\sigma$ is \emph{block-diagonal}, where the diagonal block corresponding to $x$ contains the non-normalized density matrix $p(x)\sigma_x$.  The classical state is recoverable as
$\rho = \tr_A \sigma,$ while the average quantum state is $\tr_X\sigma = \sum_{x\in\CX}\sigma_x$.  The classical-quantum formalism is not only of interest in its own right; information quantities evaluated on cq states play an important role in characterizing what is possible in quantum information theory.

\section{Dynamics} \label{section:dynamics}
We we have already seen an example of quantum dynamics; namely, the measurement process.  In this section, we introduce the most general types of dynamical processes we will consider in this dissertation.  The approach taken here will be to consider quantum channels whose inputs and/or outputs are classical-quantum systems.  But first, let us review the notion of classical channels.  
\subsection{Classical channels} \label{section:classicalchannels}
A discrete classical channel with input symbols belonging to a finite alphabet $\CX$ and output symbols from a finite alphabet $\CY$ is modeled by a collection of \emph{transition probabilities}
$p(y|x)$.  These probabilities comprise a \emph{stochastic matrix} $[p(y|x)]_{yx}$, because the following two conditions are satisfied:
\[p(y|x)\geq 0 \,\,\text{for each} \,\, (x,y)\in \CX\times\CY\]
and 
\[\sum_y p(y|x) = 1 \,\,\text{for each} \,\, x\in \CX\]
ensuring that to each input symbol $x$, there corresponds a conditional probability mass function on the output symbols $\CY$.  

Given an $\CX$-valued random variable $X$ with probability mass function $p(x)$, the action of the channel then defines another random variable $Y$, jointly distributed with $X$ according to 
\[p(x,y) = p(x)p(y|x).\] 
Alternatively, we may view $p(y|x)$ a linear map from the simplex of probability mass functions on $\CX$ to the simplex of probability mass functions on $\CY$, via
\[p(x) \mapsto p(y) = \sum_x p(x)p(y|x).\]
In this sense, a classical channel is a model for a device which allows a sender to ``prepare probability mass functions" at the output.  This way of looking at classical channels leads to our first ``partial" quantum generalization, described in the next section.

\subsection{Classical $\rightarrow$ quantum (c $\rightarrow$ q) channels}
This generalization of classical channels consists of channels with a classical input and a quantum output.  However, instead of preparing probability mass functions at the output, the sender prepares density matrices.  A c $\rightarrow$ q channel $\CX\rightarrow B$ is specified by a collection of \emph{conditional density matrices} $\{\rho^B_x\}_{x\in \CX}$, labeled by the elements of a finite set $\CX$.  As with classical channels, such maps extend to mappings from the simplex of probability mass functions on the input alphabet $\CX$ to the density matrices on the output quantum system $B$ via 
\[p(x) \mapsto \sum_x \rho^B_x.\]
Such channels were implicitly considered in Section~\ref{section:cqsystem}, where we saw that if the input is modeled by a random variable $X$ distributed according to $p(x)$, the combined input-output is a cq system with cq state
\[\rho^{XB} = \sum_x p(x) \proj{x}\otimes\rho^B_x.\]
The collection of c $\rightarrow$ q channels with the same input set $\CX$ and output quantum system $B$ has the structure of a compact convex set. Given two such channels with conditional density matrices $\{\rho_x\}_{x\in \CX}$ and $\{\sigma_x\}_{x\in \CX}$, if $0\leq \lambda \leq 1$, their corresponding convex combination has conditional density matrices $\{\lambda\rho_x + (1-\lambda)\sigma_x\}_{x\in \CX}$.  The extremal points of this convex set have conditional density matrices which are extremal in the convex set of density matrices on $B$.  In other words, the extremal points consist of channels which prepare pure states.  This fact will be important when we discuss classical capacities of quantum channels in Section~\ref{section:hsw}.

\subsection{Unitary quantum channels} \label{section:unitarychannels}
The simplest quantum channel is a unitary transformation.  For a closed quantum system $A$, this is the kind of evolution predicted by the Schrodinger equation
\[\ket{\psi}^A\mapsto \ket{\psi'}^A = U\ket{\psi}^A.\]
We will write
\[U\colon A\rightarrow A\] 
to reflect the fact that $U \in \bC^{|A|\times |A|}$ is a square matrix mapping
\[U\colon \bC^{|A|}\rightarrow \bC^{|A|}.\] 
In this thesis, we will be exploring the consequences for processing quantum information which result from the ability to cause \emph{any} unitary evolution to occur to a given quantum system.  Ensuring that a quantum system undergoes a particular unitary evolution is generally a difficult engineering task, since it involves influencing the system in just the right way, from the outside, so as to inhibit its natural tendency to evolve in the way that it would have without any influence.  To say that this will be of no concern to us here would be somewhat untrue.  In fact, the central goal of this thesis is to show that, under that assumption that error-free processing of quantum information is possible, one can in fact \emph{protect} and \emph{correct} quantum information from this natural tendency to interfere with other quantum information and with the environment.  Indeed, we will assume that it is possible to process quantum information \emph{fault tolerantly}. 
If the state of $A$ is specified by a density matrix $\rho^A$, the unitary channel acts as 
\[U\colon \rho^A \mapsto \rho'^A = U\rho U^\dag.\]
In other words, $\rho$ transforms according to the adjoint map associated to $U$.  We will frequently abbreviate this map as 
\[U(\rho) \equiv U\rho U^\dag.\]
It will often be useful for us to speak of unitaries \emph{between} quantum systems.  For example, we may think of a quantum system $A$ at some time $t$, being turned into another quantum system $B$ at a later time $t'$, where $|A| = |B|$.  If this process acts unitarily, we will write
\[U\colon A\rightarrow B\] 
for the associated unitary channel.  As an example, consider a physical scenario in which an electron placed at a position $x$ at time $t$ is transferred to some other position $x'$ by some later time $t+T$, after having been rotated by $180^\circ$ about its $z$ axis.  The quantum system $A$ thus represents the original preparation of the electron at $x$, while $B$ represents the evolved electron, $T$ seconds later, with its new state at position $x'$.  

\subsection{Quantum channels} \label{section:quantumchannels}
Quantum channels represent a physical process which transfers quantum states forward in time.  The state at the output of the channel will be some noisy version of what was put in.  Examples include an optical fiber over which the polarization of an input photon may become corrupted by noise, or a quantum dot which will hold a single electron for an uncertain amount of time. 

Here, we will give a precise mathematical definition of quantum channels as functions from the density matrices of an \emph{input} quantum system to the density matrices of an \emph{output} quantum system, generalizing the notion of discrete memoryless classical channels described in Section~\ref{section:classicalchannels}, which map probability mass functions on the input alphabet to probability mass functions on the output.  The mathematical properties which we require a channel to satisfy are from the standard literature on open quantum systems and quantum information theory, so much of the content here is presented without proof.  Some standard references for this material include \cite{cn,op}.

By a \emph{quantum channel} $\CN\colon A\rightarrow B$, we mean a mathematical object which maps density matrices on $A$ to density matrices on $B$, while satisfying the following three physically motivated properties described below.  
\begin{property}[Linearity] 
\[\CN\colon \bC^{|A|\times|A|}\rightarrow  \bC^{|B|\times |B|}\] 
is a linear map, so that 
\[\CN\Big(\sum_i p_i \rho_i\Big) = \sum_i p_i \CN(\rho_i).\]
\end{property}
\begin{property}[Trace preservation]
$\CN$ preserves the trace of the input density operator 
\[\tr \rho = \tr \CN(\rho).\]
\end{property}
This technical requirement will sometimes be relaxed to the requirement that $\CN$ only be \emph{trace-non-increasing}
\[\tr \rho \geq \tr \CN(\rho).\]
With a slight loss in pedantry, we will generally refer to such maps as \emph{trace-reducing}.
In such a case, $\CN$ can be interpreted as a channel which is executed with some probability less than one.  
To introduce the third property, let us show that there is a unique way in which $\CN$ acts on the $A$ part of a composite quantum system $AC$.  It is sufficient to see what happens when acting upon part of a pure state 
\[\ket{\Psi}^{AC} = \sum_{ac} d_{ac} \ket{a}^A\ket{c}^C.\]
Here, we obtain 
\begin{eqnarray*}
(\CN\otimes 1^C)(\Psi) &=& (\CN\otimes 1^C)\Big(\sum_{a'ac'c} d_{a'c'}d_{ac}^* \ket{a'}\bra{a}\otimes\ket{c'}\bra{c}\Big) \\
&=&\sum_{a'ac'c} d_{a'c'}d_{ac}^* \CN\big(\ket{a'}\bra{a}\big)\otimes\ket{c'}\bra{c}.
\end{eqnarray*} 
The action of $\CN\otimes 1^C$ is then uniquely defined on any density matrix $\omega^{AC}$ by first writing any pure state decomposition 
\[\omega^{AC} = \sum_i p_i \Psi^{AC}_i.\]
Then by linearity, 
\[(\CN\otimes 1^C)(\omega) = \sum_i p_i (\CN\otimes 1^C)(\Psi_i).\] 
Now, we can mention the third characteristic property of a quantum channel. 
\begin{property}[Complete positivity]
The channel must be \emph{completely positive}, meaning that not only must $\CN:A\rightarrow B$ take nonnegative definite matrices on $A$ to nonnegative definite matrices on $B$, but for any $C$ it must take nonnegative definite matrices on $AC$ to nonnegative definite matrices on $BC$.
\end{property}
A physically satisfying consequence of these three properties is that if a quantum channel acts on part of a convex combination of density matrices, the resulting operator will be a density matrix.   
Quantum channels also obey the following locality properties, which can be derived from the above three.  
We will later invoke these (quite frequently) without reference.   
\begin{property}[Locality I]
Given a bipartite density matrix $\rho^{AB}$ and two quantum channels 
\[\CN\colon A\rightarrow C \,\,\text{and}\,\,\CM\colon B\rightarrow D,\]
the actions of $\CN$ and $\CM$ commute with one another, i.e. 
\[(\CN\otimes 1^B)\circ(1^A\otimes\CM) = (1^A\otimes\CM)\circ(\CN\otimes 1^B) = 
\CN\otimes\CM.\]
These equations are summarized by the leftmost commutative diagram below.  The rightmost diagram is to remind the reader of the subsystems on which the corresponding states are defined. 
\[\xymatrix@R=2cm @C=2cm @H=0cm
{ \rho \ar @{->} [r] ^{\CN} 
       \ar @{->} [rd] ^{\CN\otimes\CM} 
       \ar @{->} [d] ^{\CM}
& (\CN\otimes 1^B)(\rho) \ar @{->} [d] ^{\CM} \\ 
(1^A\otimes\CM)(\rho) \ar @{->} [r] ^{\CN}  & (\CN\otimes\CM)(\rho)
}\hspace{.5in}
\xymatrix@R=1.5cm @C=1.5cm @H=0cm
{ AB \ar @{->} [r] ^{\CN} 
       \ar @{->} [rd] ^{\CN\otimes\CM} 
       \ar @{->} [d] ^{\CM}
& CB \ar @{->} [d] ^{\CM} \\ 
AD \ar @{->} [r] ^{\CN}  & CD
}
\]
\end{property}

\begin{property}[Locality II]
Given a bipartite density matrix $\rho^{AB}$, a local operation on $B$ will not affect the reduced density matrix on $A$, i.e.  given a quantum channel
$\CN\colon B\rightarrow C$,  we have 
\[\tr_C(1^A\otimes\CN)(\rho) = \tr_B\rho.\] 
This is summarized by the commutative diagram on the left below.  On the right, we remind the reader of the subsystems involved.
\[\xymatrix@R=2cm @C=2cm @H=0cm{ 
\rho \ar @{->} [r] ^-{\CN}
%     \ar @{->} [rd] ^{\CN\otimes\CM} 
     \ar @{->} [rd] _{\tr_B}
& (1^A \otimes \CN)(\rho) \ar @{->} [d] ^{\tr_C} \\
&  \tr_B\rho
}\hspace{.5in}
\xymatrix@R=1.5cm @C=1.5cm @H=0cm
{ AB \ar @{->} [r] ^{\CN} 
       \ar @{->} [rd] ^{\tr_B} 
& AC \ar @{->} [d] ^{\tr_C} \\ 
& A
}
\]
\end{property}
This last property can be paraphrased as stating that $\tr_B\rho^{AB}$ is independent of any physical process which is carried out on $B$.  
Throughout this dissertation, we will often omit identity maps in expressions such as 
$1^A\otimes \CN$, so that $\CN\colon B\rightarrow C$ will be interpreted as the map 
$\CN\colon AB\rightarrow AC$ whenever necessary.  An advantage of this approach is that it allows long expressions to be simplified.  This leaves no room for ambiguity, as the action of a channel on part of a larger system is always uniquely defined.

\subsection{Representing quantum channels}\label{section:represent}
In this section, we review two useful representation theorems for quantum channels.  The first, due to Stinespring, shows how unitary processes can give rise to quantum channels.  The second, due to Kraus, shows how quantum channels can be viewed as measuring devices which ``forget", or ``keep secret", the measurement result. 
 
Suppose that a quantum system $A$ is prepared and allowed to evolve unitarily 
with some extra system $E$ which is promised to be prepared in some known pure state $\ket{1}^E$ according to a unitary $U\colon AE\rightarrow AE$. 
Since the state of $E$ is guaranteed to be in the same state before the application of $U$, some of the elements of $U$ are irrelevant to the dynamics.  For example, fixing bases $\{\ket{a}^A\}$ and $\{\ket{e}^E\},$ suppose that $U$ is given by 
\[U = \sum_{ae}\ket{\phi_{ae}}^{AE}\bra{a}^A\bra{e}^E\]
for some other orthogonal basis $\{\ket{\phi_{ae}}^{AE}\}$ of the combined system $AE$.
Then, an arbitrary pure state of $A$
\[\ket{\phi}^A = \sum_a \al_a\ket{a}^A\]
will be mapped to 
\begin{eqnarray*}U\ket{\phi}^A\ket{1}^E &=& \sum_{ae}\ket{\phi_{ae}}^{AE}\bra{a}^A\bra{e}^E
\big(\sum_{a'}\al_{a'}\ket{a'}^A\big)\ket{1}^E \\
&=& \sum_{aa'e}\al_{a'}\braket{a}{a'}\braket{e}{1}\ket{\phi_{ae}}^{AE}\\
&=& \sum_a \al_a \ket{\phi_{a1}}^{AE}.
\end{eqnarray*}
Thus, only the first $|A|$ columns of $U$ are relevant to this situation.  Keeping only this ``chunk" of the unitary $U$ defines an \emph{isometry} $\CU\colon A\rightarrow AE$.  
Mathematically, $\CV\colon A\rightarrow B$ is an isometry if and only if it satisfies one (and thus both) of the following conditions:
\[\CV^\dag\circ\CV = 1^A \,\text{and}\,\CV\circ\CV^\dag = \Pi_A.\]
Above, $\Pi_A$ is a projection matrix on $B$ satisfying $\tr \Pi_A = |A|$.  
In other words, an isometry is a length-preserving matrix whose range is a subspace of the target space, giving an image of the input space on the output space.  

Returning to the isometry $\CU\colon A\rightarrow AE$, consider what will happen if the extra system is disregarded.  Given a density matrix $\rho^A$, a mapping $\tr_E\CU = \CN\colon A\rightarrow A$ results.  This map $\CN$ is a quantum channel, and the map $\CU$ will be called an \emph{isometric extension} of $\CN$.  We will generally use a subscript to identify the channel which is being extended, saying that $\CU_\CN$ \emph{isometrically extends} $\CN$.  This way of representing a quantum channel is often referred to as the \emph{Stinespring} representation, and we will use it almost exclusively throughout this dissertation.  To be precise, we will often invoke the following proposition.   
\begin{prop*}[Isometric extension representation]
A map $\CN\colon A\rightarrow B$ is a quantum channel if and only if there exists an isometric extension
$\CU_\CN\colon A\rightarrow BE$ of $\CN$. 
\end{prop*}
\begin{rem}
In general, an isometric extension $\CU_\CN$ of $\CN$ is \emph{not unique}.  This can be seen by defining $\CU'_\CN = \CV\circ\CU_\CN$, where $\CV\colon E\rightarrow E'$ is any isometry into a (potentially different) environment $E'$.  Since $\tr_E\CU_\CN = \tr_{E'} \CV\circ\CU_N$, these extend the same channel $\CN$. 
\end{rem}

%\subsection{Operator sum representation}\label{section:osr}
Another way to represent a quantum channel is due to Kraus, and is called the operator sum representation (OSR).  The following proposition was first proved in [?].
\begin{prop*}[Operator sum representation]
A map $\CN\colon A\rightarrow B$ is a quantum channel if and only if it can be written as 
\[\CN(\rho) = \sum_{i=1}^k N_i\rho N_i^\dag\]
for matrices $\{N_i\in \bC^{|B|\times |A|}\}$ which satisfy 
\[\sum_{i=1}^k N_i^\dag N_i = 1^A.\]
\end{prop*}
The matrices $\{N_i\}$ are called the operator sum matrices (OSR matrices) of the representation.  Such a representation of $\CN$ is generally not unique.  
It should be mentioned that this representation bears a strong resemblance to the measurement model of POVMs given in Section~\ref{section:povm}. 

For a given channel, the two representations given above are intimately related, and having at hand one representation immediately gives the other as follows.  If the action of $\CN$ can be expressed in terms of OSR matrices $\{N_i\}_{i=1}^k$, 
an isometric extension $\CU_\CN\colon A\rightarrow BE$ into an environment of size 
$|E| = k$ can be constructed as 
\[\CU_\CN = \sum_{i=1}^k \ket{i}^E\otimes N_i.\]
This is perhaps easier expressed by writing $\CU_\CN$ as a block matrix (in the flattened representation), with blocks given by the OSR matrices as 
\[\CU_\CN = \begin{pmatrix}
    N_1 \\ N_2 \\ \vdots \\ N_k
            \end{pmatrix}.
\]
Note that the dimensions match up; namely, $\CU_\CN \in \bC^{|E|\cdot |B| \times |A|}$.
The reverse is also true, and the construction just involves identifying the OSR matrices with the corresponding blocks of a given isometric extension $\CU_\CN$.  

\begin{rem}
As the nonuniqueness of isometric extensions is due to the isometric freedom in describing the environment, the operator sum representation inherits this freedom as well.
\end{rem}

\subsection{Complementary channels} \label{section:complementarychannels}
Suppose that a channel $\CN\colon A\rightarrow B$ is given.  Fixing an isometric extension $\CU_\CN\colon A\rightarrow BE$ of $\CN$, define the channel 
$\CN^c \colon A\rightarrow E$ via $\CN^c = \tr_B\CU_\CN$.  We will say that the channel $\CN^c$ is \emph{complementary} to $\CN$.  If the channel acts on a density matrix $\rho^A$, the state $\CN^c(\rho)$ on $E$ can be thought of as the disturbance induced into an initially pure environment by the action of the channel.  
\begin{rem}
While the choice of complementary channel is generally not unique, it is unique up to isometries on $E$, inheriting this freedom from the choice of isometric extension.    
\end{rem}

\subsection{Controlled quantum channels and cq $\rightarrow$ q channels}
Consider a collection of quantum channels $\{\CM_x\colon A\rightarrow B\}_{x\in \CX}$, labeled by a finite set $\CX$.  Introducing a controlling classical system $X$, available at the input and output, the collection of channels can be represented by a \emph{controlled channel}
$\CM\colon XA\rightarrow XB$.  This channel acts on a cq state
\[\sigma^{XB} = \sum_{x\in\CX} p(x) \proj{x}^X\otimes\sigma^B_x\]
as
\[\CM(\sigma) = \sum_{x\in\CX}p(x)\proj{x}^X\otimes\CM_x(\sigma_x).\]
If the controlling system $X$ is not available at the output, 
the action of the channel is modified to 
\[\CM'(\sigma) = \tr_X\CM(\sigma) = \sum_{x\in\CX}p(x)\CM_x(\sigma_x).\]

We will show next that for \emph{any} quantum channel 
$\CN\colon XB\rightarrow C$ which is only intended to act on cq states, less data is required to specify the action of the channel.  In such a case, the channel can be represented in the same fashion as $\CM'$, in the sense that 
%\[\CN = \sum_{x\in \CX}\proj{x}\otimes\CN_x.\]
the action of $\CN$ on $\sigma^{XB}$ decomposes as   
\[\CN(\sigma) = \sum_{x\in\CX}p(x)\CN_x(\sigma_x)\]
for some channels $\{\CN_x\colon A\rightarrow B\}_{x\in \CX}.$
To see this,  
suppose that $\CN\colon XB\rightarrow C$ has an operator sum decomposition 
\[\CN\colon\tau \rightarrow \sum_{i=1}^d N_i\tau N_i^{\dagger},\]
where the $|C|\times |\CX|\cdot|B|$-dimensional matrices $N_i$ satisfy 
$\sum_{i=1}^d N_i^\dagger N_i = 1^{XB}$.  Consider each $N_i$ to be composed of $|\CX|$ blocks of size $|C|\times |B|$, as 
\[N_i = \begin{pmatrix} N_{i1} & N_{i2} & \cdots & N_{i|\CX|}\end{pmatrix}. \]
The action of $\CN$ on $\sigma$ then simplifies as
\begin{eqnarray*}
\CN(\sigma) &=& \sum_{i=1}^d N_i\sigma N_i^\dagger \\
&=& \sum_{i=1}^d \sum_{x\in\CX} p(x) N_{ix} \sigma_x N_{ix}^\dagger \\
&=& \sum_{x\in\CX} p(x)\sum_{i=1}^d N_{ix} \sigma_x N_{ix}^\dagger \\
&\equiv& \sum_{x\in\CX} p(x) \CN_x(\sigma_x),
\end{eqnarray*}
where in the last step we identify, for each $x$, the matrices $\{N_{ix}\}_{x\in\CX}$ as the components of a trace preserving map $\CN_x$.  

\subsection{Quantum instruments (q $\rightarrow$ cq channels)} \label{section:instruments}
A \emph{quantum instrument} \cite{davies} $\boldsymbol{\CN}\colon A\rightarrow BX$ is a quantum channel whose output is a cq system.  Mathematically, it is specified by collection of completely positive, trace-reducing channels $\{\CN_x\colon A\rightarrow B\}_{x\in \CX}$, labeled by a finite set $\CX$, such that the sum $\CN = \sum_x \CN_x$, which acts on an arbitrary input state $\rho^A$ as 
\[\CN(\rho) = \sum_x \CN_x(\rho),\] 
is trace preserving (and is thus a quantum channel).  The action of the instrument on $\rho^A$ is given by  
\[\boldsymbol{\CN}(\rho) = \sum_x \proj{x}\otimes \CN_x(\rho).\]
The measurement process can be modeled by a quantum instrument as follows.  Given a POVM $\{\Lambda_x\}_{x\in \CX},$ consider the quantum instrument 
$\boldsymbol{\CN}\colon A\rightarrow AX$ with components acting as 
\[\CN_x(\rho) = \sqrt{\Lambda_x}\rho\sqrt{\Lambda_x}.\]
Then, the action of $\boldsymbol{\CN}$ is just 
\begin{eqnarray*}
\boldsymbol{\CN}(\rho) &=& \sum_x \proj{x}\otimes \sqrt{\Lambda_x}\rho\sqrt{\Lambda_x} \\
&=&\sum_x \tr\Lambda_x\rho \proj{x}\otimes \frac{\sqrt{\Lambda_x}\rho\sqrt{\Lambda_x}}{\tr\Lambda_x\rho} \\
&\equiv& \sum_x p(x) \proj{x}\otimes\rho_x,
\end{eqnarray*}
where the $\{\rho_x\}$ are the post-measurement states.  In other words, 
$\CN_x(\rho)$ is an unnormalized density matrix satisfying 
\[\tr\CN_x(\rho) = p(x)\]
which is proportional to the post-measurement state.  We will later utilize such a quantum instrument in order to simultaneously decode classical and quantum information which have been transmitted over a quantum multiple access channel. 

It is also possible to use an instrument to model a measurement which ignores the post-measurement state.  This is done with a \emph{measuring instrument} $\boldsymbol{\CM}\colon \rightarrow X$, which can be defined in terms of the previous instrument as $\tr_A\boldsymbol{\CN}$.  This simpler instrument acts as 
\begin{eqnarray*}
\boldsymbol{\CM}(\rho) &=& \tr_A \sum_x \proj{x}\otimes \sqrt{\Lambda_x}\rho\sqrt{\Lambda_x} \\
&=& \sum_x (\tr_A\Lambda_x)\proj{x}  \\
&=& \sum_x p(x) \proj{x},
\end{eqnarray*}
which is exactly as one would expect a measuring device to act. 

As an instrument is also a channel, it makes sense to speak of an isometric extension and complementary channel to an instrument.  In the appendix (Section~\ref{section:instrumentcomplement}), we will demonstrate that any channel complementary to an instrument is another instrument with similar structure.  Namely, the components of the complementary instrument are obtained as complements of the components of the original instrument.

\pagebreak

\chapter{Entropy and information quantities}
In this chapter, we review the notion of quantum entropy, as well as some related information theoretical quantities which characterize the capacities to be introduced later.   
\section{Entropy} \label{section:entropy}
Let $\CX$ be a finite set, and let $X$ be a $\CX$-valued random variable, distributed according to $p(x)$. 
The \emph{Shannon entropy} of $X$ is defined as
\[H(X) = -\sum_x p(x)\log p(x).\]
All logarithms in this dissertation will be to the base 2 ($\log \equiv \log_2$).  Also, note that we will always take 
$0 \log 0 = 0$, as $\lim_{x\rightarrow 0} x\log x = 0$ by continuity.
Further note that $H(\cdot)$ does not depend on the values taken by $X$.
Rather, it is a functional of the probability mass function $p(x)$ of $X$.  Indeed, $\CX$ is merely abstract set whose elements are merely labels for events.  For example, $X$ may be taken to represent the result of a fair coin flip, whereby 
$\CX = \{\text{heads},\text{tails}\}$
and $p(\text{heads}) = p(\text{tails}) = \frac{1}{2}$.  In this 
case, $H(X) = 1 \text{ bit}$.  One interpretation to be gained from this example is that we obtain a bit of information by learning the result of a fair coin flip.  In this sense, the coin flip example defines a ``unit of information" equal to 1 bit.  

$H(X)$ can also be interpreted as the number of bits, on average, required to represent the random variable $X$.  Intuitively, entropy may be thought of as a measure of the amount of ``information contained in" the random variable $X$.  By definition, this is a statement concerning the asymptotic statistics of sequences of i.i.d.\ random variables $X^n = (X_1,\dotsc,X^n)$.  Such an operational definition has its roots in the source coding theorem, which dates to Shannon's original paper \cite{shannon}, where the entropy was established as the fundamental limit on the compressibility of information.  As this dissertation will focus on the closely related problem of \emph{channel coding}, we will not pursue this interpretation further.

Suppose that a quantum system is prepared with density matrix $\rho$.  We define the \emph{von Neumann entropy} of $\rho$ as
\[H(\rho) \equiv -\tr\rho\log\rho.\]
Note that we overload the letter $H$ to mean both Shannon and von Neumann entropy.
Writing an eigendecomposition of $\rho$ as 
\[\rho = \sum_x p(x)\proj{x}\]
we obtain an ensemble of orthogonal pure states $\{p(x),\ket{x}\}$ which also gives rise to the density matrix $\rho$.  
The von Neumann entropy of $\rho$ is then equal the Shannon entropy of the eigenvalues of $\rho$.  Indeed,
\begin{eqnarray*}
H(\rho) &=& - \tr\Big(\sum_x p(x)\proj{x}\Big)\Big(\sum_x \log p(x)\proj{x}\Big) \\
&=& - \tr\Big(\sum_x p(x)\log p(x)\proj{x}\Big) \\
&=& - \sum_x p(x)\log p(x) = H(X)
\end{eqnarray*}
where $X$ is a random variable with probability mass function $p(x)$.

% Many familiar properties of Shannon entropy are also enjoyed by von Neumann entropy.  If $X$ is a random variable taking $d$ possible values, the Shannon entropy of $X$ satisfies the simple inequalities
% \[0\leq H(X)\leq\log d.\]
% These bounds are tight, as the lower bound is achieved iff $X$ is deterministic ($\Pr\{X=x\}\in\{0,1\}$), while the upper bound is saturated iff
% $X\sim \text{unif}(\{1,\dotsc,d\})$.
% Similar bounds hold in the quantum case, as
% \[0\leq H(\rho)\leq\log d.\]
% The lower bound is saturated iff the quantum system is in a pure state, i.e. $\rho = \proj{\psi}$, for some $\ket{\psi}\in\mathbb{C}$.  Tightness of the upper bound is obtained iff $rho=\frac{1}{d}1_d \equiv \pi_d$ [DEFINE $\pi$ EARLIER (ENTANGLEMENT SECTION)].
% 
% For any two random variables $X$ and $Y$, jointly distributed according to $p(x,y)$, their joint entropy $H(XY)$ satisfies
% \[H(X) \leq H(XY)\leq H(X)+H(Y).\]
% The lower bound is saturated iff $Y = f(X)$, for some deterministic function $f$.  Equality is obtained in the upper bound iff $X$ and $Y$ are independent, which means that their joint distribution factorizes as $p(x,y)= p(x)p(y)$.
% This latter inequality reflects \emph{subadditivity of Shannon entropy}.

If $\rho^A$ is associated with system $A$, we will often write $H(\rho) = H(A)_\rho,$ omitting the subscript when the state is apparent from the context. Given some multipartite state $\Om^{AB}$, the above notation gives a useful way to denote entropies of partial traces of $\Om$.  For example, $H(A)_\Om = H(\tr_B\Om)$, while $H(AB)_\Om = H(\Om)$.  We now state the following elementary properties of entropy.  These are proved in many introductory textbooks such as \cite{cn}.
\begin{property}[Entropy is nonnegative]
\[H(A) \geq 0\]
\end{property}
This bound is saturated if and only if $A$ is in a pure state $\ket{\phi}^A$.
\begin{property}[Entropy is bounded]
\[H(A) \leq \log|A|\]
\end{property}
This bound is saturated if and only if $A$ is prepared in a \emph{maximally mixed state}
\[\pi^A \equiv \frac{1}{|A|}1^A.\]
\begin{property}[Entropy is subadditive]
\[H(AB) \leq H(A) + H(B)\]
\end{property}
This bound is saturated if and only if $AB$ is prepared in a \emph{product state} $\rho^A\otimes\sig^B$.
\begin{property}[Lieb's inequality]
\[|H(A) - H(B)| \leq H(AB)\]
\end{property}
Let us compute the entropy of a generic cq state 
\begin{equation}
\rho^{XA} = \sum_x p(x)\proj{x}\otimes\rho_x^A. \label{cqexample}
\end{equation}
To do so, we first diagonalize each $\rho_x$ as 
\begin{equation}
\rho_x = \sum_y p_x(y) \proj{y_x}, \label{rhospec}
\end{equation}
where for each $x$, the vectors $\big\{\ket{y_x}\big\}_{y_x=1}^{|A|}$ form (generally) different orthonormal bases for $A$. 
Then, we write 
\begin{eqnarray*}
H(XA) &=& -\tr\Big(\sum_x p(x)\proj{x}\otimes \rho_x\Big)\log\Big(\sum_x p(x)\proj{x}\otimes \rho_x\Big) \\
&=& - \tr \sum_x p(x) \proj{x}\otimes \Big(\rho_x\log \big(p(x) \rho_x\big)\Big) \\
&=& - \sum_x p(x) \tr \Big(\rho_x\log \big(p(x) \rho_x\big)\Big) \\
&=& - \sum_{xy} p(x) p_x(y) \log(p(x) p_x(y)) \\
&=& - \sum_{x} p(x) \Big(\log p(x) + \sum_y p_y(x) \log p_y(x)\Big) \\
%&=& - \sum_x p(x)\log p(x) + \sum_x\Big(-\sum_y p_y(x) \log p_y(x)\Big) \\
&=& H(X) + \sum_x p(x)H(\rho_x).  
\end{eqnarray*}

Together with subadditivity, the calculation of the joint entropy of a cq state allows a simple proof of the convexity of entropy \cite{cn}.
\begin{property}[Convexity of entropy]
\[\sum_x p_x H(\rho_x) \leq H\Big(\sum_x p_x\rho_x\Big).\]
\end{property}
\begin{proof}
Consider the cq state $\rho^{XA}$ from (\ref{cqexample}).  Beginning with subadditivity, 
we have
\begin{eqnarray*}
H(X) + \sum_x p(x)H(\rho_x) &=& H(XA)_\rho \\
&\leq& H(X) + H(A) \\
&=& H(X) + H\Big(\sum_x p_x(\rho_x)\Big).
\end{eqnarray*}
Subtracting $H(X)$ from each side completes the argument.
\end{proof}
\begin{property}[Invariance of entropy]
For any density matrix $\rho^A$ and any isometry 
$\CV\colon A\rightarrow B$,
\[H(A)_\rho = H(B)_{\CV(\rho)}.\]
\end{property}
\begin{proof}
The eigenvalues of $\rho$ and of $\CV(\rho)$ are the same.
\end{proof}

%\section{Maximal entanglement}
%Let us take a slight detour 

\section{Conditional entropy} \label{section:conditionalentropy}
Let us begin by making the following formal definition for the \emph{conditional entropy}
\[H(A|B) = H(AB) - H(B).\]
By the calculation of $H(XA)$ for the cq state (\ref{cqexample}) from the previous section,  
\[H(A|X) = \sum_x p(x)H(\rho_x).\]
Observe that $H(A|X)$ is equal to the the average entropy of $A$, averaged over the classical part of the cq state.  In classical information theory, conditional entropy is often \emph{defined} as
\[H(Y|X) = -\sum_{xy} p(x,y) \log p(y|x).\]
It interesting to note that if we start with the cq state $\rho^{XA}$ from (\ref{cqexample}), we may define a random variable $Y$ which is jointly distributed with $X$ in accordance with the conditional distribution 
$p(y|x)\equiv p_x(y)$, using the notation from (\ref{rhospec}).  The equality
$H(A|X\!=\!x) = H(Y|X\!=\!x)$ holds, and thus  
 $H(A|X) = H(Y|X)$ holds as well.  

However, for an arbitrary state on $AB$, this interpretation of $H(A|B)$ as an average entropy is not valid.  In particular, suppose that $|A|=|B|=2$, and that $AB$ is in a pure state  
\[\ket{\Psi}^{AB} = \frac{1}{\sqrt{2}}\big(\ket{0}^A\ket{0}^B + \ket{1}^A\ket{1}^B\big).\]
Since $\tr_A \Psi = \pi^B$, it follows that for this state,
\[H(A|B) = H(\Psi) - H(\pi^B) = 0 - 1 = -1.\]
Defined in this formal way, conditional entropy can in fact be negative!  
As we will see in Sections~\ref{section:coherentinformation} and \ref{section:lsd}, the negative of $H(A|B)$, referred to as the \emph{coherent information}, plays a role in characterizing the quantum capacity of a quantum channel. 

Let us conclude our discussion by noting the following property of conditional entropy.  A proof can be found in \cite{cn}.
\begin{property}
$H(A|B)_\rho$ is concave as a function of $\rho^{AB}$.
\end{property}

\section{Mutual Information} \label{section:mutualinformation}
Given two random variables $X$ and $Y$, jointly distributed according to $p(x,y)$, the \emph{mutual information} $I(X;Y)$ measures the amount of correlation between the two random variables.  $I(X;Y)$ is typically defined as an expected log likelihood ratio
\[I(X;Y) = \sum_{xy}p(x,y)\log\frac{p(x,y)}{p(x)p(y)}.\]
Simple algebraic manipulations yield the following alternative formulas for $I(X;Y)$.
\begin{eqnarray*}
I(X;Y) &=& H(X) + H(Y) - H(XY) \\
&=& H(X) - H(X|Y) \\
&=& H(Y) - H(Y|X).
\end{eqnarray*}
% The last two expressions give rise to a particularly satisfying interpretation of mutual information.  For instance, since $H(X)$ represents the total amount of uncertainty associated with $X$, and $H(X|Y)$ is the reduction in uncertainty of $X$ obtained by knowledge of $Y$, $I(X;Y)$ can then be thought of as quantifying how much is learned about $X$ when only $Y$ is known.
Given a stochastic matrix $p(y|x)$ of conditional probabilities, further denotation of an input distribution $p(x)$ determines a joint distribution $p(x,y)$ for the random variables $X$ and $Y$.  
In Section~\ref{section:classicalcap}, we will see that the capacity of a classical channel with transition matrix $p(y|x)$ is given by the expression
\[C = \max_{p(x)} I(X;Y).\]
A similar expression can be given for the capacity of a 
c~$\rightarrow$~q channel, in terms of the \emph{quantum mutual information} evaluated on cq states.

Rather than define quantum mutual information in terms of a log-likelihood ratio, we opt here to give the following algebraic definition, valid for any composite quantum system $AB$.
\[I(A;B) = H(A) + H(B) - H(AB).\]
Using the formal definition of conditional quantum entropy from 
the previous section, we could have equivalently defined $I(A;B)$ as
\begin{eqnarray*}
I(A;B) &=& H(A) - H(A|B)
\end{eqnarray*}
or as
\begin{eqnarray*}
I(A;B) &=& H(B) - H(B|A).
\end{eqnarray*}
Most relevant to this dissertation is the evaluation of mutual information on a cq state such as 
\[\rho^{XB} = \sum_x p(x) \proj{x}^X\otimes\rho_x^B.\]
With respect to $\rho^{XB},$ let us evaluate
\begin{eqnarray*}
I(X;B) &=& H(B) - H(B|X) \\
&=& H\left(\sum_x p(x)\rho_x\right) -  \sum_x p(x) H(\rho_x).
\end{eqnarray*}
Together with the cq channel $\CX\rightarrow B$ defined by the conditional density matrices $\{\rho^B_x\}$, the cq state $\rho^{XB}$ represents the joint distribution on the input and output of the channel, serving the same purpose that $p(x,y) = p(x)p(y|x)$ did in the purely classical case.
In fact, an analogous capacity formula is obtainable as well
\[C = \max_{p(x)} I(X;B).\]
This capacity is easily computable, as a consequence of the first of the following two convexity properties enjoyed by $I(X;B)$.
\begin{property}
For a fixed cq channel $\CX\rightarrow B$ defined by the conditional density matrices $\{\rho^B_x\}$, $I(X;B)$ is a concave function of $p(x)$.
\end{property}
\begin{proof}
As the $\rho^B$ is linear in $p(x)$, and $H(B)$ is concave in $\rho^B$, $H(B)$ is concave in $p(x)$.  But $H(B|X)$ is linear in $p(x)$, completing the argument.
\end{proof}
\begin{property}
For a fixed input distribution $p(x)$, $I(X;B)$ is a convex function of the cq channel $\CX\rightarrow B$. 
\end{property}
\begin{proof}
This follows because $H(X|B)$ is a concave function of $\rho^{XB}$, which is itself linear in the conditional density matrices $\{\rho_x^B\}$.
\end{proof}

% once an input distribution $p(x)$ is designated, we may write down the joint cq density matrix $\rho^{XB}$, having the form used above.  The quantum mutual information can be written in the same way as we did classically, via
% \begin{eqnarray*}
% I(X;B)_\omega &=& H(X) + H(B) - H(XB) \\
% &=& H(B) - H(B|X) \\
% &=& H(X) - H(X|B).
% \end{eqnarray*}
% While we have not yet discussed the form $H(X|B)$, where the conditioning variable is quantum but the main one is classical, this expression is actually known to yield an achievable rate for classical data compression with quantum side information as was demonstrated by Devetak and Winter [CITE], who studied distributed data compression of classical quantum states.  The second expression is the form of the ``Holevo $\chi$ quantity"
% \[\chi\big(\{p(x),\omega_x\}\big) = H\left(\sum_x p(x)\omega_x\right) -  \sum_x p(x) H(\omega_x),\]
% which, in [CITE],  Holevo proved to be an upper bound to the amount of classical information about a the preparation of a quantum system which could extracted by quantum measurement.  Holevo [CITE], and subsequently Schumacher and Westmoreland [CITE], proved that this bound was in fact achievable, by utilizing random coding arguments showing that for large blocks of preparations, the Holevo bound could be asymptotically achieved.  More specifically, they proved that the classical capacity of a cq channel is given by the maximum of the quantum mutual information over all input distributions.
For an arbitrary quantum channel $\CN:A\rightarrow B$, specification of a collection of input states $\{\rho^{A}_x\}$, or equivalently, of a cq channel with those conditional density matrices, yields a new cq channel $\CX\rightarrow B$ with conditional density matrices $\{\CN(\rho_x)\}$.  This channel
is mathematically equivalent to the composed actions of the cq and quantum channels.  By the discussion above, optimization over input distributions $p(x)$ then gives the classical capacity of the newly constructed cq channel.  However, the ultimate capacity of the quantum channel involves an optimization over collection of input states.  Concavity of quantum mutual information in the input ensemble implies that extremal ensembles maximize capacity; such are ensembles of pure states.  However, whether or not a single-letter converse can be obtained in this case remains a very important open problem in quantum information theory.  As a result, the best known characterization of the capacity of a quantum channel for the transmission of classical information is 
\[C(\CN) = \lim_{k\rightarrow \infty} \frac{1}{k}
\max_{XA^k} I_c(X;B^k)_\sig\]
where for each $k$, the maximization is over all pure state ensembles 
$\{p(x),\ket{\phi_x}^{A^k}\}$ consisting of 
$|\CX|\leq \min\{|A|,|B|\}^{2k}-1$ states. The mutual information is evaluated with respect to the corresponding cq states
\[\sig^{XB^k} = \sum_x p(x)\proj{x}\otimes \CN^{\otimes k}(\phi_x^{A'^k}).\]
A state such as $\sig$ will be said to \emph{arise from the channels} $\CN^{\otimes k}$ in the above sense.

\section{Coherent Information} \label{section:coherentinformation}
Suppose a channel $\CN\colon A'\rightarrow B$ is given. Fix an isometric extension $\CU_\CN\colon A'\rightarrow BE$, and let $\CN^c = \tr_B\CU_\CN$ the associated complementary channel.
For a given input density operator $\rho^{A'}$, the \emph{coherent information} is defined as 
\[I_c(\rho,\CN) = H(\CN(\rho)) - H(\CN^c(\rho)).\]
Since any two complementary channels are equivalent up to an isometry on $E$, 
and since isometries preserve entropy, 
this quantity is independent of the particular complementary channel $\CN^c$ chosen for the calculation.  $H(\CN^c(\rho))$ is frequently referred to as the $\emph{entropy exchange}$ associated with sending a system with density matrix $\rho$ over the channel $\CN$.

Coherent information can be used to characterize the capacity of a quantum channel for transmitting quantum information as 
\[Q(\CN) = \lim_{k\rightarrow \infty} \frac{1}{k}
\max_{\rho^{A'^k}} I_c(\rho,\CN^{\otimes k}).\]
In Section~\ref{section:lsd} we will give an operational definition of quantum capacity, as well a discussion of the proof of this capacity formula.
It should be noted that this multi-letter characterization is the most general expression known for an arbitrary quantum channel. However, as we illustrate in Sections~\ref{section:degradable} and \ref{section:dephasing}, there are classes of channels for which a single-letter expression suffices. 

%These points will be discussed in detail in Section [?], where we give a number of equivalent ways of defining the task of sending quantum information, as well as a brief outline of the proof.

%Due to the obvious analogy between coherent information and the classical mutual information, is desirable to write it in a form which resembles its classical counterpart. 
Let us now explore other ways of writing $I_c(\rho,\CN)$.  With respect to the joint output-environment state $\CU_\CN(\rho)$ on $BE$, observe that \[H(\CN(\rho)) = H(B)\,\,\text{ and } \,\,H(\CN^c(\rho)) = H(E).\]  Then, 
\[I_c(\rho,\CN) = H(B) - H(E).\]
It is possible to write this quantity without making explicit mention of the environment.  To do this, first fix any purification $\ket{\Psi}^{AA'}$ of $\rho^{A'}$.  Then, use this to write a global pure state 
\[\ket{\Om}^{ABE} = \CU_\CN\ket{\Psi}^{AA'}.\]
Since $\ket{\Om}^{ABE}$ is pure, it follows that $H(E) = H(AB)$. This allows us to rewrite 
\[H(B) - H(E) = H(B) - H(AB) = -H(A|B).\]
\begin{rem}
Written this way, it is clear that coherent information can be positive or negative.  However, $Q(\CN) \geq 0$ for every channel $\CN$, as $I_c(\ket{\phi}^{A'},\CN) = 0$ for every pure state $\ket{\phi}^{A'}$.
\end{rem}

Observe that since any two purifications of $\rho^{A'}$ are the same up to local unitaries on $A$, and such unitaries preserve $H(AB)$, this last expression is independent of the particular purification $\ket{\Psi}^{AA'}$ chosen for $\rho^{A'}$.  Further note that to compute $-H(A|B)$, it suffices to consider the joint state 
\[\omega^{AB} = \tr_E\Om^{ABE} = \CN(\Psi^{AA'}).\] 
%\[\tr_{BE}\Psi = \rho,\,\, \tr_E\Psi = \omega, \,\,
%\tr_{AE}\Psi = \CN(\rho), \,\, \tr_{AB}\Psi = \CN^c(\rho)\]
It is common to write 
\[I_c(A\,\rangle B)_\omega \equiv -H(A|B)_\omega\]
acknowledging the directionality of coherent information \emph{from} $A$ \emph{to} $B$.   While we will freely interchange the two notations for coherent information throughout this dissertation, we will generally write $I_c(A\,\rangle B)$ when characterizing capacity regions and proving the converses for the main theorems, while the notation $I_c(\rho,\CN)$ will be utilized more frequently in the coding theorems.

Let us review a few facts concerning coherent information.
\begin{property}
For a maximally entangled state 
\[\ket{\Phi}^{AB} = \frac{1}{\sqrt{k}}\sum_{i=1}^k\ket{i}^A\ket{i}^B\]
we have 
\[I_c(A\,\rangle B)_\Phi = \log k.\]
\end{property}
\begin{proof}
\[H(B)_\Phi - H(AB)_\Phi = H(\pi_k) - H(\Phi) = \log k - 0.\]
\end{proof}

\begin{property}
For any state on $AB$, 
\[I_c(A\,\rangle B) \leq \min\{H(A),H(B)\}.\] 
\end{property}
\begin{proof}
We begin by observing that
\begin{eqnarray*}
I_c(A\,\rangle B) &=& H(B) - H(AB) \leq H(B)\\
&\leq& H(B).
\end{eqnarray*}
To see that $I_c(A\,\rangle B) \leq H(A)$, we start with Lieb's inequality
\[|H(B) - H(A)| \leq H(AB).\]
Getting rid of the absolute value and subtracting $H(B)$ from each side yields
\[-H(A) \leq H(A|B).\]
Multiplying the sides by $-1$ completes the argument.
\end{proof}
\begin{property}
For any channel $\CN\colon A'\rightarrow B$ and any $\rho^{A'}$, 
\[I_c(\rho,\CN) \leq \log|A'|.\]
\end{property}
\begin{proof}
Fix a purification $\ket{\Phi}^{AA'}$ of $\rho^{A'}$.  Then 
\begin{eqnarray*}
I_c(\rho,\CN) &=& I_c(A\,\rangle B)_{\CN(\Phi)} \\
&\leq& H(A)_\Phi \\
&=& H(A')_\Phi \\
&\leq& \log|A'|.
\end{eqnarray*}
\end{proof}
%We mentioned in Section~\ref{section:conditionalentropy} that it is possible for $H(A|B)$ to be negative if the joint state of $AB$ is entangled.  
\begin{property}
For fixed $\rho^{A'}$,
$I_c(\rho,\CN)$ is a convex function of $\CN$. 
\end{property}
\begin{proof}
Fixing a purification $\ket{\Psi}^{AA'}$ of $\rho^{A'}$,
observe that the state $\omega^{AB} = \CN(\Psi)$ is linear function of $\CN$.  But $H(A|B)$ is concave in $\omega^{AB}$, and thus in $\CN$, so $I_c(\rho,\CN) = -H(A|B)$ is convex in $\CN$. 
\end{proof}
\begin{rem}
This property is in close agreement to the corresponding statement that $I(X;Y)$ is convex in $p(y|x)$.  However, $I(\rho,\CN)$ is not generally concave or convex in $\rho$, for a given fixed $\CN$.
\end{rem}
\section{Conditional coherent information}
\label{section:condcoherentinformation}
In the appendix (Section~\ref{section:instrumentcomplement}), we show that if $\boldsymbol{\CN}\colon A'\rightarrow BX$ is an instrument with components $\{p(x)\CN_x\}$, then any
isometric extension $\CU\colon A'\rightarrow BEX$ of $\boldsymbol{\CN}$ can be expressed as 
\[\CU = \sum_x \sqrt{p(x)}\ket{x}^X\ket{x}^{X'}\otimes\CU_x,\]
where the $\CU_x\colon A'\rightarrow BE'$ are isometric extensions of the $\CN_x$, and $E\equiv E'X'$.  We also verify there that $\tr_E\CU = \boldsymbol{\CN}$, while $\tr_{BX} \CU = \boldsymbol{\CN^c}$, where
\[\boldsymbol{\CN^c} = \sum_x p(x)\proj{x}^{X'}\otimes \CN^c_x.\]
Above, each component $p(x)\CN_x^c$ is formed from a complement $\CN_x^c\colon A'\rightarrow E'$ of the corresponding normalized component $\CN_x$ of $\boldsymbol{\CN}$.
The main observation here is that the environment $E=E'X'$ of the instrument includes the common environment $E'$ to the component channels $\CN^c_x$ \emph{as well as} a part $X'$ which purifies the classical component $X$ of $\boldsymbol{\CN}.$

For any $\rho^{A'}$, the coherent information over $\boldsymbol{\CN}$ can thus be expressed as 
\begin{eqnarray*}
I_c(\rho,\boldsymbol{\CN}) 
&=& H\big(\boldsymbol{\CN}(\rho)\big) - H\big(\boldsymbol{\CN}^c(\rho)\big)  \\
&=& H\Big(\sum_x p(x)\proj{x}^X\otimes\CN_x(\rho)\Big)
- H\Big(\sum_x p(x)\proj{x}^{X'}\otimes\CN^c_x(\rho)\Big) \\
&=& H(X) + \sum_xp(x)H\big(\CN_x(\rho)\big)
- H(X) -  \sum_xp(x)H\big(\CN^c_x(\rho)\big) \\
&=& \sum_x p(x) I_c(\rho,\CN_x).
\end{eqnarray*}
In the third line, we mirror the calculation of the entropy of a cq state performed in Section~\ref{section:conditionalentropy}.
The coherent information over $\boldsymbol{\CN}$ is thus just the average of the coherent information over each $\CN_x$. 
Another way to see this is to note that 
\begin{eqnarray*}
I_c(\rho,\boldsymbol{\CN}) 
&=& H\big(\boldsymbol{\CN}(\rho)\big) - H\big(\boldsymbol{\CN}^c(\rho)\big)  \\
&=& H(BX) - H(E'X') \\
&=& H(B|X) - H(E'|X).
\end{eqnarray*}
A third derivation 
fixes a purification $\ket{\Psi}^{AA'}$ of $\rho^{A'}$
and defines the state 
\begin{eqnarray*}
\ket{\Omega}^{ABEX} &\equiv& \ket{\Omega}^{ABE'X'X} \\
&=& \CU\ket{\Psi^{AA'}},
\end{eqnarray*}
noting that  
\begin{eqnarray*}
H(BX) - H(E) &=& H(BX) - H(ABX) \\
&=& -H(A|BX)\\
&=& I_c(A\,\rangle BX).
\end{eqnarray*}

\chapter[Single-user capacity theorems]{Capacity theorems for single-user channels} \label{chapter:capacity}
In this chapter we recall various existing capacity theorems from the literature.  After reviewing the proof of the capacity theorem for a classical channel, we will see that the main ingredients of that proof have counterparts for quantum channels, both for the transmission of classical and of quantum information.  The common element to all of the situations is as follows.  Each assumes that the sender and receiver are able to transmit an unlimited number of times over a collection of identical channels. It is useful to think of these channels as acting in parallel, as sequential transmissions can be thought of as parallel transmissions ``in time".  After giving an operational definition of a set of rates at which the sender can communicate to the receiver arbitrarily well, the capacity is then \emph{defined} to be the supremum, or least upper bound, of those achievable rates, representing the ultimate rate at which arbitrarily reliable communication can occur, provided that the channel can be used any number of times. The capacity is then described, or \emph{characterized}, in terms of some optimization of entropic quantities
over a well-defined collection of classical probabilities or quantum states.
Later, when we characterize various capacity regions for quantum multiple access channels, we will invoke the single-user coding theorems for quantum channels introduced in this chapter.

\section{Classical capacities of classical channels} \label{section:classicalcap}
Suppose that two parties, Alice and Bob, are connected by a large number of identical classical channels with probability transition matrix $p(y|x)$.  This is to be interpreted as follows.  At any given time, Alice can choose to send a symbol $x\in\CX$ to Bob.  Because of noise, Bob ``hears" a corrupted version of the symbol $x$.  Specifically, he receives the symbol $y\in \CY$ with the conditional probability $p(y|x)$. 
Fixing a probability distribution $p(x)$ on Alice's input symbols defines a random variable $X$.  Together with the conditional probabilities $p(y|x)$, this yields a joint distribution $p(x,y)$ of a pair of correlated random variables $X$ and $Y$.  The classical capacity of the channel $p(y|x)$ is the logarithm of the number of distinguishable inputs, whereby Alice uses the channel many times to send Bob a message which he can ascertain arbitrarily well.  Shannon \cite{shannon} gave the following formula for the capacity:
\begin{equation}
C = \max_{p(x)} I(X;Y).\label{ccap}
\end{equation}
Mathematically, he proved that this expression equals a certain operationally defined capacity which we now review.
Suppose Alice tries use the channel $n$ times to send information to Bob at a rate of $R$ bits per channel use.  To this end, she selects a collection of \emph{codewords}, consisting of $2^{nR}$ sequences of input symbols $x^n(m)$, one sequence for each message she would like to send, and reveals them to Bob. 
This can be modeled by an \emph{encoding function} 
\[f\colon 2^{nR} \rightarrow \CX^n.\]
Since the channel is noisy, Bob will receive a noisy version of Alice's message, denoted $Y^n(m)$.  Let the decoding function 
\[g\colon \CY^n\rightarrow 2^{nR}\]
describe some scheme by which Bob attempts to decide which message Alice had intended for him to receive.  Using this scheme, Alice and Bob have effectively created a new channel 
\[Q(\h{m}|m) = \sum_{y^n\in g^{-1}(\widehat{m})}p(y^n|f(m)),\] whereby each message $m\in 2^{nR}$ Alice may choose to send induces a distribution on the possible messages Bob may decode.  We might allow Alice to use a stochastic encoder $p(x^n|m),$ in which case the effective channel would be 
\[Q(\h{m}|m) = \sum_{y^n\in g^{-1}(\widehat{m})}
\sum_{x^n}p(y^n|x^n)p(x^n|m).\]
If Alice sends the message $m\in 2^{nR}$, the probability Bob decodes the message incorrectly can be expressed in a number of ways: 
\begin{eqnarray*}
P_e(m) &\equiv& \Pr\{\h{M} \neq m|M=m\} \\
&=& \Pr\{g(Y^n(m)) \neq m\} \\
&=& 1-Q(m|m) \\
&=& \sum_{\stackrel{\h{m}\in 2^{nR}}{\h{m}\neq m}} Q(\h{m}|m).
\end{eqnarray*}
Associated to the coded channel $Q(\h{m}|m)$ is its \emph{maximal probability of error}
\[P_{\text{max}} = \max_{m\in{2^{nR}}} P_e(m)\]
and its \emph{average probability of error}
 \[P_{\text{ave}} = 2^{-nR} \sum_{m\in 2^{nR}} P_e(m).\]

One may phrase the goal of successful communication as that of \emph{simulating} a fictitious identity channel 
id$\colon 2^{nR} \rightarrow 2^{nR}$ from Alice to Bob, 
where id$(\h{m}|m) = \delta_{\h{m},m}$.  Perfect simulation would amount to using a zero-error code.  Approximate simulation can be gauged in a number of ways.  For example, one could require that either $P_{\text{ave}}$ or $P_{\text{max}}$ is small.  Clearly, the former will imply the latter.  

Suppose that Alice chooses her message $M$ randomly according to the distribution $P(m)$.  If she sends her message through the identity channel to Bob, the two will hold a perfectly correlated pair of random variables $(M,M),$ distributed as
\[\text{dist}(M,M)_P(m,\widehat{m}) = P(m)\delta_{m,\widehat{m}}.\]
However, Alice will actually be sending through the coded channel $Q(\h{m}|m)$, generating a pair of noisy correlated random variables 
$(M,\h{M})$ distributed as 
\[\text{dist}(M,\h{M})_P(m,\widehat{m}) = P(m)Q(\widehat{m}|m).\]
One way to judge the success of the simulation is to consider the
the $\ell_1$~norm $\Delta(P)$ between the two distributions dist$(M,\widehat{M})_P$ and dist$(M,M)_P$.  This is calculated as 
\begin{eqnarray*}
\Delta(P) 
&=& \left|\text{dist}(M,M)_P - \text{dist}(M,\widehat{M})_P\right|_1 \\
&=& \sum_{m,\widehat{m}=1}^{\mu} 
 \big| P(m)\delta_{m,\widehat{m}} 
      - P(m)Q(\widehat{m} |m)\big| \\
&=& \sum_{m,\widehat{m}=1}^{\mu} P(m)
 \big|\delta_{m,\widehat{m}} 
      - Q(\widehat{m} | m)\big| \\
&=& \sum_{m=1}^{\mu} P(m)\Big( 
 ( 1 - Q(\widehat{m}|m)) 
   + \sum_{\stackrel{\widehat{m}=1}{\widehat{m}\neq m}}^{\mu} 
   Q(\widehat{m}| m)\Big) \\
&=& 2\sum_{m=1}^{\mu} P(m)P_e(m) \\
&=& 2\E_P P_e(M).
\end{eqnarray*}
In other words, the $\ell_1$ distance between the ideal and the actual joint distributions is precisely equal to twice the expected error 
probability.   Observe that 
\[\Delta(\text{unif}(2^{nR})) = 2P_\text{ave} \text{ and }\,
\Delta(\delta_m) = 2P_e(m).\]
Further note that requiring that the maximal error probability be less than $\epsilon$ is equivalent to demanding that 
$\Delta(\delta_m) \leq 2\epsilon$ for each $m$, where $\delta_m$ is a point distribution at $\{M=m\}$.  It is worth noting that the latter requirement is also equivalent to requiring that  $\Delta(P) \leq 2\epsilon$ for all distributions $P(m)$.

So, communication can be viewed in the light of generating near perfect common randomness over noisy quantum channels.  We have phrased things in this way as it makes the road to quantum communication a bit easier.  Rather than asking the sender and receiver to end up with classical correlations, we will see later in Section~\ref{section:lsd} that they attempt to build \emph{quantum correlations}.

Any code $(f,g)$ which encodes $2^{nR}$ messages using $n$ instances of a channel $p(y|x)$ such that $P_e(m) \leq \epsilon$ for all $m\in 2^{nR}$ will be called an $(R,n,\epsilon)$ maximal error code for the channel $p(y|x)$.  A rate $R$ is said to be \emph{achievable} if there exists a sequence of $(R,n,\epsilon_n)$ maximal error codes with $\epsilon_n\rightarrow 0$.  The (operational) capacity of the channel $p(y|x)$ is then defined to be the supremum of the set of achievable rates.  Shannon's capacity theorem states that this operationally defined capacity is equal to the 
number $C$, defined in (\ref{ccap}).

\begin{figure}
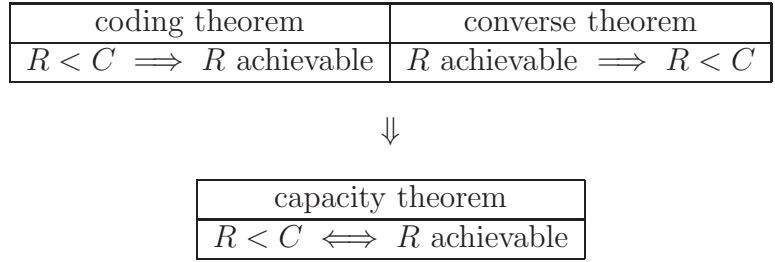
 \label{fig:captheocomp}
\center
%\begin{tabular}{c}
  \begin{tabular}{|c|c|}
  \hline coding theorem & converse theorem \\ \hline 
   $R<C  \implies R$ achievable & $R$ achievable $\implies R < C$ \\ \hline 
  \end{tabular}  
  \[\Downarrow\] 
\begin{tabular}{|c|}
\hline capacity theorem \\ \hline  $R<C \iff R$ achievable \\ \hline
\end{tabular}
\caption{Components of a capacity theorem}
\end{figure}

The channel capacity theorem is proved in two main parts.  First, it is proven that for any rate $R < C$, $R$ is achievable. This is provided by a \emph{coding theorem}, which is generally structured as follows.  Given $\epsilon > 0$ and some rate $R < C$, it is shown that there is a long enough blocklength $n$ so that there exists an $(R,n,\epsilon)$ code.  As $\epsilon$ was arbitrary, this immediately implies the existence of a sequence of such codes which achieves the rate $R$, corresponding to any sequence of error probabilities which go to zero.  The second component is called the \emph{converse}.  In this part, it is shown that every achievable rate $R$ satisfies $R<C$.  These components are summarized in Figure~\ref{fig:captheocomp}.

One route to proving the coding theorem involves first showing that codes with a weaker error constraint exist.  Rather than requiring that every message have a low error probability, it is sufficient to show that the error probability, averaged over all codewords $m\in 2^{nR}$ is small.
A code satisfying this weaker constraint will be called an \emph{average error code.}  
A way to prove such a coding theorem is through the technique of random coding.  
For an arbitrary distribution $p(x)$, define the product distribution 
$p(x^n) = \prod_{i=1}^n p(x_i)$.  A rate $R$ random encoder is then defined by randomly selecting $2^{nR}$ codewords 
\[\CC = \{X^n(1),\dotsc,X^n(2^{nR})\}\]
i.i.d.\ according to $p(x^n)$.  The following coding proposition, or some variant thereof, is proved in many textbooks on information theory, such as in \cite{coverthomas,ck}.
\setcounter{prop}{-1}
\begin{prop}[Classical channel coding theorem]
Given is a channel $p(y|x)$, an input distribution $p(x)$, and a number 
$0 \leq R < I(X;Y)$, where $I(X;Y)$ is computed with respect to $p(x,y) = p(x)p(y|x)$.  For every $\epsilon > 0$, there is $n$ sufficiently large so that if $2^{nR}$ codewords $\CC = \{X^n(1),\dotsc,X^n(2^{nR})\}$  are chosen i.i.d.\ according to the product distribution $p(x^n) = \prod_i p(x_i)$, there exists a decoding function $g\colon \CY^n\rightarrow 2^{nR}$
which depends on the random choice of codebook $\CC$ and correctly identifies the input message with expected average probability of error less than $\epsilon$, in the sense that 
\[\E_\CC 2^{-nR}\sum_{m\in 2^{nR}} \Pr\{g(Y^n(m)) = m\} \geq 1-\epsilon.\]
\end{prop}
Observe that, because of the symmetry in the code construction, the expectation of each term in the above summation is the same.  It is thus possible to reexpress that error condition as 
\[\E_\CC  \Pr\{g(Y^n(1)) = 1\} \geq 1-\epsilon,\]
showing that at the level of random codes, one may assume that the message $m=1$ has been sent without losing any generality.

It is a simple task to ``derandomize" any code which is guaranteed to exist by Proposition~0.  Suppose that Alice chooses a message uniformly distributed on the set $\{1,\dotsc,2^{nR}\}$, represented by the random variable $M$, to send to Bob.  Then 
\begin{eqnarray*}
\E_\CC\Pr\{g(Y^n(M) = M)\}
&=& 2^{-nR}\sum_{m\in 2^{nR}} \E_\CC \Pr\{g(Y^n(m) = m)\} \\
&\geq & 1-\epsilon. 
\end{eqnarray*}
It is then immediate that there must exist a particular deterministic code yielding an average probability of success at least as large as $1-\epsilon$.

So far, this is enough to conclude that every input distribution $p(x)$ yields a lower bound to the average error capacity of $p(y|x)$.  This is because each $p(x)$ corresponds to a set of achievable rates $\{R : 0\leq R < I(X;Y)\},$  and the largest such set is given by optimizing over all $p(x)$.

Recall that we have defined the operational capacity $C$ in terms of the maximal probability of error constraint.  However, we have only outlined how to show that codes with low average error exist.  By Markov's inequality from probability theory, if the average error probability is less than $\epsilon$, then at least half of the codewords have an error probability less than $\sqrt{\epsilon}$.  By only using these codewords, 
a rate $R-\frac{1}{n}$ code with maximal error probability $\sqrt{\epsilon}$ is obtained, and thus every rate less than $R$ is achievable with maximal error, showing that the maximal and average error capacities are the same.   

While the coding proposition implies the existence of sequences of codes achieving any rate less than capacity, it remains to prove that no such sequences exist for rates above capacity.  Rather than reproduce the entire converse theorem, we outline the basic structure of the theorem.  First, one assumes that $R$ is an achievable rate.  This means that there should exist a sequence of $(2^{nR},n,\epsilon_n)$ codes with $\ep_n\rightarrow 0$.  For any $n$, let 
$p(x^n,y^n) = p(x^n)\prod_i p(y_i|x_i)$ be the joint distribution on $X^n$ and $Y^n$ induced by selecting codewords uniformly at random from the corresponding code in the sequence.  An initial step in the proof shows that 
\[R < \frac{1}{n}I(X^n;Y^n) + \epsilon'_n\]
where $\ep'_n\rightarrow 0$  as $\ep_n\rightarrow 0$, and $I(X^n;Y^n)$
is evaluated with respect to the induced distribution $p(x^n,y^n)$.  
For \emph{any} joint distribution on $X^n$ and $Y^n$, the following can be easily proved:
\[\frac{1}{n}I(X^n;Y^n) \leq \frac{1}{n}\sum_{i=1}^n I(X_i;Y_i)
\leq \max_i I(X_i;Y_i).\]
If $i^*$ achieves the maximum on the right hand side, the marginal distribution $p(x_{i^*})$ provides a ``witness" to the fact that the rate $R$ is in fact achievable ($R$ is thus less than the maximum mutual information over all input distributions).  This proves that the capacity formula is \emph{additive}, and thus that every achievable rate is upper bounded by the solution of a ``single-letter" optimization problem.  For this reason, this second conceptual step in the converse is known as single-letterization.  Without it, one would only be able to write the capacity as 
\[C = \lim_{k\rightarrow \infty} \frac{1}{k} \max_{p(x^k)} I(X^k;Y^k)\]
a result which follows by applying Proposition~0 to extensions of the channel \[p(y^k|x^k) = \prod_{i=1}^k p(y_i|x_i).\]  Such an expression has become known as a ``regularized" expression for the capacity.
Actually, this is a persistent problem in quantum information theory.  The best known expressions characterizing the capacities of an arbitrary quantum channel to transmit classical or quantum information are regularized maximizations of information quantities over appropriate sets of input states.

\section{Classical capacities of quantum channels}
\label{section:hsw}
Suppose that Alice and Bob are connected via some large number $n$ of instances of a quantum channel $\CN$, and that Alice wishes to transmit classical messages to Bob.  The overall maximal rate at which is this is possible is the \emph{classical capacity} $C(\CN)$ of the channel $\CN$, which is the logarithm of the number of physical input preparations Alice can make, per channel use, so that Bob can distinguish them arbitrarily well by measuring the induced states at the outputs of the channels.
The best known expression for the classical capacity of a quantum channel,
due to Holevo \cite{holcap} Schumacher and Westmoreland \cite{sw2}, is the following regularized formula, known as the HSW Theorem:
\[C(\CN) = \lim_{k\rightarrow \infty}\frac{1}{k}\max_{XA'^k}I(X;B^k)_\omega.\]
Here, the maximization is over all pure state input ensembles 
$\{p(x),\ket{\phi_x}^{A'^k}\}$ of states for Alice to prepare at the inputs to $k$ parallel instances of the channel $\CN$.  For a given ensemble, the mutual information is computed relative to the corresponding cq state
\[\omega^{XB^k} = \sum_m \proj{x}^X\otimes\CN^{\otimes k}(\phi_x).\]   
Operationally, the classical capacity of $\CN$ is defined in analogy to that of a classical channel.  A $(2^{nR},n)$ code consists of 
$2^{nR}$ message states $\{\ket{\phi_1}^{A'^k},\dotsc,\ket{\phi_{2^{nR}}}^{A'^k}\}$ for Alice and a corresponding measurement for Bob, mathematically modeled as POVM with $2^{nR}$ outcomes $\{\Lambda_m\}_{m\in 2^{nR}}$.  We call this code an 
$(2^{nR},n,\epsilon)$ code if the following constraint on success probability, averaged over all messages, is satisfied:
\[2^{-nR}\sum_{m\in 2^{nR}} \tr \Lambda_m\CN^{\otimes n}(\phi_m) \geq 1-\epsilon.\]
A rate $R$ is achievable if there exists a sequence of $(2^{nR},n,\epsilon_n)$ codes with $\epsilon_n\rightarrow 0$, and the capacity $C(\CN)$ is the supremum of all achievable rates. 

As with the capacity of a classical channel, the proof that $C(\CN)$ 
can be expressed in such a regularized form has two parts, a coding theorem and a converse.  The following coding theorem is attributed to Holevo \cite{holcap}, Schumacher and Westmoreland \cite{sw2}.
\begin{prop}[HSW Theorem] \label{prop:hsw}
Given is a cq state
$\sigma^{XB}=\sum_x p(x)\proj{x}^X\otimes\rho^B_x$ and a number $0\leq R < I(X;B)_\sigma.$ For every $\epsilon > 0$, there
is $n$ sufficiently large so that if $2^{nR}$ codewords $\CC=\{X^n(m)\}$ are chosen i.i.d.\ according to the product distribution $p(x^n) = \prod_{i=1}^n p(x_i)$, corresponding to input preparations
\[\rho_{x^n} = \rho_{x_1}\otimes\cdots\otimes\rho_{x_n},\] there exists a decoding POVM $\{\Lambda_m\}$ on $B^n$ which depends on the random choice of codebook $\CC$ and correctly identifies the index $m$ with average probability of error less than $\epsilon,$ in the sense that
\begin{eqnarray}
\E_\CC 2^{-nR}\sum_{m=1}^{2^{nR}} \tr\rho_{X^n(m)} \Lambda_m
\geq 1-\epsilon.
\end{eqnarray}
\end{prop}
Due to the symmetry of the distribution of $\CC$ under codeword permutations, it is clear that the expectations of each term in the above sum are equal.  In other words,
\begin{eqnarray}
\E_\CC 2^{-nR}\sum_{m=1}^{2^{nR}} \tr\rho_{X^n(m)} \Lambda_m
= \E_\CC \tr\rho_{X^n(1)} \Lambda_1, \label{symmetry}
\end{eqnarray}
The arguments for derandomization and for obtaining a good maximal error code are identical to those used for classical channels in the previous section.  

A proof of the converse begins, as before, by assuming that $R$ is an achievable rate.  Taking a cq state $\omega^{XB^n}$ induced by an $(R,n,\epsilon_n)$ code in the achieving sequence, Fano's inequality (Lemma~\ref{lemma:fano}) and the Holevo Bound (Lemma~\ref{lemma:holevo}) are used
\footnote{These details are given more explicitly in the converse proofs of the main theorems (Section~\ref{section:converses}).}
to show that \[R < \frac{1}{n}I(X;B^n)_\omega,\] 
where again $\ep_n'\rightarrow 0$ as $\ep_n\rightarrow 0$.
However, it is an important open problem as to whether a single-letterization step can be proved.  No counterexample to additivity is known, and it is widely believed that none exists. 

\section{Quantum capacities of quantum channels}
\label{section:lsd}
The quantum capacity $Q(\CN)$ of a quantum channel $\CN\colon A'\rightarrow B$ is the answer to a number of physical questions regarding the possibilities of performing various operational information processing tasks over many parallel instances of the channel $\CN$.  $Q(\CN)$  
is the logarithm of various quantities:
\begin{itemize}
\item the amount of entanglement that can be created
(entanglement generation)
\item the amount of entanglement that can be sent
(entanglement transmission)
\item the size of a Hilbert space all of whose states can be reliably transmitted 

(subspace transmission)
\item the size of a Hilbert space all of whose entangled states can be reliably transmitted (strong subspace transmission).
\end{itemize}
All of these quantities have units of \emph{qubits per channel use}, and as the rates at which these tasks are possible all coincide, it is justifiable to say that they all represent ``sending quantum information,"
and hence to speak of a single quantum capacity $Q(\CN).$  The best known characterization of the quantum capacity is a regularized maximization of the coherent information 
\[Q(\CN) = \lim_{k\rightarrow \infty}\frac{1}{k}\max_{XA'} I_c(A\,\rangle B^k)_\omega,\] 
where for each $k$, the maximization is over all states of the form
\[\omega^{AB^k} = \CN^{\otimes k}(\Psi^{AA'^k}).\]
Such a state $\omega$ will be said to \emph{arise from} $\CN^{\otimes k}$ or rather, to arise from the action of $\CN^{\otimes k}$ on the bipartite pure state $\ket{\Phi}^{AA'^k}$.
Here, the regularization is known to be necessary for a general quantum channel, as opposed to the case with the classical capacity $C(\CN)$, 
where the existence of a single-letterization step in the converse is an open problem.  The existence of a counterexample to additivity is known \cite{ss}.

%Using the equivalent representation of the coherent information mentioned in [?], another way to write this characterization is as 
%\[Q(\CN) = \lim_{k\rightarrow \infty}\frac{1}{k}\max_{\rho^{A'^k}} I_c(\rho,\CN^{\otimes k}).\]

Of the different operational definitions of $Q(\CN)$, the simplest to describe is entanglement generation, since it can defined without explicit mention of encodings.  Suppose that a large number $n$ of channels $\CN\colon A'\rightarrow B$ are available from Alice to Bob.  Alice and Bob will use the channels to build a large maximally entangled state between degrees of freedom of some physical systems located in their respective laboratories. 
To this end, Alice prepares some bipartite pure state 
$\ket{\Upsilon}^{AA'^n}$, entangled between some system $A$ of dimension $|A| = 2^{nQ}$ in her laboratory, and the inputs $A'^n$ of the channels.  After the actions of the channels, Alice's system $A$ is correlated with the outputs $B^n$ of the channels quantum mechanically.  Bob 
then performs some post-processing procedure, modeled by a quantum operation $\CD\colon B^n\rightarrow \widehat{A}$, to transfer the quantum correlations from the outputs $B^n$ of the channels to an ``output" physical system $\widehat{A}$, also of dimension $|\h{A}| = 2^{nQ}$ in his laboratory.  Their goal is to produce a state which is close to some target maximally entangled state $\ket{\Phi}^{A\widehat{A}}$.  More specifically, we say that they generate entanglement at rate $Q$ if they produce a maximally entangled state of the form 
\[\ket{\Phi}^{A\widehat{A}} = \frac{1}{\sqrt{2^{nQ}}}\sum_{a\in 2^{nQ}}
\ket{a}^A\ket{a}^{\widehat{A}}.\] 
We will call such a state a \emph{rate $Q$ maximally entangled state}.  The blocklength $n$ will always be apparent from the context.

$(\ket{\Upsilon}^{AA'^n},\CD)$ will be called a $(Q,n,\epsilon)$ \emph{entanglement generation code} for the channel $\CN$ if, for the rate $Q$ maximally entangled state $\ket{\Phi}^{A\h{A}}$, we have
\[F\big(\ket{\Phi}^{A\h{A}},\CD\circ\CN^{\otimes n}(\Upsilon^{AA'^n})\big) \geq 1-\epsilon.\]
A rate $Q$ is an \emph{achievable rate for entanglement generation} over the channel $\CN$ if there exists a sequence of $(Q,n,\epsilon_n)$
entanglement generation codes with $\ep_n\rightarrow 0$.  
The \emph{entanglement generating capacity} $Q^{\text{eg}}(\CN)$ of $\CN$ 
is then defined operationally as the supremum of all such achievable rates.  

%The corresponding capacity theorem characterizes $Q(\CN)$ as a regularized maximization of coherent information 
%\[Q(\CN) = \lim_{k\rightarrow \infty} \frac{1}{k}\max_{\rho^{A'^k}}
%I_c(\rho,\CN^{\otimes k}).\]

We will now introduce a number of coding propositions from \cite{dev}, each a more refined version of the previous one.  While the first is sufficient to prove achievability for single-user channels, the others have additional properties which we will need later when we characterize various capacity regions of quantum multiple access channels. 

\begin{prop*}[Entanglement generation coding theorem]
Given is a channel $\CN\colon A'\rightarrow B$, a density matrix $\rho^{A'}$, and a number $0\leq Q < I_c(\rho,\CN).$  For every $\epsilon>0$, there is $n$ sufficiently large so that there is a $(Q,n,\epsilon)$ entanglement generation code $(\ket{\Upsilon}^{AA'^n},\CD)$ for $\CN$.
\end{prop*}

Recall the discussion in Section~\ref{section:coherentinformation} regarding the two different ways of expressing coherent information.  Given an input density operator $\rho^{A'}$, if $\ket{\Psi}^{AA'}$ is any purification of $\rho$, then the identity
\[I_c(\rho,\CN) = I_c(A\,\rangle B)_{\CN(\Psi)}\]
holds.  This proposition then guarantees that for every state $\omega^{AB} = \CN(\Psi)$ arising from the action of $\CN$ on a state $\ket{\Psi}^{AA'}$, every rate $0\leq Q< I_c(A\,\rangle B)_\om$ is an achievable rate.  This works by applying the coding theorem to the input state $\rho^{A'} = \tr_A \Psi$.

As with the classical capacity, it is also true that for each integer $k>0$, if $\omega'$ arises from 
$\CN^{\otimes k}$, then every rate $0\leq Q < \frac{1}{k}I_c(A\,\rangle B^k)_{\omega'}$ is achievable as well.  We then conclude that 
\[Q(\CN) \geq \lim_{k\rightarrow \infty}\frac{1}{k}\max I_c(A\,\rangle B^k).\]

The usual Shannon-theoretic prescription for converse theorems applies here as well, although as mentioned above, it known that a single-letterization step cannot be proved for arbitrary $\CN$.  Suppose that $Q$ is achievable, and fix a $(Q,n,\epsilon_n)$ entanglement generation code $(\ket{\Upsilon}^{AA'^n},\CD)$ in the achieving sequence of codes.  The encoding $\ket{\Upsilon}$ gives rise to the state 
$\omega^{AB^n} = \CN^{\otimes n}(\Upsilon)$.  
It is a simple consequence of the quantum data processing inequality (Lemma~\ref{lemma:dataproc}) and continuity of coherent information in the input density operator (Lemma~\ref{lemma:continuity}) that \footnote{these details are given more explicitly in the converse proofs of the main theorems (Section~\ref{section:converses}).}
\[Q \leq \frac{1}{n}I_c(A\,\rangle B^n)_\omega + \ep'_n\]
where $\ep'_n\rightarrow 0$.  By standard arguments we then conclude that 
\[Q(\CN) \leq \lim_{k\rightarrow \infty}\frac{1}{k}\max I_c(A\,\rangle B^k).\]

The state $\tr_A \Upsilon$ which is induced by Alice's encoding at the inputs $A'^n$ of $\CN^{\otimes n}$ is called the \emph{code density operator} of the entanglement generation code.  With randomization, it is possible to make this operator arbitrarily close to the product state 
$\rho^{\otimes n}$, where $\rho^{A'}$ is the input density matrix used when invoking the proposition.  If Alice and Bob have access to a shared source of randomness, they may utilize an ensemble of codes to this end.  
This is very useful for our multiple access coding theorems, as it guarantees that if one sender codes randomly, the induced channel seen by the other sender is close to a product channel, allowing coding theorems for product channels to be invoked.

A $(Q,n,\epsilon)$ \emph{random entanglement generation code} consists of a collection of deterministic $(Q,n,\epsilon)$ entanglement transmission codes
$(\ket{\Upsilon^\beta}^{AA'^n},\CD^\beta)$ and a probability distribution $P_\beta$, corresponding to a source of shared common randomness available to both sender and receiver.  We will often omit the subscript, once the randomness of the code has been clarified, and it will be understood that $\ket{\Upsilon}$ and $\CD$ constitute a pair of classically correlated random objects.
Associated to a random code is its expected, or average code density operator 
\[\varrho^{A'^n} = \E_\beta\tr_A\Upsilon = \sum_\beta P_\beta \tr_A\Upsilon^\beta\] 
which is the expectation, over the shared randomness, of the state at the channel inputs $A'^n$.  
The following extension of the previous coding proposition pertains to these random codes and is also proved in \cite{dev}.
\begin{prop*}[Random entanglement generation coding theorem]
Given is a channel $\CN\colon A'\rightarrow B$, a density matrix $\rho^{A'}$, and a number $0\leq Q < I_c(\rho,\CN).$  For every $\epsilon>0$, there is $n$ sufficiently large so that there is a $(Q,n,\epsilon)$ random entanglement generation code $(P_\beta,\ket{\Upsilon^\beta}^{AA'^n},\CD^\beta)$ for $\CN$
with average code density operator 
\[\varrho^{A'^n} = \E_\beta\tr_A\Upsilon = 
\sum_\beta P_\beta \tr_A\Upsilon^\beta\] satisfying
\[|\varrho - \rho^{\otimes n}|_1\leq\epsilon.\]
\end{prop*}

Finally, there are certain features of the decoder structure of random entanglement generation codes that are necessary for proofs which utilize quantum side information at the decoder.  This final form of the coding proposition is the most powerful, utilizing features which are implicit from the proof of the coding theorem of \cite{dev}.
This will be the proposition which is invoked later in the dissertation.
\begin{prop} \label{prop:lsd}
Given is a channel $\CN\colon A'\rightarrow B$, a density matrix $\rho^{A'}$, and a number $0\leq Q < I_c(\rho,\CN).$  For every $\epsilon>0$, there is $n$ sufficiently large so that there is a random $(Q,n,\epsilon)$ entanglement generation code $(P_\beta,\ket{\Upsilon^\beta}^{AA'^n},\CD^\beta)$ for $\CN$
with average code density operator 
\[\varrho^{A'^n} = \E_\beta\tr_A\Upsilon = 
\sum_\beta P_\beta \tr_A\Upsilon^\beta\] satisfying
\[|\varrho - \rho^{\otimes n}|_1\leq\epsilon.\]
Furthermore, given any particular isometric extension 
$\CU_\CN \colon A'\rightarrow BE$ of $\CN$, it is possible to choose isometric extensions $\CU_{\CD}^\beta\colon B^n\rightarrow \widehat{A}F$ of the deterministic decoders so that 
\[F\big(\ket{\Phi}^{A\widehat{A}}\ket{\lambda}^{E^nF},\CU_{\CD}^\beta\circ
\CU_\CN^{\otimes n}\ket{\Upsilon^\beta}^{AA'^n}\big)\geq 1-\epsilon\] 
for every $\ell$ and the same fixed pure state $\ket{\lambda}^{E^nF}$. 
\end{prop}

% \com{elaborate on controlled operators and coherent random coding...  kind of like how you did it during the talk.  Also, try to briefly characterize the proof of the coding theorem, saying why the added clauses of the proposition are true.}

\chapter{Main results} \label{chapter:mainresults}
\section{Quantum multiple access channels}
For this dissertation, a quantum multiple access channel will have two senders and a single receiver.  While many-sender generalizations of the theorems which appear here are readily obtainable, we focus on the case with two senders for simplicity.  Such a channel $\CN\colon AB'\rightarrow C$ will generally be one in which Alice and Bob simultaneously transmit to Charlie.  We will assume throughout that no other resources are available to the three parties.  Namely, none of the parties share any prior classical or quantum correlations between themselves, nor do they have access to any other auxiliary channels. 
If Alice inputs a physical system with density matrix $\rho^{A'}_1$, while Bob's input has density matrix $\rho^{B'}_2$, Charlie will receive the state
$\CN(\rho_1\otimes\rho_2)$.  
%If Alice instead inputs half of a bipartite pure state $\ket{\Psi_1}^{AA'}$, while Bob does the same with a state $\ket{\Psi_2}^{BB'}$, this gives rise to the following global state
%\[\omega^{ABC} =\CN(\Psi_1\otimes\Psi_2).\]

In the next section, we give an operational definition of the four-dimensional region $\CS(\CN)$, which consists of the rates at which each sender can simultaneously send classical and quantum information to Charlie.  Sections~\ref{section:cqchar} and \ref{section:qqchar} state the main results of this dissertation.  These results characterize the two-dimensional shadows of $\CS(\CN)$ corresponding to the situation where Alice sends classically while Bob sends quantum information (Theorem~1), and that where each sends quantum information (Theorem~2).

These theorems will be proved by first showing in Chapter~\ref{chapter:proofs} that the characterizations given in Sections~\ref{section:cqchar} and \ref{section:qqchar} describe other sets of operationally defined rates, corresponding to weaker constraints on good codes than those to be introduced in this chapter.  In Chapter~\ref{chapter:transmission}, it will ultimately be shown that the other sets of operationally defined rates equal those introduced in this chapter.

\section{$\CS(\CN)$ - the general problem}
\label{section:generalproblem}
Assume that Alice and Bob are connected to Charlie by $n$ instances of a multiple access channel  $\CN\colon A' B' \rightarrow C$, where Alice and Bob respectively have control over the $A'^n$ and $B'^n$ inputs. 
We will describe a scenario in which Alice wishes to transmit classical information at a rate of $R_a$ bits per channel use, while simultaneously transmitting quantum information at a rate of $Q_a$ qubits per channel use.  At the same time, Bob will be transmitting classical and quantum information at rates of $R_b$ and $Q_b$ respectively.  Alice attempts to convey any one of $2^{nR_a}$ messages to Charlie, while Bob tries to send him one of $2^{nR_b}$ such messages.  We will also assume that the senders are presented with systems $\tilde{A}$ and $\tilde{B}$,
where $|\tilde{A}| = 2^{nQ_a}$ and $|\tilde{B}| = 2^{nQ_b}$.  Each will be required to complete the following two-fold task. Firstly, they must individually transfer the quantum information embodied in $\tilde{A}$ and $\tilde{B}$ to their respective inputs $A'^n$ and $B'^n$ of the channels, in such a way that it is recoverable by Charlie at the receiver.  Second, they must simultaneously make Charlie aware of their independent messages $M_a$ and $M_b$. % which are 
%respectively chosen uniformly from sets of sizes $2^{nR_a}$ and $2^{nR_b}$.
Alice and Bob will encode with maps 
from the cq systems holding their classical and quantum messages to their respective inputs of $\CN^{\otimes n}$, which we denote
\[\CE_1\colon M_a \tilde{A} \rightarrow A'^n \,\text{ and }\,
\CE_2\colon M_b \tilde{B} \rightarrow B'^n.\]
%from the cq system holding their classical and quantum messages to their respective inputs of $\CN^{\otimes n}$.  
Charlie decodes with a quantum instrument 
\[\boldsymbol{\CD}\colon C^n\rightarrow\h{M}_a\h{M}_b\h{A}\h{B}.\]  
The output systems are assumed to be of the same sizes and dimensions as their respective input systems.  
For the quantum systems, we assume that there are pre-agreed upon unitary correspondences id$_a\colon \tilde{A} \rightarrow \h{A}$ and id$_b\colon \tilde{B}\rightarrow \h{B}$ between the degrees of freedom in the quantum systems presented to Alice and Bob which embody the quantum information they are presented with and the target systems in Charlie's laboratory to which that information should be transferred.  
The goal for quantum communication will be to, in the strongest sense, simulate the actions of these corresponding identity channels.  We similarly demand low error probability for each pair of classical messages.  Formally, $(\CE_1,\CE_2,\boldsymbol{\CD})$ will be said to comprise an $(R_a,R_b,Q_a,Q_b,n,\epsilon)$ \emph{strong
subspace transmission code} for the channel $\CN$ if for all $m_a \in 2^{nR_a}$, $m_b \in 2^{nR_b}$, $\ket{\Psi_1}^{A\tilde{A}}$, $\ket{\Psi_2}^{B\tilde{B}}$, where $A$ and $B$ are purifying systems of arbitrary dimensions,  
%\setlength{\arraycolsep}{0.0em}
% \bIE{lCL}
%  F\Big(\ket{\ell}\ket{m}\ket{\Psi_1}\ket{\Psi_2},& & \nn  \\ 
%  \hspace{.25in}\CD\circ\CN^{\otimes n}\big(\CE_a(\proj{\ell}\otimes\Psi_1) \nn\\
%  \hspace{.715in}
%  \otimes\,\\CE_b(\proj{m}\otimes\Psi_2)\big)\Big) %\\ 
%  & \geq & 1-\epsilon.  \nn
% \eIE
\[F\Big(\ket{m_a}^{\widehat{M}_a}\ket{m_b}^{\widehat{M}_b}
\ket{\Psi_1}^{A\widehat{A}}\ket{\Psi_2}^{B\widehat{B}},
\Omega_{m_am_b}\Big) 
\geq 1-\epsilon\]
where 
%\[\Omega^{\widehat{M}_a\widehat{M}_bA\widehat{A}B\widehat{B}}
%=\sum_{m_a m_b}\proj{m_a}\otimes\proj{m_b}
%\otimes\Omega_{m_a m_b}\]
%and 
\begin{eqnarray*}
\Omega_{m_a m_b}^{\h{M}_a\h{M}_bA\h{A}B\h{B}} =  
\boldsymbol{\CD}\circ\CN^{\otimes n}
\Big(\CE_1\big(\proj{m_a}^{M_a}\otimes\Psi_1^{A\tilde{A}}\big) 
\otimes\,\CE_2\big(\proj{m_b}^{M_b}\otimes\Psi_2^{B\tilde{B}}\big)
\Big). 
\end{eqnarray*}
We will say that a rate vector $(R_a,R_b,Q_a,Q_b)$ is \emph{achievable} 
if there exists a sequence of  $(R_a,R_b,Q_a,Q_b,n,\epsilon_n)$ strong subspace transmission codes with $\epsilon_n \rightarrow 0$.  The simultaneous capacity region $\CS(\CN)$ is then defined as the closure of the collection of achievable rates.  Setting various rate pairs equal to zero uncovers six two-dimensional rate regions.  The next section contains our first theorem, which gives a multi-letter characterization of the two shadows relevant to the situation where one user only sends classical information, while the other only sends quantum information.  The following section contains a theorem which describes the rates at which each sender can send quantum information via a multi-letter formula.

\section{$\CC\CQ(\CN)$ - classical-quantum capacity region} \label{section:cqchar}
Suppose that Alice only wishes to send classical information at a rate of $R$ bits per channel use, while Bob will only send quantum mechanically at $Q$ qubits per use of the channel. The rate pairs $(R,Q)$ at which this is possible comprise a classical-quantum (cq) region $\CC\CQ(\CN)$ consisting of rate vectors in $\CS(\CN)$ of the form $(R,0,0,Q)$.  Our first theorem gives a characterization of $\CC\CQ(\CN)$ as a regularized union of rectangles.
\begin{theo} \label{theo:cq}
 $\CC\CQ(\CN)$ = the closure of the union of pairs of nonnegative rates 
 $(R,Q)$ satisfying
\begin{eqnarray*}
R  &\leq& \frac{1}{k}I(X;C^k)_\omega \\
Q  &\leq& \frac{1}{k}I_c(B\,\rangle C^kX)_\omega
\end{eqnarray*}
for some $k$, some pure state ensemble $\{p(x),\ket{\phi_x}^{A'^k}\}$ 
and some bipartite pure state $\ket{\Psi}^{BB'^k}$ giving rise to the state
\begin{eqnarray}
\omega^{XBC^k} = \sum_x p(x) \proj{x}^X\otimes \CN^{\otimes k}(\phi_x\otimes\Psi)). \label{th1arise}
\end{eqnarray}
Further, it is sufficient to consider ensembles for which  
\[|\CX| \leq \max\{|A'|,|C|\}^{2k}.\]
\end{theo}

It should also be noted that this characterization does not apparently lead to a finite computation for determining the capacity regions, as it does not admit a single-letter characterization in general.  
However, as an application, the following example contains a channel for which this region is additive.
 
\begin{ex}
Consider an erasure channel into which Alice inputs a classical bit (or rather, a qubit that will be dephased into the $\ket{0}^{A'},\ket{1}^{A'}$ basis), while Bob inputs a qubit. 
If Alice inputs $\ket{0}^{A'}$, Charlie receives Bob's qubit without error.  If Alice inputs $\ket{1}^{B'}$, Charlie receives a pure erasure state $\ket{e}^C$ which is orthogonal to the degrees of freedom of Bob's input state.  The cq capacity region of this channel is equal to the collection of pairs of nonnegative cq rates $(R,Q)$ which satisfy
\begin{eqnarray*}
R &\leq& H(p) \\
Q &\leq& 1-2p
\end{eqnarray*}
for some $0\leq p\leq \frac{1}{2}$.  This region is pictured in Figure~\ref{erasure}.  
\begin{figure}
 \centering
     \includegraphics[width=.4\textwidth]{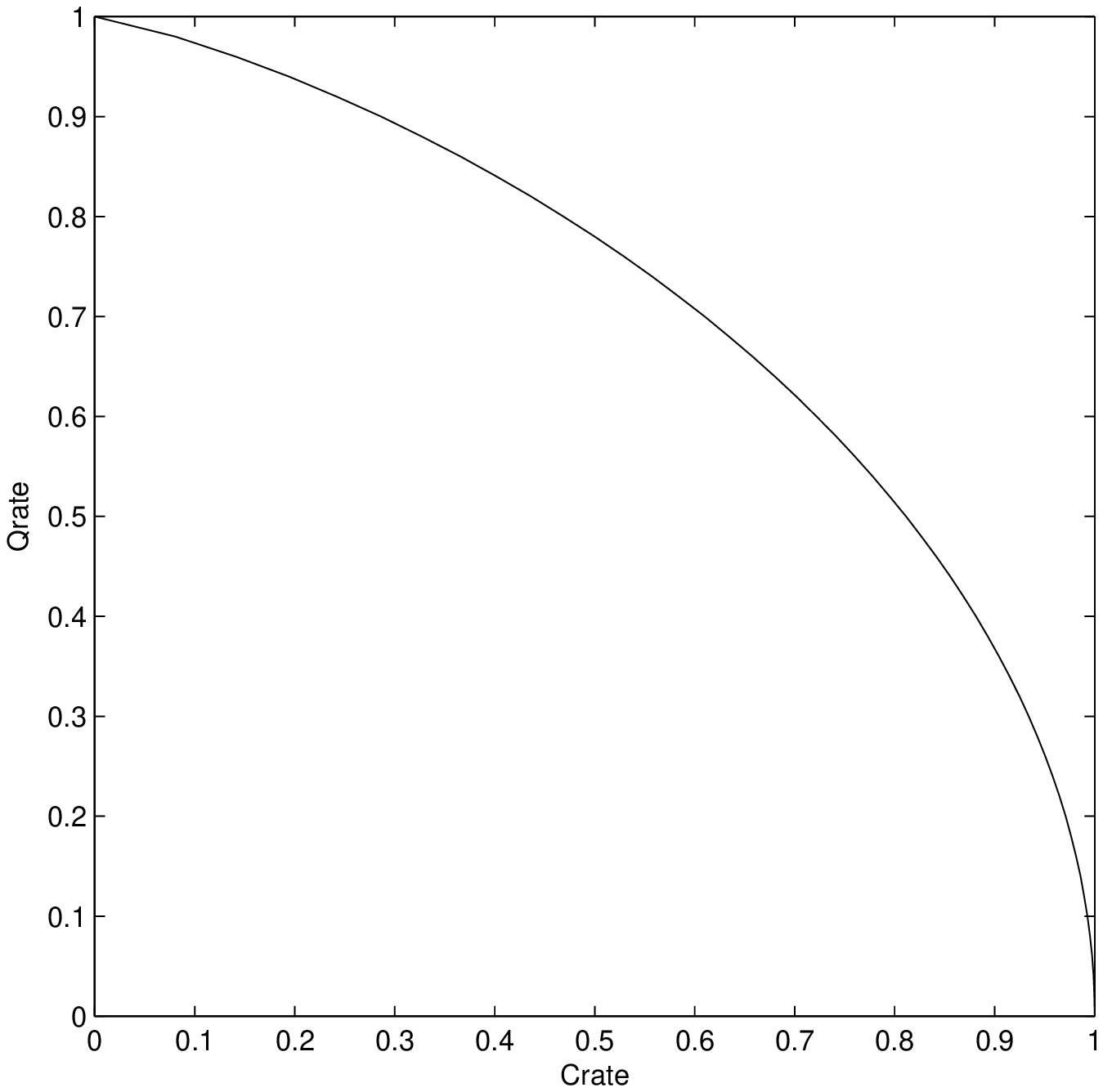}
     \caption{$\CC\CQ($erasure channel$)$}
     \label{erasure}
\end{figure}
\end{ex}
\begin{proof}
In Section~\ref{section:erasure}, we prove this for the more general case where Bob inputs a $d$-level quantum system.
\end{proof}
\begin{rem}
It is also possible to characterize $\CC\CQ(\CN)$ as a regularized union of pentagons, a form which is analogous to the result of \cite{ahlswede,liao} for classical multiple access channels.  As we do not yet know an example of a channel for which this characterization is single-letter (and not equivalent to the rectangle region above), we defer further consideration of this characterization until Chapter~\ref{chapter:discussion}.
\end{rem}
% \begin{theo1'}
% $\CC\CQ(\CN)$ = the closure of the union of pairs of nonnegative rates 
%  $(R,Q)$ satisfying
% \begin{eqnarray*}
% R  &\leq& \frac{1}{k}I(X;BC^k)_\omega \\
% Q  &\leq& \frac{1}{k}I_c(B\,\rangle C^kX)_\omega \\
% R+Q &\leq& \frac{1}{k}\left(I(X;C^k)_\omega + I_c(B\,\rangle C^kX)_\omega\right)
% \end{eqnarray*}
% for some $k$, some pure state ensemble $\{p(x),\ket{\phi_x}^{A'^k}\}$ 
% and some bipartite pure state $\ket{\Psi}^{BB'^k}$ giving rise to the state
% \begin{eqnarray}
% \omega^{XBC^k} = \sum_x p(x) \proj{x}^X\otimes \CN^{\otimes k}(\phi_x\otimes\Psi)). \label{th1'arise}
% \end{eqnarray}
% Further, it is sufficient to consider ensembles for which  
% \[|\CX| \leq \max\{|A'|,|BC|\}^{2k}.\]
% \end{theo1'}
% \begin{rem}  This characterization will not be explicitly proved in this dissertation.  The proof of achievability requires two coding theorems.  The first one required is that for the corner point of the rectangles of Theorem~1.  The second follows from a trivial modification of the coding theorem of the corner points of the pentagons of Theorem~2
% below.  It should be pointed out the the sum rate bound satisfies 
% \[I(X;C) + I_c(B\,\rangle CX) = I(X;BC) + I_c(B\,\rangle C).\]
% The proof of the bound on $|\CX|$ is also a trivial extension of that of Theorem~1.
% \end{rem}

\begin{rem}
The proof of the bound on $|\CX|$ is found in the appendix  (Section~\ref{section:cardinality}).
\end{rem}

\section{$\CQ(\CN)$ - quantum-quantum capacity region}   \label{section:qqchar}
The situation in which each sender only attempts to convey quantum infomation to Charlie is described by the quantum-quantum (qq) rate region $\CQ(\CN)$ which consists of rate vectors in $\CS(\CN)$ of the form $(0,0,Q_a,Q_b)$.  Our second theorem gives a characterization of $\CQ(\CN)$ as a regularized union of pentagons. 
\begin{theo} \label{theo:qq}
$\CQ(\CN)$ = the closure of the union of pairs of nonnegative rates $(Q_a,Q_b)$ satisfying
\begin{eqnarray*}
Q_a &\leq& \frac{1}{k}I_c(A\,\rangle BC^k)_\omega \\
Q_b &\leq& \frac{1}{k}I_c(B\,\rangle AC^k)_\omega  \\
Q_a+Q_b &\leq& \frac{1}{k}I_c(AB\,\rangle C^k)_\omega 
\end{eqnarray*}
for some $k$ and some bipartite pure states $\ket{\Psi_1}^{AA'^k}$, $\ket{\Psi_2}^{BB'^k}$ giving rise to 
\begin{eqnarray}
\omega^{ABC^k} = \CN^{\otimes k}(\Psi_1\otimes\Psi_2). \label{th2arise}
\end{eqnarray}
\end{theo}

% The following characterization appeared in an early draft of [?].  The reader is referred to Chapter~? for a discussion of events which resulted in the extension of the region below to the more accurate Theorem~2.  
% 
% \begin{theo2'}
% $\CQ(\CN)$ = the closure of the union of pairs of nonnegative rates $(Q_a,Q_b)$ satisfying
% \begin{eqnarray*}
% Q_a &\leq& \frac{1}{k}I_c(A\,\rangle C^k)_\omega \\
% Q_b &\leq& \frac{1}{k}I_c(B\,\rangle C^k)_\omega 
% \end{eqnarray*}
% for some $k$ and some bipartite pure states $\ket{\Psi_1}^{AA'^k}$, $\ket{\Psi_2}^{BB'^k}$ giving rise to 
% \begin{eqnarray}
% \omega^{ABC^k} = \CN^{\otimes k}(\Psi_1\otimes\Psi_2). \label{th2'arise}
% \end{eqnarray}
% \end{theo2'}
\begin{ex}
An example of a channel for which this region is single-letter is a channel into which Alice and Bob each input a qubit.  With probability $p$, each of their qubits undergoes a phase flip, or $180^\circ$ rotation about the $z$-axis, before being received by Charlie.  Otherwise, Charlie receives both qubits without error. 
The qq capacity region of this channel is given by a single pentagon, consisting of the pairs of nonnegative qq rates $(Q_a,Q_b)$ which satisfy 
\begin{eqnarray*}
 Q_a &\leq& 1 \\
 Q_b &\leq& 1 \\
 Q_a + Q_b &\leq& 2-H(p).
\end{eqnarray*}
%This region is pictured in Figure~?.
\end{ex}
\begin{proof}
See Section~\ref{section:bitflip}.
\end{proof}
\begin{rem}
There does not appear to be any obstacle preventing application of the methods used in this paper to prove many-sender generalizations of Theorems~1 and 2.  For simplicity, we have focused on the situations with two senders.
\end{rem}
\begin{rem}
Contrary to the corresponding result for classical multiple access channels, the regions of Theorems 1 and 2 do not require convexification.  That this follows from the multi-letter nature of the regions will be demonstrated in the appendix (Section~\ref{section:convex}). 
\end{rem}

\chapter{Supplementary results}
In this chapter, we collect a number of auxiliary results which will be used to prove the main theorems.  The first section contains some relationships satisfied by the distance measures of trace distance and fidelity which will comprise the machinery used to prove the coding theorems.  The main novel contribution of that section is the statement and proof of Lemma~\ref{lemma:transitivity}.  The next section contains other lemmas, proved elsewhere, which will needed later.  In the third section we review strong subadditivity of quantum entropy, and explore a number of its consequences.  These include quantum versions of the classical data processing inequality, as well as the fact that conditioning decreases conditional quantum entropy or equivalently, increases coherent information.  We also obtain a particularly elegant proof of the Holevo bound on the accessible information of an ensemble of quantum states. 
\section{Further properties of distance measures} \label{section:distance}
We first collect some relevant results which will be used in what follows, starting with some relationships between our distance measures.  If $\rho$ and $\sigma$ are density matrices defined on the same (or isomorphic) Hilbert spaces, set
\[F = F(\rho,\sigma) \text{ and } T = |\rho - \sigma|_1.\]
Then, the following inequalities hold (see e.g.\ \cite{cn})
\begin{eqnarray}
1-\sqrt{F} &\leq \;\;T/2\;\; \leq&  \sqrt{1-F}, \label{trfid}\\
1-T &\leq \;\;F\;\; \leq& 1-T^2/4. \label{fidtr}
\end{eqnarray}
From these inequalities, we can derive the following more useful relationships
\begin{eqnarray}
F > 1-\epsilon &\Rightarrow& T \leq 2\sqrt{\epsilon} \label{trfide}\\
T \leq \epsilon &\Rightarrow& F > 1-\epsilon, \label{fidtre}
\end{eqnarray}
which are valid for $0\leq\epsilon\leq 1.$
Uhlmann \cite{uhl} has given the following characterization of fidelity
\[F(\rho,\sigma) = \max_{\ket{\Psi_\rho},\ket{\Phi_\sigma}} |\braket{\Psi_\rho}{\Phi_\sigma}|^2
= \max_{\ket{\Psi_\rho}}|\braket{\Psi_\rho}{\Phi_\sigma}|^2\]
where the first maximization is over all purifications of each state, and the second maximization holds for any fixed purification $\ket{\Phi_\sigma}$ of $\sigma.$
This characterization is useful in two different ways.  First, for any two states, it guarantees the existence of purifications of those states whose squared inner product equals the fidelity.  Second, one can derive from that characterization the following monotonicity property \cite{bcfjs} associated with an arbitrary trace-preserving channel $\CN$,
\begin{eqnarray}
F(\rho,\sigma) &\leq& F(\CN(\rho),\CN(\sigma)) \label{fmon}
\end{eqnarray}
An analogous property is shared by the trace distance \cite{rustrace},
\begin{eqnarray}
|\rho-\sigma|_1 &\geq& \left|\CN(\rho)-\CN(\sigma)\right|_1, \label{trmon}
\end{eqnarray}
which holds even if $\CN$ is trace-reducing.  A simple proof for the trace-preserving case can be found in \cite{cn}.
These inequalities reflect the fact that completely-positive maps are \emph{contractive} and cannot improve the distinguishability of quantum states; the closer states are to each other, the harder it is to tell them apart.  Another useful property will be the multiplicativity of the fidelity under tensor products
\begin{eqnarray}
F(\rho_1\otimes \rho_2,\sigma_1\otimes \sigma_2)
= F(\rho_1,\sigma_1)F(\rho_2,\sigma_2). \label{fmult}
\end{eqnarray}
Since the trace distance comes from a norm, it satisfies the triangle inequality.  The fidelity does not come from a norm, but it is possible to derive the following analog by applying (\ref{trfid}) and (\ref{fidtr}) to the triangle inequality for the trace distance
\begin{eqnarray}
F(\rho_1,\rho_3) \geq 1 - 2\sqrt{1-F(\rho_1,\rho_2)} - 2\sqrt{1-F(\rho_2,\rho_3)}.
\label{ftri}
\end{eqnarray}
It will be possible to obtain a sharper triangle-like inequality as a consequence of the following lemma, which states that if a measurement succeeds with high probability on a state, it will also do so on a state which is close to that state in trace distance.
\begin{lem}\label{lemma:special}
Suppose that $\rho,\sigma,\Lambda \in \bC^{d\times d},$ where $\rho$ and $\sigma$ are density matrices, and $0\leq \Lambda \leq 1.$  Then, $\tr\Lambda\sigma \geq \tr\Lambda\rho - |\rho-\sigma|_1.$

\end{lem}
\begin{proof}
\begin{eqnarray*}
\tr\Lambda\sigma &=& \tr\Lambda\rho - \tr\Lambda(\rho - \sigma) \\
&\geq& \tr\Lambda\rho - \max_{0\leq\Lambda\leq 1} 2\tr\Lambda(\rho - \sigma)\\
&=& \tr\Lambda\rho - |\rho-\sigma|_1,
\end{eqnarray*}
where the last equality invokes a characterization of the trace distance between density matrices given in Section~\ref{section:tr}.
\end{proof}

Since $F(\phi,\rho) = \tr\phi\rho$ when $\phi$ is a pure state, a corollary of Lemma~\ref{lemma:special} is a fact we will refer to as the ``special triangle inequality."
\begin{cor*}[Special triangle inequality]
\begin{eqnarray*}
F(\phi,\sigma) \geq F(\phi,\rho) - |\rho - \sigma|_1,
\end{eqnarray*}
\end{cor*}

The following lemma can be thought of either as a type of transitivity property inherent to any bipartite state with a component near a pure state, or as a partial converse to the monotonicity of fidelity.
\begin{lem}\label{lemma:transitivity}
For arbitrary quantum systems $A$ and $B$, let
$\ket{\phi}^A$ be a pure state, $\rho^B$ a density matrix,
and $\Omega^{AB}$ a density matrix of the composite system $AB$ with partial traces $\Omega^A = \tr_B\Omega$ and $\Omega^B = \tr_A\Omega.$  Then 
%\begin{eqnarray}
%F(\phi,\Omega^A) \geq 1-\epsilon. \label{l4f1}
%\end{eqnarray}
\[F(\phi\otimes \rho, \Omega) \geq 1-|\rho - \Omega^B|_1 - 3\big(1-F(\phi,\Omega^A)\big).\]
\end{lem}
\begin{proof}
We begin by defining the subnormalized density matrix $\tilde{\omega}$ via the equation
\begin{eqnarray}
(\phi\otimes 1)\Omega (\phi\otimes 1) = \phi\otimes \tilde{\omega}, \label{l4po}
\end{eqnarray}
which we interpret as the upper-left block of $\Omega$, when the basis for $\bC^{|A|}$ is chosen in such a way that $\ket{\phi} = (1,0,\ldots,0)^T.$  Notice that $F(\phi,\tr_B\Omega) = \tr\tilde{\omega}\equiv (1-\ep).$
Writing the normalized state $\omega = \tilde{\omega}/(1-\ep),$ we see that it
is close to $\tilde{\omega}$ in the sense that
\begin{eqnarray}
|\omega - \tilde{\omega}|_1 &\leq& \epsilon|\tilde{\omega}|_1
  \nonumber \\
&\leq& \epsilon. \label{l4oo}
\end{eqnarray}
Now we write
\begin{eqnarray}
\sqrt{F(\phi\otimes\rho,\Omega)}
&=& \tr\sqrt{\sqrt{(\phi\otimes\rho)}\Omega\sqrt{(\phi\otimes\rho)}}
  \nonumber\\
&=& \tr\sqrt{(1\otimes\sqrt{\rho})
  (\phi\otimes 1)\Omega(\phi\otimes 1)
  (1\otimes\sqrt{\rho})}\nonumber\\
&=& \tr\sqrt{(1\otimes\sqrt{\rho})(\phi\otimes\tilde{\omega})(1\otimes\sqrt{\rho})}
  \nonumber\\
&=& \tr\sqrt{\phi\otimes(\sqrt{\rho}\,\tilde{\omega}\sqrt{\rho})}\nonumber\\
&=& \tr\sqrt{\sqrt{\rho}\,\tilde{\omega}\sqrt{\rho}}\nonumber\\
&=& \sqrt{F(\tilde{\omega},\rho)}\nonumber\\
&=& \sqrt{(1-\ep) F(\omega,\rho)}\nonumber\\
&\geq& \sqrt{(1-\ep) (1-|\omega - \rho|_1)}. \label{l4F}
\end{eqnarray}
The first line is the definition of fidelity and the third follows from (\ref{l4po}).
The last equality relies on the fact that the fidelity, as we've defined it, is linear in either of its two inputs,
while the inequality follows from (\ref{fidtr}).

Noting that $\Omega^B \geq \tilde{\omega}$, we define another positive operator
$\omega' = \Omega^B - \tilde{\omega},$ which satisfies $\tr\omega' \leq \epsilon$ and can be interpreted as the sum of the rest of the diagonal blocks of $\Omega.$
The trace distance in the last line above can be bounded via double application of the triangle inequality as
\begin{eqnarray}
|\rho - \omega|_1
&\leq& |\rho - (\rho-\omega')|_1 + |(\rho-\omega') - \tilde{\omega}|_1 +
   |\tilde{\omega} -\omega|_1 \nonumber \\
&\leq& \tr\omega' + \left|\rho - \Omega^B\right|_1 + \epsilon \nonumber\\
&\leq& \left|\rho - \Omega^B\right|_1 + 2\epsilon \label{l4ro},
\end{eqnarray}
where the second line follows from (\ref{l4oo}).
Combining (\ref{l4F}) with (\ref{l4ro}), we obtain
\begin{eqnarray*}
F(\phi\otimes\rho,\Omega)
&\geq& (1-\epsilon)(1-|\rho - \Omega^B|_1-2\epsilon)\\
&\geq& 1-\left|\rho - \Omega^B\right|_1-3\epsilon. %\qedhere
\end{eqnarray*}%\qedhere
\end{proof}

\section{Other useful lemmas} \label{section:lemmas}
This continuity lemma from \cite{alickifannes} shows that if two bipartite states are close to each other, the difference between their associated coherent informations is small.
\begin{lem}[Continuity of coherent information] \label{lemma:continuity}
Let $\rho^{AB}$ and $\sigma^{AB}$ be two states of a
finite-dimensional bipartite system $AB$ satisfying
$|\rho-\sigma|_1 \leq \epsilon$. Then
\[|I_c(A\,\rangle B)_\rho - I_c(A\,\rangle B)_\sigma|
   \leq 2H(\epsilon) + 4\log|A|\epsilon,\]
where $H(\epsilon)$ is the binary entropy function.
\end{lem}

Next is Winter's ``gentle measurement" lemma \cite{gentle}, which implies that a measurement which is likely to be successful in identifying a state tends not to significantly disturb that state.
\begin{lem}[Gentle measurement]\label{lemma:gentle}
Let a density matrix $\rho^A$ be given, where $|A|$ is finite.  If $\Lambda\in\bC^{|A|\times |A|}$ is nonnegative with spectrum bounded above by 1,  then
\[\tr \rho\Lambda \geq 1-\epsilon\]
implies
\[\left|\sqrt{\Lambda}\rho\sqrt{\Lambda} - \rho\right|_1 \leq \sqrt{8\epsilon}.\]
\end{lem}

We will also need a lemma from classical information theory which bounds the conditional entropy of two random variables with the same support in terms of the probability they are different.
\begin{lem}[Fano's inequality] \label{lemma:fano}
Let $M$,$\widehat{M}$ be $\CM$-valued random variables, and write $P_e = \Pr\{M\neq \widehat{M}\}$.  Then 
\[H(M|\widehat{M}) \leq H(P_e) + P_e\log|\CM|.\]
\end{lem}
\begin{proof}
See \cite{coverthomas}.
\end{proof}
\section{Strong subadditivity and its consequences}
In this section, we recall an inequality which holds for \emph{any} tripartite quantum system $ABC$.  This inequality goes by the name \emph{strong subadditivity}, and was originally proved in \cite{ssuborig}, stating that
\begin{equation} 
H(AB) + H(BC) \geq H(B) + H(ABC). \label{ssad}
\end{equation}
As much has been written about the proof of strong subadditivity of quantum entropy (see e.g. \cite{ssub}),  we will not discuss the proof of the theorem here.
% although we pause to mention that it can be proved as a consequence of the monotonicity of the quantum relative entropy (?)
%\[D(\rho||\sigma) = \tr \rho\log\rho - \tr\rho\log\sigma\]
%with respect to the action of any quantum channel,
%a result whose proof requires the consideration of \emph{operator convex functions}, the introduction of which is beyond the scope of this discussion.   
Rather, we will endeavor to show here how strong subadditivity can be used as a mathematical ``hammer of Thor," enabling short and elegant proofs of many known entropy inequalities
in quantum information theory.  In fact, many of these results will turn out to be \emph{equivalent} to strong subadditivity, in the sense that the latter is easily derivable from many of them. 

%Many of these proofs require only reexpressing and reinterpreting (\ref{ssad}) in terms of the shorthand expressions for conditional entropy, coherent information and mutual information. 

Begin by subtracting $H(B) + H(BC)$ from either side of (\ref{ssad}) to yield 
\begin{equation}
H(A|B) \geq H(A|BC). \label{Hcond}
\end{equation}
This inequality can be interpreted as a demonstration that \emph{conditioning reduces entropy}.  Collecting the terms on a single side yields the compact formula
\[I(A;B|C) \geq 0,\]
showing that the quantum conditional mutual information is always positive.  Note that in the classical case, these inequalities are quite simple to prove \cite{coverthomas}.  For instance, positivity of $I(X;Z|Y)$  follows from positivity of mutual information which, in turn, is a consequence of positivity of the Kullback-Leibler distance $D(P||Q)$. 

Classically, it is simple to show that $I(X;Z|Y) = 0$ if and only if $X-Y-Z$ forms a Markov chain in that order.  Necessary and sufficient conditions for saturation of quantum strong subadditivity were recently determined in \cite{hjpw}, who showed that $I(A;C|B) = 0$ if and only if 
\[\rho^{ABC} = \bigoplus_x p_x \rho^{AB^A_x}_x \otimes \rho^{B^C_x C}_x\]
where $B = \oplus_x B^A_x B^C_x$.  In other words, if and only if there is a local measurement that can be performed on $B$ which determines $x$ without disturbing the global state.  Conditioned on knowing $x$, the global system is in a product state.  Such a measurement is commonly known as a ``which path measurement."  
%We will see later, in Section~\ref{chapter:discussion}, that the sequence of global states induced by the encodings from any achieving sequence of qq entanglement generation codes asymptotically saturates strong subadditivity.

Recall the definition of coherent information as 
\[I_c(A\,\rangle B) = - H(A|B).\]
Simply reexpressing (\ref{Hcond}) in terms of coherent information yields the inequality 
\begin{equation}
I_c(A\,\rangle BC) \geq I_c(A\,\rangle C), \label{Iccond}
\end{equation}
which can be interpreted in light of the quantum capacity theorem as saying that losing access to part of the output of a quantum channel can only decrease capacity.  
Observe that a similar property is obeyed by the classical mutual information, namely that 
\[I(X;YZ) \geq I(X;Y).\]
More generally, coherent information can be shown to obey an analog of the classical data processing inequality (see e.g. \cite{coverthomas}), which says that if $X-Y-Z$ is a Markov chain, then 
\[I(X;Y) \geq I(X;Z).\]
A quantum version of the data processing inequality \cite{sn96} can be proved easily from strong subadditivity.  
\begin{lem}[Quantum data processing inequality] \label{lemma:dataproc}
Let a bipartite density matrix $\rho^{AB}$ and a channel $\CN\colon B\rightarrow C$ be given.  Then 
\[I_c(A\,\rangle B)_\rho \geq I_c(A\,\rangle C)_{\CN(\rho)}.\]
\end{lem}
\begin{proof} Choose any isometric extension $\CU_\CN\colon B\rightarrow CE$ of $\CN$.  Then  
\begin{eqnarray*}
I_c(A\,\rangle B)_\rho &=& I_c(A\,\rangle CE)_{\CU_\CN(\rho)} \\
&\geq& I_c(A\,\rangle C)_{\CN(\rho)},
\end{eqnarray*}
where the first step is because isometries preserve entropy, while the second is by 
(\ref{Iccond}).
\end{proof}

It is thus apparent that post-processing of $B$ can never increase coherence with $A$.  It is also possible to derive strong subadditivity from data processing, by taking $\CN = \tr_C$, so the data processing inequality is another equivalent way to express strong subadditivity.  

The quantum data processing inequality can be used to derive a more direct analog \cite{ahlswedelober} of the classical data processing inequality, dealing with quantum mutual information rather than coherent information.  A simple corollary of Lemma~\ref{lemma:dataproc} is 
\begin{cor*}[Quantum mutual information data processing inequality]
With the same conditions as in Lemma~\ref{lemma:dataproc}, 
\[I(A;B)_\rho \geq I(A;C)_{\CN(\rho)}.\]
\end{cor*}
\begin{proof}  The conclusion of Lemma~\ref{lemma:dataproc} can be rewritten in terms of conditional entropies as 
\[-H(A|B)_\rho \geq - H(A|C)_{\CN(\rho)}.\]
Adding $H(A)$ to each side yields the required inequality. 
\end{proof}
As a simple consequence of this corollary, we obtain a completely elementary proof of the Holevo bound \cite{holevobound}, an essential step in the converse part of the HSW capacity theorem.
\begin{lem}[Holevo bound] \label{lemma:holevo}
Let a cq state
\[\rho^{XB} = \sum_x p(x)\proj{x}^X\otimes\rho_x^B\] be given.  For any measurement on $B$ with POVM $\{\Lambda_y\}_{y\in \CY}$, the following inequality holds.
\[I(X;B)_\rho \geq I(X;Y).\] 
\end{lem}
\begin{proof}
Construct a measuring instrument $\boldsymbol{\CN}\colon B\rightarrow Y$ (as in Section~\ref{section:instruments}), acting as 
\[\boldsymbol{\CN}(\tau^B) = \sum_y \big(\tr \Lambda_x \tau\big)\proj{y}^Y.\]
Application of the previous version of the data processing inequality proves the result. 
\end{proof}

The following inequality from \cite{cn} will be useful in Section~\ref{section:degradable}, where we give a proof that \emph{degradable} quantum channels have single-letter quantum capacities.
\begin{lem}[Joint subadditivity of conditional entropy] \label{lemma:CEsubadditivity}
For any quadripartite state on $ABCD$, the following entropy inequality applies
\begin{equation}
H(AB|CD) \leq H(A|C) + H(B|D).
\end{equation}
\end{lem}
\begin{proof}
Using the original formulation of strong subadditivity (\ref{ssad}), we may write the following two inequalities:
\begin{eqnarray*}
H(ABCD) + H(C) &\leq& H(AC) + H(BCD) \\
H(BCD) &\leq&  H(BD) + H(CD) - H(D).
\end{eqnarray*}
Combining these gives 
\[H(ABCD) + H(C) \leq H(AC) + H(BD) + H(CD) - H(D).\]
Rearranging terms gives the required result.
\end{proof}

Observe that this lemma can equivalently be expressed in terms of coherent information as 
\begin{equation}
I_c(A\,\rangle C) + I_c(B\,\rangle D) \leq I_c(AB\,\rangle CD).
\label{Icsuperadditivity}
\end{equation}
Note that if (\ref{Icsuperadditivity}) is computed on a state of the form 
\[\Omega^{ABCD} = \proj{\phi}^B \otimes \omega^{ACD},\]
it follows that 
\[I_c(B\,\rangle D) = H(D) - H(BD) = H(D) - H(D) = 0\]
and 
\[I_c(AB\,\rangle CD) = H(CD) - H(ABCD) = H(CD) - H(ACD) = I_c(A\,\rangle CD),\]
implying that  
\[I_c(A\,\rangle C) \leq I_c(A\,\rangle CD),\]
which is just the original strong subadditivity inequality we started with.  
So we see that strong subadditivity is equivalent to Lemma~\ref{lemma:CEsubadditivity}, as well as to the fact that coherent information is superadditive.

\chapter{Entanglement generation capacities} \label{chapter:proofs}
As a first step towards proving the theorems stated in Chapter~\ref{chapter:mainresults}, we introduce a less restrictive communication scenario, \emph{entanglement generation}.  While the criterion of strong subspace transmission is analogous to a classical requirement that the maximal error probability be small, the entanglement generation criterion will rather be related to an average error constraint on good codes.  
%\section{Entanglement generation}

\paragraph{classical-quantum scenario}
Alice sends classical information to Charlie at rate $R$, while Bob sends quantum information at rate $Q$.  Rather than being required to transmit half of \emph{any} quantum state Bob is presented with, Bob will only need to \emph{create} near maximal quantum correlations with Charlie at rate $Q$.  To this end, Bob begins by preparing a bipartite pure state
$\ket{\Upsilon}^{BB'^n},$ entangled between a physical system $B$ located in his laboratory, and the $B'^n$ part of the inputs of $\CN^{\otimes n}$.

At the same time, Charlie will only need to identify Alice's classical message with a low average error probability, averaged over all of Alice's classical messages.  As with strong subspace transmission, Charlie's post-processing procedure will be modeled by a quantum instrument.  While the outer bound provided by our converse theorem will apply to any decoding modeled by an instrument, the achievability proof will require a less general approach, consisting of the following steps.

In order to ascertain Alice's message $M$, Charlie first performs some measurement on $C^n$, whose statistics are given by a POVM $\{\Lambda_{m}\}_{m\in 2^{nR}}$.  We let the result of that measurement be denoted $\widehat{M},$ his declaration of the message sent by Alice.  Based on the result of that measurement, he will perform one of $2^{nR}$ decoding operations $\CD'_m\colon C^n\rightarrow \widehat{B}.$  These two steps can be mathematically combined to define a quantum instrument $\boldsymbol{\CD}\colon C^n\rightarrow \widehat{M}\widehat{B}$ with (trace-reducing) components
\[\CD_m\colon \tau\mapsto\CD'_m(\sqrt{\Lambda_m}\tau\sqrt{\Lambda_m}).\]
The instrument acts as
\[\boldsymbol{\CD}\colon \tau\mapsto \sum_{m=1}^{2^{nR}}
\proj{m}^{\widehat{M}}\otimes\CD_m(\tau),\]
and induces the trace preserving map $\CD\colon C^n\rightarrow \widehat{B}$, acting according to
\[\CD\colon \tau\mapsto\tr_{\widehat{M}}\boldsymbol{\CD}(\tau) = \sum_{m=1}^{2^{nR}}\CD_m(\tau).\]
We again remark that this is the most general decoding procedure required of Charlie.  Any situation in which he were to iterate the above steps by measuring, manipulating, measuring again, and so on, is asymptotically just as good as a single instance of the above mentioned protocol.  This is because the inner and outer bounds provided by the coding theorem and converse coincide.
$(\{\phi_m\}_{m\in 2^{nR}},\Upsilon^{BB'^n},\boldsymbol{\CD})$ will be called an
$(R,Q,n,\epsilon)$ \emph{cq entanglement generation code} for the channel $\CN$ if 
\begin{eqnarray}
2^{-nR}\sum_{m=1}^{2^{nR}} P_s^{\text{eg}}(m,\Upsilon) \geq 1-\epsilon, \label{cqIs}
\end{eqnarray}
where
\begin{eqnarray}
P_s^{\text{eg}}(m,\Upsilon) = F\left(\ket{m}\ket{\Phi}^{B\widehat{B}},
\boldsymbol{\CD}\circ\CN^{\otimes n}
(\phi_m^{A'^n}\otimes\Upsilon^{BB'^n})\right). \label{cqIPs}
\end{eqnarray}

We will say that $(R,Q)$ is an \emph{achievable cq rate pair for entanglement generation} if there exists a sequence of $(R,Q,n,\epsilon_n)$ cq entanglement generation codes
with $\epsilon_n\rightarrow 0$.
The capacity region $\CC\CQ_{\text{eg}}(\CN)$ is defined to be the closure of the collection of all achievable cq rate pairs for entanglement generation.

\paragraph{quantum-quantum scenario}
As above, Alice and Bob are no longer required to transmit arbitrary quantum correlations with which they are presented.  Rather, each has the goal of creating near-maximal entanglement with Charlie.  For encoding, Alice and Bob respectively prepare the states
$\ket{\Upsilon_1}^{AA'^n}$ and $\ket{\Upsilon_2}^{BB'^n},$
entangled with the $A'^n$ and $B'^n$ parts of the inputs of 
$\CN^{\otimes n}$.  Their goal is to do this in such a way 
so that Charlie, after applying a suitable decoding operation
$\CD\colon C^n\rightarrow \widehat{A}\widehat{B}$, can hold the $\widehat{A}\widehat{B}$ part of a state which is close to $\ket{\Phi_1}^{A\widehat{A}}\ket{\Phi_2}^{B\widehat{B}}$.   
Formally, $(\Upsilon_1^{AA'^n}, \Upsilon_2^{BB^n},\CD)$ is an $(R,Q,n,\epsilon)$ \emph{qq entanglement generation code} for the channel $\CN$ if
\begin{eqnarray}
F(\Phi_1\otimes\Phi_2,
\CD\circ\CN^{\otimes n}(\Upsilon_1\otimes\Upsilon_2)) \geq 1-\epsilon. \label{qqIs}
\end{eqnarray}
$(R,Q)$ is an \emph{achievable qq rate pair for entanglement generation} if there is a sequence of $(R,Q,n,\epsilon_n)$ qq entanglement generation codes with $\epsilon_n\rightarrow 0$.  The capacity region $\CQ_\text{eg}(\CN)$ is the closure of the collection of all such achievable rates.

\section{The coding theorems}
For any quantum multiple access channel $\CN\colon A'B'\rightarrow C$, we first prove that the single-letter regions $\CC\CQ^{(1)}(\CN)$ and $\CQ^{(1)}(\CN)$, defined as the restrictions to $k=1$ of the respective characterizations from Sections~\ref{section:cqchar} and \ref{section:qqchar}, are respectively contained in $\CC\CQ_{\text{eg}}(\CN)$ and in $\CQ_{\text{eg}}(\CN)$.  It will then follow that 
\[\bigcup_{k=1}^\infty\frac{1}{k}\CC\CQ^{(1)}(\CN^{\otimes k})\subseteq\CC\CQ_{\text{eg}}(\CN) \,\,\,\text{and}\,\,\,
\bigcup_{k=1}^\infty\frac{1}{k}\CQ^{(1)}(\CN^{\otimes k})\subseteq\CQ_{\text{eg}}(\CN)\]
by applying the coding theorems to extensions $\CN^{\otimes k}$ of $\CN$.
\begin{proof}[Proof of Theorem~1 (coding theorem)]
Our method of proof for the coding theorem will work as follows.  We will employ random HSW codes and random entanglement generation codes to ensure that the average state at the input of $\CN^{\otimes n}$ is close to a product state.  Each sender will utilize a code designed for the product channel induced by the other's random input, whereby
existing coding theorems for product channels will be invoked.
The quantum code used will be one which achieves the capacity of a modified channel, in which the classical input is copied, without error, to the output of the channel.
As the random HSW codes will exactly induce a product state input, the existence of these quantum codes will follow directly from Proposition~\ref{prop:lsd}.

The random HSW codes will be those which exist for product channels.  As random entanglement generation codes exist with average code density matrix arbitrarily close to a product state, this will ensure that the resulting output states are distinguishable with high probability.  Furthermore, obtaining the classical information will be shown to cause but a small disturbance in the overall joint quantum state of the system.  As we will show, it is possible to mimic the channel for which the quantum code is designed by placing the identities of the estimated classical message states into registers appended to the outputs of each channel in the product.

The decoder for the modified channel will then be shown to define a quantum instrument which satisfies the success condition for a cq entanglement transmission code, on average.  This feature will then be used to
infer the existence of a particular, deterministic code which meets the same requirement.

Fix a pure state ensemble $\{p(x),\ket{\phi_x}^{A'}\}$ and a bipartite pure state $\ket{\Psi}^{BB'}$ which give rise to the cq state
\[\omega^{XBC} =
\sum_x p(x)\proj{x}^X\otimes(1^B\otimes \CN)
(\phi_x^{A'}\otimes \Psi^{BB'}),\]
which has the form of (\ref{th1arise}).
Define $\rho_1^{A'} = \sum_x p(x)\phi_x$ and $\rho_2^{B'} = \tr_B\Psi$.   We will demonstrate the achievability of the corner point $(I(X;C),I_c(B\,\rangle CX))_\omega$ by showing that for every $\epsilon,\delta>0,$ if $R = I(X;C)_\omega-\delta$ and $Q = I_c(B\,\rangle CX)_\omega-\delta$,
there exists an $(R,Q,n,\epsilon)$ cq entanglement generation code for the channel $\CN$, provided that $n$ is sufficiently large.
The rest of the region will follow by timesharing.

For encoding, Alice will choose $2^{nR}$ sequences $\CC=\{X^n(m)\}_{m\in 2^{nR}}$, i.i.d.\ according to the product distribution $p(x^n) = \prod_{i=1}^n p(x_i)$.  As each sequence corresponds to a preparation of channel inputs
$\ket{\phi_m}^{A'^n}=
\ket{\phi_{X_1(m)}}\otimes\cdots\otimes\ket{\phi_{X_n(m)}},$ the expected  average density operator associated with Alice's input to the channel is precisely
\[
\E_\CC 2^{-nR}\sum_{m=1}^{2^{nR}}\proj{\phi_m} = \sum_{x^n} p(x^n) \proj{\phi_{x^n}}
=\rho_1^{\otimes n}.
\]
Define a new channel $\boldsymbol{\CN}'\colon B'\rightarrow C\widehat{X}$ (which is also an instrument) by
\[\boldsymbol{\CN}'\colon \rho\mapsto
\sum_x p(x)\CN(\phi_x\otimes\rho)\otimes\proj{x}^{\widehat{X}},\]
This can be interpreted as a channel which reveals the identity of Alice's input state to Charlie, with the added assumption that Alice chooses her inputs at random.  Alternatively, one can view this as a channel with state information available to the receiver, where nature is randomly choosing the ``state" $x$ at Alice's input.  By Proposition \ref{prop:lsd}, there exists a $(Q,n,\epsilon)$ random entanglement generation code $\{q_\be,\ket{\Upsilon^\be}^{AA'^n},\CD^\be\}$ for the channel $\boldsymbol{\CN}'$ with average code density operator
$\varrho^{B'^n} = \sum_\be q_\be \tr_A\Upsilon^\be$ satisfying \[|\varrho - \rho_2^{\otimes n}|_1\leq \epsilon.\]
In what follows, we will use the shorthand $\ket{\Upsilon}$ for the random vector which takes the value $\ket{\Upsilon^\be}$ with probability $q_\be$.  We further abbreviate 
\[\E_\be \CM(\Upsilon) \equiv \sum_\be q_\be \CM(\Upsilon^\be),\]
where $\CM$ is any function of the random vector $\Upsilon$.

Now, by Proposition \ref{prop:hsw}, for the channel $\CN_1\colon \rho\mapsto \CN(\rho\otimes\rho_2)$ which would result if Bob's average code density operator were \emph{exactly} equal to $\rho_2^{\otimes n},$ there exists a decoding POVM $\{\Lambda_m\}_{m\in 2^{nR}}$ which would identify Alice's index $m$ with expected average probability of error less than $\epsilon$,
in the sense that
\[\E_\CC 2^{-nR}\sum_{m=1}^{2^{nR}}\tr\Lambda_m\tau'_m \geq 1-\epsilon,\]
where
\[\tau'_m = \CN^{\otimes n}(\phi_m\otimes\rho_2^{\otimes n}).\]
By the symmetry of the random code construction, we utilize (\ref{symmetry}) to write this as
\[\E_\CC\tr\Lambda_1\tau'_1 \geq 1-\epsilon.\]
Define the \emph{actual} output of the channel corresponding to $M=m$ as
\[\tau_m = \CN^{\otimes n}(\phi_m\otimes\tr_A\Upsilon),\]
as well as its extension
\[\xi_m^{BC^n} = \CN^{\otimes n}(\phi_m\otimes\Upsilon),\]
where $\ket{\Phi}^{B\tilde{B}}$ is the maximally entangled state which Bob is required to transmit.
Note that
\[\E_\beta\tau_m = \E_\beta\tr_B\xi_m = \CN^{\otimes n}(\phi_m\otimes\varrho).\]
It follows from monotonicity of trace distance that
\[\left|\E_\beta\tau_1 - \tau_1'\right|_1\leq \epsilon,\]
which, together with Lemma~\ref{lemma:special}, implies that
\begin{eqnarray*}
\E_\CC 2^{-nR}\sum_{m=1}^{2^{nR}}\tr\Lambda_m \E_\beta\tau_m
= \E_{\CC\beta}\tr\Lambda_1 \tau_1
\geq 1-2\epsilon.
\end{eqnarray*}
This allows us to bound the expected probability of correctly decoding Alice's message as
\begin{eqnarray}
\E_{\CC\beta}\tr(1\otimes\Lambda_1)\xi_1 \geq 1-2\epsilon. \label{EP}
\end{eqnarray}

In order to decode, Charlie begins by performing the measurement $\{\Lambda_m\}_{m\in 2^{nR}}.$  He declares Alice's message to be $\widehat{M} = m$ if measurement result $m$ is obtained.  Charlie will then attempt to simulate the channel $\boldsymbol{\CN}'^{\otimes n}$, by associating a separate classical register $\widehat{X}_i$ to each channel $\CN\colon A'_i\rightarrow C_i$ in the product, preparing the states $\ket{X_i(m)}^{\widehat{X}_i}$, for each $1\leq i \leq n$.
Additionally, he stores the result of the measurement in the system $\widehat{M}$, his declaration of the message intended by Alice.
This procedure results in the global state
\[\Gamma^{BC^n\widehat{X}^n\widehat{M}} =
\sum_{m=1}^{2^{nR}}\left(1\otimes\sqrt{\Lambda_m}\right)\xi_1\left(1\otimes\sqrt{\Lambda_m}\right)
\otimes \proj{X^n(m)}^{\widehat{X}^n}\otimes \proj{m}^{\widehat{M}}.\]
Let $\Theta^{BC^n\widehat{X}^n} = \tr_{\widehat{M}} \Gamma$.
If Charlie was able to perfectly reconstruct Alice's classical message, $\Gamma$ would instead be
\[\Gamma' = \xi_1 \otimes \proj{X^n(1)}^{\widehat{X}^n}\otimes\proj{1}^{\widehat{M}},\]
with $\Theta' = \tr_{\widehat{M}}\Gamma'$.
When averaged over Alice's random choice of HSW code, $\Theta'$
is precisely equal to the state which would arise via the action of the modified channel $\boldsymbol{\CN}'.$ This is because
\begin{eqnarray}
\E_\CC \Theta' &=&
\sum_{x^n} p(x^n) \xi_{x^n}\otimes\proj{x^n}^{\widehat{X}^n} \nonumber \\
&=& \boldsymbol{\CN}'^{\otimes n}(\Upsilon), \label{ECU}
\end{eqnarray}
where we have written the joint state which results when Alice prepares  $\phi_{x^n}$ as
\[\xi_{x^n}^{BC^n} = \CN^{\otimes n}(\phi_{x^n}\otimes \Upsilon).\]
However, our choice of a good HSW code ensures that he can almost perfectly reconstruct Alice's message.  A consequence of this will be that the two states $\Theta$ and $\Theta'$ are almost the same, as we will now demonstrate.

In what follows, we will need to explicitly keep track of the randomness
in our codes, by means of superscripts which are to be interpreted as indexing the deterministic codes which occur with the probabilities $p_\CC$ and $q_\beta$.
Rewriting (\ref{EP}) as
\[\sum_{\CC\beta} p_\CC q_\beta
 \tr\left(1\otimes\Lambda_1^\CC\right)\xi_1^{\CC\beta}\geq 1-2\epsilon,\]
it is clear that we may write
\[\tr\left(1\otimes\Lambda_1^\CC\right)\xi_1^{\CC\beta} \geq  1-\epsilon_{\CC\beta},\]
for positive numbers $\{\epsilon_{\CC\beta}\}$ chosen to satisfy
\[\sum_{\CC\beta} p_\CC q_\beta \epsilon_{\CC\beta} = 2\epsilon.\]
By the gentle measurement lemma,
\[\left| \left(1\otimes\sqrt{\Lambda_1^\CC}\right)
    \xi_1^{\CC\beta}
   \left(1\otimes\sqrt{\Lambda_1^\CC}\right)
   - \xi_1^{\CC\beta}\right|_1 \leq \sqrt{8\epsilon_{\CC\beta}},\]
and thus, by the concavity of the square root function,
\begin{eqnarray*}
& & \hspace{-.5in}\E_{\CC\beta} \left|\left(1\otimes\sqrt{\Lambda_1}\right) \xi_1  \left(1\otimes\sqrt{\Lambda_1}\right) - \xi_1\right|_1  \\  
 & = & \sum_{\CC\beta} p_\CC q_\beta 
\left|\left(1\otimes\sqrt{\Lambda_1^\CC}\right)
    \xi_1^{\CC\beta}
   \left(1\otimes\sqrt{\Lambda_1^\CC}\right)
   - \xi_1^{\CC\beta}\right|_1 \\
& \leq & 4\sqrt{\epsilon}. 
\end{eqnarray*}
Along with (\ref{EP}) and monotonicity with respect to $\tr_{\widehat{M}}$, this estimate lets us write
\begin{eqnarray}
\E_{\CC\beta}|\Theta - \Theta'|_1
&\leq& \E_{\CC\beta}|\Gamma - \Gamma'|_1 \nonumber \\ 
&=& \E_{\CC\beta}\left|\left(1\otimes\sqrt{\Lambda_1}\right)
      \xi_1\left(1\otimes\sqrt{\Lambda_1}\right) - \xi_1\right|_1 
      \nonumber \\
& &  + \E_{\CC\beta}\sum_{m=2}^{2^{nR}}
\left|\left(1\otimes\sqrt{\Lambda_m}\right)
      \xi_1\left(1\otimes\sqrt{\Lambda_m}\right)\right|_1  \nonumber \\
&=&\E_{\CC\beta}\left|\left(1\otimes\sqrt{\Lambda_1}\right)
      \xi_1\left(1\otimes\sqrt{\Lambda_1}\right) - \xi_1\right|_1 \\
& & + \E_{\CC\beta}\sum_{m=2}^{2^{nR}} \tr(1\otimes\Lambda_m)\xi_1  \nonumber \\
&\leq& 4\sqrt{\epsilon} + 2\epsilon  \nonumber \\
&\leq& 5\sqrt{\epsilon}, \label{ECbU}
\end{eqnarray}
provided that $\epsilon\leq \frac{1}{2}.$
Since the the entanglement fidelity is linear in $\CD(\Theta),$ which is itself linear in $\Theta,$ we can also use the special triangle inequality to write
\begin{eqnarray*}
F(\ket{\Phi},\CD(\E_{\CC\beta}\Theta))
&=& F(\ket{\Phi},\E_\beta\CD(\E_\CC\Theta))\\
&\geq& F\big(\ket{\Phi},\E_\beta\CD(\E_\CC\Theta')\big)
-\big|\E_\beta\CD(\E_\CC\Theta')-\E_\beta\CD(\E_\CC\Theta)\big|_1.
\end{eqnarray*}
Using our earlier observation from (\ref{ECU}) and the definition of a $(Q,n,\epsilon)$ entanglement transmission code, we can bound the first term as
\begin{eqnarray*}
F(\ket{\Phi},\CD(\E_\CC\Theta'))
&=& F(\ket{\Phi},\CD\circ\boldsymbol{\CN}'^{\otimes n}\circ\Upsilon) \\
&\geq& 1-\epsilon.
\end{eqnarray*}
An estimate on the second term is obtained via
\begin{eqnarray*}
\left|\E_\beta\CD(\E_\CC\Theta) - \E_\beta\CD(\E_\CC\Theta') \right|_1
&\leq& \E_\beta\left|\CD(\E_\CC\Theta)-\CD(\E_\CC\Theta') \right|_1\\
&\leq& \E_\beta\left|\E_\CC\Theta - \E_\CC\Theta'\right|_1\\
&\leq& \E_{\CC\beta}\left|\Theta - \Theta'\right|_1\\
&\leq& 5\sqrt{\epsilon},
\end{eqnarray*}
where first three lines are by convexity, monotonicity, and convexity once again of the trace norm.  The last inequality follows from (\ref{ECbU}).
Putting these together gives
\begin{eqnarray}
\E_{\CC\beta}F(\ket{\Phi},\CD(\Theta))
&\geq& 1-\epsilon - 5\sqrt{\epsilon} \nonumber\\
&\geq& 1- 6\sqrt{\epsilon}. \label{EF}
\end{eqnarray}
At last, observe that the final decoded state $\Omega$ (which still depends on both sources of randomness $\CC$ and $\beta$) is equal to 
\[\Omega^{B\widehat{B}\widehat{M}} = 
\CD(\Gamma^{BC^n\widehat{X}^n\widehat{M}}) 
\equiv \boldsymbol{\CD}(\xi_1^{BC^n}),\]
implicitly defining the desired decoding instrument $\boldsymbol{\CD}\colon C^n \rightarrow \widehat{B}\widehat{M}$.
The expectation of (\ref{cqIs}) can now be bounded as   
\begin{eqnarray*}
\E_{\CC\beta} 2^{-nR}\sum_{m=1}^{2^{nR}} P_s^{\text{eg}}(m) &=&
%\E_{\CC\beta} 2^{-nR}\sum_{m=1}^{2^{nR}} F\big(\ket{m}\ket{\Phi},
%\boldsymbol{\CD}\circ\CN^{\otimes n}(\phi_m\otimes \Upsilon)\big) \\
\E_{\CC\beta} P_s^{\text{eg}}(1) \\
%&=& \E_{\CC\beta} F(\ket{1}\ket{\Phi}, \Omega) \\
&=& F(\ket{1}\ket{\Phi}, \E_{\CC\beta}\Omega) \\
&\geq& 
1 - \left|\tr_{B\widehat{B}} \E_{\CC\beta}\Gamma - \proj{1}\right|_1
- 3\big(1 - F(\ket\Phi,\CD(\Theta))\big) \\
&\geq& 1 - 2\sqrt{2\epsilon} - 18\sqrt{\epsilon} \\
&\geq& 1 - 21\sqrt{\epsilon}.
\end{eqnarray*}
The third line above is by Lemma \ref{lemma:transitivity}. The first estimate in the fourth line follows from (\ref{EP}), while the second estimate is by (\ref{EF}), together with (\ref{trfide}).
We may now conclude that there are particular values of the randomness indices $\beta$ and $\CC$ such that the same bound is satisfied for a deterministic code.
We have thus proven that $(\{\phi_m\}_{m\in 2^{nR}},\CE,\boldsymbol{\CD})$ comprises a
$(R,Q,n,21\sqrt{\epsilon})$ entanglement generation code.  This concludes the coding theorem.
\end{proof}

\pagebreak

\begin{proof}[Proof of Theorem~2 (coding theorem)]
Begin by fixing bipartite pure states $\ket{\Psi_1}^{A''A'}$ and $\ket{\Psi_2}^{B''B'}$ which give rise to the state
\[\omega^{A''B''C} = (1^{A''B''}\otimes\CN^{\otimes n})(\Psi_1\otimes\Psi_2),\]
and defining $\rho_1^{A'} = \tr_A\Psi_1$, $\rho_2^{B'} = \tr_B\Psi_2.$
Letting $\epsilon, \delta>0$ be arbitrary,
we will show that there exists a $(Q_a,Q_b,n,\epsilon)$ qq entanglement transmission code where
\[Q_a=I_c(A''\,\rangle C)_\omega-\delta \text{ and }
Q_b=I_c(B''\,\rangle A''C)_\omega-\delta\]
provided that $Q_a,Q_b \geq 0$.
Note that the rates in Theorem~2 will be implied by taking the channel to be $\CN^{\otimes k},$ with $\omega^{ABC^k}$ defined similarly.

Let us begin by choosing an isometric extension $\CU_\CN:A'B'\rightarrow CE$ of $\CN$.  
Define the ideal channel $\CN_1\colon A'\rightarrow C$ which would effectively be seen by Alice were Bob's average code density operator exactly equal to $\rho_2^{\otimes n}$  as
\[\CN_1\colon \tau\mapsto \CN(\tau \otimes\rho_2).\]
We now use $\CU_\CN$ to define a particular isometric extension 
$\CU_{\CN_1}\colon A'\rightarrow CE'$ of $\CN_1$, where $E'=B''E$, as 
\[\CU_{\CN_1}\colon\tau \mapsto \CU_\CN(\tau\otimes \Psi_2).\]
Observe that Bob's fake input $B''$ is treated as part of the environment of Alice's ideal induced channel.   
We then further define the channel $\CN_2\colon B'\rightarrow A''C$ by 
\[\CN_2\colon \tau\mapsto \CN(\Psi_1\otimes\tau).\]
In contrast to the interpretation of $\CN_1$, this may be viewed as the channel 
which would be seen by Bob if Alice were to input the $A'$ part of the purification $\ket{\Psi_2}^{A''A'}$ of $\rho_2^{A'}$ to her input of the channel and then send the $A''$ system to Charlie via a noiseless quantum channel.  As in the proof of Theorem~1, Charlie will first decode Alice's information, after which he will attempt to simulate the channel $\CN_2$, allowing a higher transmission rate for Bob than if Alice's information was treated as noise.  Since quantum information cannot be copied, showing that this is indeed possible will require different techniques than were utilized in the previous coding theorem.
Although ensembles of random codes will be used in this proof, we introduce the technique of \emph{coherent coding}, in which we pretend that the common randomness is purified.  
The main advantage of this approach will be that working with states in the enlarged Hilbert space allows monotonicity to be easily exploited in order to provide the estimates we require.  
Additionally, before we  derandomize at the end of the proof, it will ultimately be only Bob who is using a random code.  Alice will be able to use any deterministic code from her random ensemble, as Charlie will implement a decoding procedure which produces a global state which is close to that which would have been created had Alice coded with the coherent randomness.  To show this, we will first analyze the state which would result if both senders used their full ensembles of codes.  Then we show that if Alice uses any code from her ensemble, Charlie can create the proper global state himself, allowing him to effectively simulate $\CN_2$ and ultimately decode both states at the desired rates.

By Proposition~\ref{prop:lsd}, for large enough $n$, there exists a $(Q_a,n,\epsilon)$ random entanglement generation code $(p_\ell,\ket{\Upsilon_1^\ell}^{AA'^n},\CD_1^\ell)$ for the channel $\CN_1,$ where  $Q_a = I_c(\rho_1,\CN_1)-\delta = I_c(A''\,\rangle C)-\delta.$ 
There similarly exists a $(Q_b,n,\epsilon)$ random entanglement generation code $(q_m,\ket{\Upsilon_2^m}^{BB'^n},\CD_2^m)$ for $\CN_2$, with $Q_b = I_c(\rho_2,\CN_2)-\delta = I_c(B''\,\rangle A''C) - \delta$.  
Proposition~\ref{prop:lsd} further guarantees that these codes can be chosen so that their 
respective average code density operators 
\[\varrho_1^{A'^n} = \sum_\ell p_\ell \tr_A\Upsilon_1^\ell \,\text{ and }\, \varrho_2^{B'^n} = \sum_m q_m \tr_B\Upsilon_2^m\] satisfy
\begin{eqnarray}
|\varrho_i - \rho_i^{\otimes n}|_1\leq \epsilon. \label{vrhorho}
\end{eqnarray}
Recall that by Proposition~\ref{prop:lsd} we may choose isometric extensions $\CU^\ell_{\CD_1}\colon C^n\rightarrow \widehat{A}F$ implementing the $\CD_1^\ell$ from Alice's random code
which satisfy
\begin{eqnarray}
F\left(\ket{\Phi_1}^{A\widehat{A}}\ket{\lambda}^{FE'^n},\CU^\ell_{\CD_1}\circ\CU_{\CN_1}^{\otimes n}\big(\Upsilon_1^\ell\big)\right) \geq 1-\epsilon \label{UD1lcond}
\end{eqnarray}
for every random code index $\ell$ and the same fixed state $\ket{\lambda}^{FE'^n}$. 
%\footnote{The construction in \cite{dev} ensures this can be done, as all codes in the random ensemble are based on the same $\rho_1$.  In the notation of that paper, each isometric decoder can be chosen so that the fixed environment state $\theta^\CE$ is purified by the same $\ket{\Phi_\theta}^{\CQ\CE\CB'}$.  Our $F$ (resp. $E'^n$) is the other paper's $\CQ\CB'$ (resp. $\CE$), so we choose $\ket{\lambda}^{FE'^n} = \ket{\Phi_\theta}$.}

Let the code common randomness between Alice and Charlie be held between the systems $L_A$ and $L_C$, represented by the state  
\[\gamma_1^{L_AL_C} = \sum_\ell p_\ell \proj{\ell}^{L_A}\otimes \proj{\ell}^{L_C},\]
defining a similar state $\gamma_2^{M_BM_C}$ for the Bob-Charlie common randomness.  For convenience, let us further pretend that $\gamma_1$ is part of a pure state 
\[\ket{\Gamma_1}^{L_EL_AL_B} = \sum_\ell\sqrt{p_\ell} 
\ket{\ell}^{L_E}\ket{\ell}^{L_A}\ket{\ell}^{L_C}.\]
Similarly, let $\gamma_2$ by purified by $\ket{\Gamma_2}^{M_EM_BM_C}$.
Write controlled encoding isometries 
%\footnote{The construction from \cite{dev} of entanglement transmission codes constructs the $\CE_1^\ell$ and $\CE_2^m$ as isometries.}
$\CE_1\colon L_A\rightarrow~L_AA'^n$ and $\CE_2\colon M_B\rightarrow M_BB'^n$ as 
\[\CE_1 = \sum_\ell \ket{\ell}\ket{\Upsilon_1^\ell}\bra{\ell} 
\,\text{  and  }\,
\CE_2 = \sum_m \ket{m}\ket{\Upsilon_2^m}\bra{m}.\]
The states which would arise if Alice and Bob each encoded \emph{coherently} are
\begin{eqnarray*}
\ket{\Upsilon_1}^{LAA'^n} &\equiv& \CE_1\ket{\Gamma_1} = 
\sum_\ell \sqrt{p_\ell}\ket{\ell}^L\ket{\Upsilon_1^\ell} \\ 
\ket{\Upsilon_2}^{MBB'^n} &\equiv& \CE_2\ket{\Gamma_2} = 
\sum_m \sqrt{q_m} \ket{m}^M\ket{\Upsilon^m_2}.
\end{eqnarray*}
Note that we have abbreviated $L = L_EL_AL_C$ and $M = M_EM_BM_C$.
As each $\ket{\Upsilon_i}$ is a purification of $\varrho_i$, together with (\ref{vrhorho}), Uhlmann's theorem tells us that there exist unitaries $V_1\colon LA\rightarrow A''^n$ and $V_2\colon MB\rightarrow B''^n$ such that 
\begin{eqnarray}
F\left(V_i\ket{\Upsilon_i},\ket{\Psi_i}^{\otimes n}\right) \geq 1-\epsilon. \label{FUV}
\end{eqnarray}
Further define a corresponding controlled isometric decoder $\CU_{\CD_1}\colon L_CC^n\rightarrow L_C \widehat{A}F$ for Alice's code as
\[\CU_{\CD_1} = \sum_\ell \proj{\ell}^{L_C}\otimes \CU^{\ell}_{\CD_1}.\]
%Charlie will first decode Alice's state isometrically.  Then, he will invert the decoding in such a way that the resulting global states appears as though Alice had used the full random ensemble.  Charlie next uses local unitaries to create a state which appears as though Bob had sent his state through the channel $\CN_2^{\otimes n},$ allowing Charlie to complete his decoding by using the decoder $\CD_2$.  We will then use the existence of this $\CD$ to infer the existence of a good deterministic code. 
Let us now imagine that each of Alice and Bob encodes using the coherent common randomness, resulting in a joint pure state 
$\CU_\CN^{\otimes n}\ket{\Upsilon_1}\ket{\Upsilon_2}$ on $LAMBC^nE^n$.  
If Charlie then applies the full controlled decoder from Alice's code, the resulting global pure state would be  
\[\ket{\Theta}^{LA\widehat{A}MBFE^n} =  
\CU_{\CD_1}\circ\CU_{\CN}^{\otimes n}
\ket{\Upsilon_1}\ket{\Upsilon_2}.\]
For each $\ell$, let us define an isometry 
$\CO^\ell\colon B'^n\rightarrow A\widehat{A}FE^n$ as
\[\CO^\ell = \CU^\ell_{\CD_1}\circ\CU_{\CN}^{\otimes n}\big(\Upsilon_1\otimes \cdot\,\big)\]
which we use to define the pure states 
\[\ket{\theta_\ell}^{A\widehat{A}MFBE^n} =
\CO^\ell\ket{\Upsilon_2}.\] 
These definitions allow us to express
\[\ket{\Theta} = \sum_\ell \sqrt{p_\ell} \ket{\ell}^L  \ket{\theta_\ell}.\]
Further writing $\ket{\lambda'}^{FMBE^n} \equiv V_2^{-1}\ket{\lambda}^{FB''^nE^n},$
the following bound applies
\begin{eqnarray}
F\left(\ket{\Phi_1}^{A\widehat{A}}\ket{\lambda'}^{FMBE^n},
  \ket{\theta_\ell}\right) 
&=& F\left(\ket{\Phi_1}\ket{\lambda'}^{FMBE^n},
  \CO^\ell\ket{\Upsilon_2}\right) \nonumber \\
&=& F\left(\ket{\Phi_1}\ket{\lambda}^{FB''^nE^n},
  \CO^\ell\circ V_2\ket{\Upsilon_2} \right) \nonumber \\
&\geq& 1 - 2\sqrt{1-F\left(\ket{\Phi_1}\ket{\lambda}^{FB''^nE^n},
  \CO^\ell\ket{\Psi_2}^{\otimes n} ) \right)}  \nonumber \\ 
  & & \,\,\,
  - 2\sqrt{1- F\left(V_2\ket{\Upsilon_2},\ket{\Psi_2}^{\otimes n} 
  \right)} \nonumber \\
&\geq& 1-2\sqrt{1-F\left(\ket{\Phi_1}\ket{\lambda}^{FE'^n},
  \CU^\ell_{\CD_1}\circ\CU_{\CN_1}^{\otimes n}\circ\ket{\Upsilon_1^\ell}\right)} - 2\sqrt{\epsilon} \nonumber \\
&\geq& 1-4\sqrt{\epsilon}. \nonumber
\end{eqnarray}
Above, the second equality is because the actions of $\CO^\ell$ and $V_2$ commute, the first inequality is by the triangle inequality and monotonicity with respect to $\CO^\ell,$ while for the second inequality, we have just rewritten the first term and used (\ref{FUV}) for the second. 
The last bound is from (\ref{UD1lcond}).
Observe that we are still free to specify the global phases of the outputs of the $\CU_{\CD_1}^\ell$ so that the above bound further implies $\bra{\theta_\ell}\ket{\Phi_1}\ket{\lambda'} \geq (1-4\sqrt{\epsilon})^{1/2}$ for each $\ell$.   
% \begin{eqnarray*}
% \mathrm{Real}\big\{\bra{\Phi_1}\bra{\lambda'}\ket{\Theta_\ell}\big\} &\geq& \sqrt{1-2\sqrt{\epsilon}} \\
%  &\geq& 1-2\sqrt{\epsilon}
% \end{eqnarray*}
Consequently, 
\begin{eqnarray}
F(\ket{\Theta},\ket{\Gamma_1}\ket{\Phi_1}\ket{\lambda'}) 
&=& \left|\sum_{\ell\ell'} \sqrt{p_\ell p_{\ell'}}\bra{\ell}\ket{\ell'}
\bra{\theta_\ell}\ket{\Phi_1}\ket{\lambda'}\right|^2 \nonumber \\
&=& \left|\sum_{\ell}p_\ell \bra{\theta_\ell}\ket{\Phi_1}\ket{\lambda'}\right|^2 \nonumber \\
&\geq& 1-4\sqrt{\epsilon}. \nonumber 
%&\geq& 1-4\sqrt{\epsilon}. \nonumber
\end{eqnarray}
Essentially, the subsystems $L$, $A\widehat{A}$ and $MBFE^n$ of $\ket{\Theta}$ are mutually decoupled.

As mentioned earlier, it will be sufficient for Alice to use \emph{any} deterministic code from the random ensemble to encode.  Without loss of generality, we assume that Alice chooses to use the first code $(\ell = 1)$ in her ensemble.  Bob, on the other hand, will need to use randomness to ensure that Alice's effective channel is close to a product channel.  
The state on $AMBC^nE^n$ which results from these encodings is $\CU_\CN^{\otimes n}\ket{\Upsilon_1^1}\ket{\Upsilon_2}$.

We will now describe a procedure by which Charlie first decodes Alice's information, then produces a global state which is close to $\ket{\Theta}$, making it look like Alice had in fact utilized the coherent coding procedure.  This will allow Charlie to apply local unitaries to effectively simulate the channel $\CN_2$ for which Bob's random code was designed, enabling him to decode Bob's information as well.   These steps will constitute Charlie's decoding 
$\CD:M_CC^n\rightarrow M_C\widehat{A}\widehat{B}$, which depends on the Bob-Charlie common randomness.  The existence of a deterministic decoder will then be inferred. 

Charlie first applies the isometric decoder $\CU_{\CD_1}^1$, placing all systems into the state $\ket{\theta_1}$.   
He then removes his local system $\widehat{A}$ (it is important that he keep $\widehat{A}$ in a safe place, as it represents the decoder output for Alice's quantum information) and replaces it with the corresponding parts of the locally prepared pure state
$\ket{\Phi_1}^{A^\circ\widehat{A}^\circ}.$  Charlie also locally prepares the state $\ket{\Gamma_1}^L$.
%\[\ket{\Gamma_1}^{L^\circ}\ket{\Phi_1}^{A^\circ\widehat{A}^\circ}\] 
The resulting state 
\[\Theta' = \Gamma_1^{L}\otimes\Phi_1^{A^\circ\widehat{A}^\circ}
\otimes \tr_{A\widehat{A}}\theta_1,\]
satisfies 
\begin{eqnarray}
F(\Theta',\Theta) 
&\geq& 1 - 
\left|\tr_{A\widehat{A}}\theta_1 - \lambda'\right|_1
  - \left|\lambda' - \tr_{LA\widehat{A}}\Theta\right|_1 \nn\\
& & \,- 3\left(1-F\big(\ket{\Gamma}\ket{\Phi_1},\tr_{MBFE^n}\Theta\big)\right) 
\nonumber \\
%&\geq& 1 - 
%- 6\sqrt{\epsilon} \\
&\geq& 1 - 2\sqrt{4\sqrt{\epsilon}} - 2\sqrt{4\sqrt{\epsilon}}
- 12\sqrt{\epsilon} \nonumber \\
&\geq& 1 - 9\epsilon^{1/4} 
\label{FThTh}
\end{eqnarray}
whenever $\epsilon \leq 12^{-4}$.
The first line combines Lemma 2 and the triangle inequality.  The first two estimates in the second line are from 
applying (\ref{trfide}) and monotonicity with respect to $\tr_{A\h{A}}$ and $\tr_{LA\h{A}}$ to the previous two estimates.
The last estimate in that line is 
from monotonicity with respect to the map $\tr_{MBFE^n}$ applied to the previous estimate. 
Next, Charlie will 
apply $V_1\circ\CU_{D_1}^{-1}$  to $\Theta'$
\footnote{This operation only acts on Charlie's local systems, i.e.   $V_1\circ\CU_{\CD_1}^{-1}\colon L A^\circ\widehat{A}^\circ F \rightarrow A''^nC^n$.}
in order to simulate the channel $\CN_2$.  To see that this will work, 
define $\CM\colon LA\widehat{A}FE^n\rightarrow A''^nC^n$ as
$\CM\equiv\tr_{E^n}V_1\circ\CU_{\CD_1}^{-1}$
and observe that 
by monotonicity with respect to
$\CN^{\otimes n}(\,\cdot\otimes\Upsilon_2)$ and (\ref{FUV}),
the states on $MBA''^nC^n$ satisfy
\begin{eqnarray}
F\left(\CM(\Theta),\CN_2^{\otimes n}(\Upsilon_2)\right) 
&=& F\left(V_1\circ\CN^{\otimes n}(\Upsilon_1\otimes\Upsilon_2)
,\CN^{\otimes n}(\Psi_1^{\otimes n}\otimes\Upsilon_2)\right) \nonumber \\
&\geq& F\left(V_1\ket{\Upsilon_1},\ket{\Psi_1}^{\otimes n}\right) \nonumber \\
&\geq& 1-\epsilon. \nonumber
\end{eqnarray}
We may now use the triangle inequality and monotonicity with respect to 
 $\CM$ to combine our last two estimates, yielding
\begin{eqnarray}
F\left(\CM(\Theta'),\CN_2^{\otimes n}(\Upsilon_2)\right)
&\geq& 1-2\sqrt{1 - F\left(\CM(\Theta'),\CM(\Theta)\right)}   
  \nn\\
& & \, - 2\sqrt{1-F\left(\CM(\Theta),\CN_2^{\otimes n}(\Upsilon_2)\right)} \nonumber \\   
&\geq& 1 - 2\sqrt{9\epsilon^{1/4}} - 2\sqrt{\epsilon}\nonumber \\
&\geq& 1-7\epsilon^{1/8} \label{FMN2}
\end{eqnarray}
whenever $\ep \leq 2^{-8/3}$.
We have thus far shown that Charlie's decoding procedure succeeds in simulating the channel $\CN_2^{\otimes n}$, while simultaneously recovering Alice's quantum information.  
%\[\Delta^{MA\widehat{A}BA''^nC^n} = \tr_L V_1^{-1}\circ\CU_{\CD_1}^{-1}\left(\tr_{E^n}\Theta'\right).\]
%where $\xi = \tr_{LA\widehat{A}E^n} \Theta.$
%\begin{eqnarray}
%F\left(\Delta,\CN_2^{\otimes n}(\Upsilon_2)\right)
%\end{eqnarray}
Charlie now uses the controlled decoder 
$\CD_2\colon M_CA''^nC^n\rightarrow M_C\widehat{B}$ defined as
\[\CD_2 = \sum_m \proj{m}^{M_C}\otimes \CD_2^m\]
to decode Bob's quantum information.
This entire procedure has defined our decoder $\CD:M_CC^n\rightarrow M_C\widehat{A}\widehat{B}$ which gives rise to a global state $\Omega^{A\widehat{A}B\widehat{B}}$ representing the final output state of the protocol, averaged over Bob's common randomness.
This state satisfies
\begin{eqnarray}
F(\ket{\Phi_1},\tr_{B\widehat{B}} \Omega) &\geq& F(\Theta,\Theta') 
\nonumber \\
&\geq& 1-9\epsilon^{1/4}, \nonumber
\end{eqnarray}
because of monotonicity with respect to $\tr_{LMBFE^n}$ applied to the bound (\ref{FThTh}).  By using the triangle inequality, the fact that Bob's codes are $\epsilon$-good for each $m$, and monotonicity of the estimate (\ref{FMN2}) with respect to $\tr_M\CD_2$, the global state can further be seen to obey
\begin{eqnarray}
F\big(\ket{\Phi_2},\tr_{A\widehat{A}} \Omega\big) &=& F\big(\ket{\Phi_2},\tr_M\CD_2\circ\CM(\Theta')\big) \nonumber \\
&\geq& 1 - 
2\sqrt{1-F\big(\ket{\Phi_2},\tr_M\CD_2\circ\CN_2^{\otimes n}(\Upsilon_2)\big)} \nonumber \\
& & \,\,\, - 2\sqrt{1- 
F\big(\tr_M\CD_2\circ\CN_2^{\otimes n}(\Upsilon_2),
\tr_M\CD_2\circ\CM(\Theta')\big)} \nonumber \\
&\geq& 1 - 2\sqrt{\epsilon} - 2\sqrt{7\epsilon^{1/8}} \nonumber \\
&\geq& 1 - 7 \epsilon^{1/16} \nonumber
\end{eqnarray}
as long as $\ep \leq 2^{-16/7}$.
Along with (\ref{trfide}), a final application of Lemma 2 combines the above two bounds to give
\begin{eqnarray*}
F(\ket{\Phi_1}\ket{\Phi_2},\Omega) &\geq&
1 - \left|\Phi_1 - \tr_{B\widehat{B}}\Omega\right|_1
- 3\Big(1-F\big(\ket{\Phi_2},\tr_{A\widehat{A}}\Omega\big)\Big) \\
&\geq& 1 - 2\sqrt{9\epsilon^{1/4}} - 21 \epsilon^{1/16} \\
&\geq& 1 - 22 \epsilon^{1/16},
\end{eqnarray*}
provided that $\epsilon \leq 6^{-16}$.
Since this estimate represents an average over Bob's common randomness, there must exist a particular value $m^*$ of the common randomness so that the corresponding deterministic code is at least as good as the random one, thus concluding the coding theorem.
\end{proof}

\section{The converse theorems} \label{section:converses}
We will now demonstrate that
\[\CC\CQ_{\text{eg}}(\CN)\subseteq\bigcup_{k=1}^\infty\frac{1}{k}\CC\CQ^{(1)}(\CN^{\otimes k}) \,\,\,\text{and}\,\,\,
\CQ_{\text{eg}}(\CN)\subseteq\bigcup_{k=1}^\infty\frac{1}{k}\CQ^{(1)}(\CN^{\otimes k}),\]
where the single-letter regions $\CC\CQ^{(1)}(\CN)$ and $\CQ^{(1)}(\CN)$ are those defined at the beginning of the last section.

\begin{proof}[Proof of Theorem~1 (converse)]
Suppose there exists a sequence of $(R,Q,n,\epsilon_n)$
entanglement generation codes with $\epsilon_n\rightarrow 0$.  Fixing a blocklength $n$, let $\{\phi_m\}, \Upsilon^{BB'^n}, \boldsymbol{\CD}$ comprise the corresponding cq entanglement generation code.
The state induced by the encoding is
\[\omega^{MBC^n} = 2^{-nR}\sum_{m=1}^{2^{nR}}
\proj{m}^M\otimes (1^B\otimes\CN^{\otimes n})(\phi_m\otimes \Upsilon).\]
After application of the decoding instrument $\boldsymbol{\CD}\colon C^n\rightarrow \widehat{B}\widehat{M}$, this state becomes
\[\Omega^{M\widehat{M}B\widehat{B}} = (1^{MB}\otimes\boldsymbol{\CD})(\omega).\]
An upper bound on the classical rate of the code can be obtained as follows:
\begin{eqnarray*}
nR &=& H(M)_\Omega \\
&=& I(M;\widehat{M})_\Omega + H(M|\widehat{M})_\Omega \\
&\leq& I(M;\widehat{M})_\Omega + H(\epsilon_n) + nR\epsilon_n \\
&\leq& I(M;C^n)_\omega + n\epsilon'_n.
\end{eqnarray*}
The first inequality follows from Fano's inequality (Lemma~\ref{lemma:fano}) while in the second we use the Holevo bound (Lemma~\ref{lemma:holevo}) and define $\epsilon_n' = \frac{1}{n} + R\epsilon_n$.
The quantum rate of the code is upper bounded as
\begin{eqnarray*}
I_c(B\,\rangle C^nM)_\omega &\geq& I_c(B\,\rangle \widehat{B}M)_\Omega\\
&\geq& I_c(B\,\rangle \widehat{B})_\Omega \\
&\geq& I_c(B\,\rangle \widehat{B})_\Phi - 2H(\epsilon_n) - 8nQ\sqrt{\epsilon_n} \\
&=& nQ - n\epsilon_n''.
\end{eqnarray*}
Above, the first two inequalities are consequences of the data processing inequality (Lemma~\ref{lemma:dataproc}), while the last inequality applies a combination of Lemma~\ref{lemma:continuity} and (\ref{trfide}), along with the definition
$\epsilon_n'' = \frac{2}{n} + nQ\sqrt{\epsilon_n}$.
Setting $X=M$, we have thus proven that
\[R \leq \frac{1}{n}I(X;C^n) + \epsilon'_n, \,\,\,\,\,\,
Q\leq \frac{1}{n}I_c(B\,\rangle C^nX) + \epsilon''_n\]
whenever $(R,Q)$ is an achievable cq rate pair for entanglement generation, where $\epsilon'_n,\epsilon''_n\rightarrow 0$. 
It follows that for any achievable rate pair $(R,Q)$ and any $\delta>0$, we have \[(R-\delta,Q-\delta) \in 
\frac{1}{n}\CC\CQ^{(1)}(\CN^{\otimes n}) \subseteq \CC\CQ(\CN).\]
Since $\CC\CQ(\CN)$ is closed by definition, this completes the proof.
\end{proof}

\begin{proof}[Proof of Theorem~2 (converse)]
Suppose that $(Q_a,Q_b)$ is an achievable qq rate pair for entanglement generation.  By definition, this means that there must exist a sequence of $(Q_a,Q_b,n,\epsilon_n)$ entanglement generation codes with $\epsilon_n\rightarrow 0$.  Fixing a blocklength $n$,
let $\ket{\Upsilon_1}^{AA'^n}, \ket{\Upsilon_2}^{BB'^n}$ and
$\CD\colon C^n\rightarrow \widehat{A}\widehat{B}$ comprise the corresponding encodings and decodings. Define
\[\omega^{ABC^n} =
(1^{AB}\otimes\CN^{\otimes n})(\Upsilon_1\otimes\Upsilon_2)\]
to be the result of sending the respective $A'^n$ and $B'^n$ parts of $\Upsilon_1$ and $\Upsilon_2$ through the channel $\CN^{\otimes n}$.  Further defining
\[\Omega^{AB\widehat{A}\widehat{B}} = (1^{AB}\otimes\CD)(\omega)\]
as the corresponding state after decoding,
the entanglement fidelity of the code is given by
\begin{eqnarray}
F_{AB}=F(\ket{\Phi_1}\otimes\ket{\Phi_2},\Omega) \geq 1-\epsilon_n. 
\label{fabconv}
\end{eqnarray}
where $\ket{\Phi_1}^{A\widehat{A}}$ and $\ket{\Phi_2}^{B\widehat{B}}$
are the maximally entangled target states.  The sum rate can be bounded as 
\begin{eqnarray*}
I_c(AB\,\rangle C^n)_\om 
&\geq& I_c(AB\,\rangle \h{A}\h{B})_\Om \\
&\geq& I_c(AB\,\rangle \h{A}\h{B})_{\Phi_1\!\otimes\Phi_2} 
- 2H(\ep_n) - 8n(Q_a+Q_b)\sqrt{\ep_n}\\ 
%&\geq& n(R+S) - 2H(\ep_n) - 8n(R+S)\sqrt{\ep_n} \\
&\geq& n(Q_a+Q_b) - n\ep'_n.
\end{eqnarray*}
The first step is by the data processing inequality (Lemma~\ref{lemma:dataproc}).  
The second step uses 
Lemma~\ref{lemma:continuity} and (\ref{trfide}), along with monotonicity applied to (\ref{fabconv}).  The last step has defined 
$\ep'_n = \frac{2}{n} - 8(Q_a+Q_b)\sqrt{\ep_n}$ and holds because the binary entropy $H(\cdot)$ is upper bounded by 1. 
We can bound Alice's rate $Q_a$ by writing
\begin{eqnarray*}
I_c(A\,\rangle BC^n)_\omega
&\geq& 
I_c(A\,\rangle C^n)_\omega \\
&\geq& 
  I_c(A\,\rangle \widehat{A}\widehat{B})_{\Omega} \\
&{\geq}& 
  I_c(A \,\rangle \widehat{A})_{\Omega} \\
&{\geq}& 
  I_c(A \,\rangle \widehat{A})_{\Phi_1} -2H(\epsilon_n) - 8nQ_a\sqrt{\ep_n} \\
 &\geq&  nQ_a - n\ep'_n.
\end{eqnarray*}
The first three steps above are by data processing (Lemma~\ref{lemma:dataproc}).  The remaining steps hold for the same reasons as in the previous chain of inequalities.  Similarly, Bob's rate also must satisfy 
\[nQ_b\leq I_c(B\,\rangle AC^n)_\om + n\ep'_n. \]
Since $\epsilon_n\rightarrow 0$ implies $\epsilon'_n\rightarrow 0,$
this means that for every $\delta > 0$, any achievable qq rate pair $(Q_a,Q_b)$ must satisfy 
\[(Q_a-\delta,Q_b-\delta) \in \frac{1}{n}\CQ^{(1)}(\CN^{\otimes n}) \subseteq \CQ(\CN).\]
Since $\CQ(\CN)$ is closed by definition, this completes the proof.
\end{proof}

\chapter{Transmission of quantum information} \label{chapter:transmission}
%\label{section:sstequiv}
In the previous chapter, we have proven the main theorems for the restricted case in which all quantum communication has been in the sense of \emph{generating} quantum correlations between senders and receiver.  The results of this chapter will complete the proofs of the main theorems, by extending the weaker error criteria of entanglement generation (which incidentally, are analogous to a classical requirement on the \emph{average} probability of error) to the stronger requirements of strong subspace transmission in the main theorem statements.  As a first step, we demonstrate how the results of the last chapter immediately imply the ability to perform an intermediate task, \emph{entanglement transmission}, where the senders are required to transmit preexisting maximal entanglement, while still adhering to an average error criterion on the classical error.  We then show how to use a given entanglement transmission code to construct a strong subspace transmission codes achieving any rates less then those of the original code, while paying a negligible price in fidelity. 

\section{Entanglement transmission} \label{section:et}
\paragraph{Classical-quantum scenario}
%(or creation of $\epsilon$-maximal entanglement)}
In this scenario, rather than generating entanglement with Charlie, Bob will act to transmit \emph{preexisting} entanglement to him.  We assume that 
Bob is presented with the $\tilde{B}$ part of the maximally entangled state $\ket{\Phi}^{B\tilde B}.$  It is assumed that he has complete control over $\tilde{B}$, while he has no access to $B$.  He will perform a physical operation in order to transfer the quantum information embodied in his system $\tilde{B}$ to the inputs
$B'^n$ of the channel, modeled by an encoding operation $\CE\colon \tilde{B}\rightarrow B'^n$.  The goal of this encoding will be to make it possible for Charlie, via post-processing of the information embodied in the system $C^n$, to hold the $\widehat{B}$ part of a state which is close to that which would have resulted if Bob had sent his system through a perfect quantum channel $\text{id}\colon\tilde{B}\rightarrow \widehat{B}$.  Here, we imagine that $\tilde{B}$ and $\widehat{B}$ denote two distinct physical systems with the same number of quantum degrees of freedom.  The role of the identity channel is to set up a unitary correspondence, or isomorphism, between the degrees of freedom of $\tilde{B}$ in Bob's laboratory and those of $\widehat{B}$ in Charlie's.  We will often tacitly assume that such an identity map has been specified ahead of time in order to judge how successful an imperfect
quantum transmission has been.  This convention will be taken for granted many times throughout the paper, wherein specification of an arbitrary state $\ket{\Psi}^{B\tilde{B}}$ will immediately imply specification of the state $\ket{\Psi}^{B\widehat{B}} = (1^B\otimes\text{id})\ket{\Psi}^{B\tilde{B}}.$
Decoding is the same as it is for entanglement generation.  

$(\{\phi_m\}_{m\in 2^{nR}},\CE,\boldsymbol{\CD})$ will be called an $(R,Q,n,\epsilon)$ \emph{cq entanglement transmission code} for the channel $\CN$ if
\begin{eqnarray}
 2^{-nR}\sum_{m=1}^{2^{nR}} P_s^{\text{et}}(m) \geq 1-\epsilon, \label{cqIIs}
\end{eqnarray}
where
\begin{eqnarray}
P_s^{\text{et}}(m) = F\left(\ket{m}\ket{\Phi}^{B\widehat{B}},
\boldsymbol{\CD}\circ\CN^{\otimes n}
(\phi_m^{A'^n}\otimes\CE(\Phi^{B\tilde{B}})\right). \label{cqIIPs}
\end{eqnarray}
Achievable rate pairs and the capacity region $\CC\CQ_{\text{et}}(\CN)$ are defined analogous to those for entanglement generation.

\paragraph{Quantum-quantum scenario}
Alice and Bob each respectively have control over the $\tilde{A}$ and $\tilde{B}$ parts of the separate maximally entangled states
$\ket{\Phi_1}^{A\tilde{A}},\ket{\Phi_2}^{B\tilde{B}}$, while neither has access to $A$ or $B$.  Alice transfers the correlations in her system to the $A'^n$ parts of the inputs of $\CN^{\otimes n}$ with an encoding operation
$\CE_1\colon\tilde{A}\rightarrow A'^n$. Bob acts similarly with $\CE_2\colon \tilde{B}\rightarrow B'^n$.  Their goal is to preserve the respective correlations, 
so that Charlie can apply a decoding operation
$\CD\colon C^n\rightarrow \widehat{A}\widehat{B}$, in order to end up holding the $\widehat{A}\widehat{B}$ part of a state which is close to $\ket{\Phi_1}^{A\widehat{A}}\ket{\Phi_2}^{B\widehat{B}}$. 
Formally,
$(\CE_1,\CE_2,\CD)$ is a $(Q_a,Q_b,n,\epsilon)$ \emph{qq entanglement transmission code} for the channel $\CN$ if
\begin{eqnarray}
F(\ket{\Phi_1}\ket{\Phi_2},\CD\circ\CN^{\otimes n}\circ(\CE_1\otimes\CE_2)(\Phi_1\otimes \Phi_2))\geq 1-\epsilon. \label{qqIIs}
\end{eqnarray}
Achievable qq rate pairs for entanglement generation and the capacity region $\CQ_{\text{et}}(\CN)$ are defined as in the previous scenario.

\section[Equivalence of ET and EG]{Equivalence of entanglement transmission and entanglement generation} \label{section:eteg}
\subsection{$\CC\CQ_{\text{eg}} \subseteq \CC\CQ_{\text{et}}$ and 
$\CQ_{\text{eg}} \subseteq \CQ_{\text{et}}$} \label{section:eteg1}
\begin{proof}
This essentially follows as an artifact of the entanglement generation coding theorem from \cite{dev}.
There, the input preparation $\ket{\Upsilon}^{AA'^n}$ for a $(Q,n)$ entanglement generation code is constructed with the particular form
\[\ket{\Upsilon}^{AA'^n} = \frac{1}{\sqrt{2^{nQ}}}
\sum_{a\in 2^{nQ}} \ket{a}^A\ket{\phi_a}^{A'^n},\]
where the $\{\ket{\phi_a}\}$ are orthogonal.  Observe that the if the encoder acts on the $\tilde{A}$ part of the maximally entangled state
\[\ket{\Phi}^{A\h{A}} =  \frac{1}{\sqrt{2^{nQ}}}
\sum_{a\in 2^{nQ}} \ket{a}^A\ket{a}^{\tilde{A}}\]
with an encoding isometry $\CE\colon \tilde{A}\rightarrow A'^n$ defined via 
\[\CE = \sum_{a\in 2^{nQ}} \ket{\phi_a}^{A'^n}\bra{a}^{\tilde{A}},\]
the identity 
$\CE\ket{\Phi}^{A\tilde{A}} = \ket{\Upsilon}^{AA'^n}$
holds trivially.  It is thus a simple task to modify the proofs of Chapter~\ref{chapter:proofs} to instead prove the existence of the entanglement transmission codes described in the previous section.
Indeed, if $(\ket{\Upsilon}, \{\phi_m\},\boldsymbol{\CD})$ is a 
$(R,Q,n,\epsilon)$ cq entanglement generation code, there then exists an encoder $\CE$ so that $(\CE, \{\phi_m\},\boldsymbol{\CD})$ is a 
$(R,Q,n,\epsilon)$ cq entanglement transmission code.   Identical reasoning shows that to every qq entanglement generation code, there a qq entanglement transmission code with the same parameters.
\end{proof}

\subsection{$\CC\CQ_{\text{et}} \subseteq \CC\CQ_{\text{eg}}$}
\begin{proof}
Suppose there exists an $(R,Q,n,\epsilon)$ cq entanglement transmission code, consisting of classical message states
$\{\ket{\phi_m}^{A'^n}\}_{m\in 2^{nR}},$ a quantum encoding map $\CE\colon \tilde{B}\rightarrow \widehat{B}$, and a decoding instrument
$\boldsymbol{\CD}\colon C^n\rightarrow \widehat{M}\widehat{B}.$
Write any pure state decomposition of the encoded state
\[(1^B\otimes\CE)(\Phi) = \sum_i p_i \proj{\Upsilon_i}.\]
Then, the success condition (\ref{cqIIs}) for a cq entanglement transmission code can be rewritten as
\begin{eqnarray}
1-\epsilon &\leq& 2^{-nR}\sum_{m=1}^{2^{nR}} P_s^{\text{et}}(m) \\
&=& 2^{-nR}\sum_{m=1}^{2^{nR}}
 F\Big(\ket{\Phi}^{B\widehat{B}},
\CD_m\circ\CN^{\otimes n}\big(\phi_m^{A'^n}\otimes
\big(\sum_i p_i \Upsilon_i\big)\big)\Big) \\
&=& \sum_i p_i \left(2^{-nR}\sum_{m=1}^{2^{nR}}
F\left(\ket{\Phi}^{B\widehat{B}},
\CD_m\circ\CN^{\otimes n}(\phi_m^{A'^n}\otimes\Upsilon_i)\right)\right) \\
&=& \sum_i p_i \left(2^{-nR}\sum_{m=1}^{2^{nR}} P_s^{\text{eg}}(m,\Upsilon_i)\right),
\end{eqnarray}
so that there is a particular value $i^*$ of $i$ for which
\[2^{-nR}\sum_{m=1}^{2^{nR}} P_s^{\text{eg}}(m,\Upsilon_{i^*}))\geq 1-\epsilon.\]
Hence, $\left(\{\ket{\phi_m}\}_{m\in 2^{nR}},\ket{\Upsilon_{i^*}},\boldsymbol{\CD}\right)$ comprises an
$(R,Q,n,\epsilon)$ cq entanglement generation code.
\end{proof}

\subsection{$\CQ_{\text{et}} \subseteq \CQ_{\text{eg}}$}
\begin{proof}
Suppose there exists a $(Q_a,Q_b,n,\epsilon)$ entanglement transmission code  $\left(\CE_1,\CE_2, \CD\right)$ which transmits the maximally entangled states $\ket{\Phi_1}, \ket{\Phi_2}$.  As in the cq case, the encoded states can be decomposed as
\[(1^A\otimes \CE_1)(\Phi_1) = \sum_i p_i\Upsilon_{1i}\]
and
\[(1^B\otimes \CE_2)(\Phi_2) = \sum_j q_j\Upsilon_{2i}.\]
The reliability condition (\ref{qqIIs}) can then be rewritten as
\[\sum_{ij} p_iq_jF(\ket{\Phi_1}\ket{\Phi_2},
\CD\otimes\CN^{\otimes n}(\Upsilon_{1i}\otimes\Upsilon_{2j}))
\geq 1-\epsilon,\]
which implies the existence of a particular pair $(i^*,j^*)$ of values of $(i,j)$ such that
\[F(\ket{\Phi_1}\ket{\Phi_2},
\CD\otimes\CN^{\otimes n}(\Upsilon_{1i^*}\otimes\Upsilon_{2j^*}))
\geq 1-\epsilon.\]
Hence, $\left(\ket{\Upsilon_{1i^*}},\ket{\Upsilon_{2j^*}},\CD\right)$ comprises a $(Q_a,Q_b,n,\epsilon)$ qq entanglement generation code.
\end{proof}

\section{Strong subspace transmission revisited}
%\section{Scenario III - Strong subspace transmission}
\label{section:sst}
%(or $\epsilon$-preservation of arbitrary bipartite states)}
The criteria of entanglement generation and transmission, both in the cq and qq cases, are directly analogous to the requirement in classical information theory that the average probability of error, averaged over all codewords, be small.  
However, the requirements imposed in Section~\ref{section:generalproblem}
are analogous to the stronger classical condition that the \emph{maximal} probability of error be small, or that the probability of error for \emph{each} pair of codewords be small.  There are examples of classical multiple access channels for which, when
each encoder is a deterministic function from the set of the messages to the set of input symbols, the maximal error capacity region is \emph{strictly} smaller than the average error region \cite{dueck}. However, it is known that if stochastic encoders are allowed (see Problem 3.2.4 in \cite{ck}), the maximal and average error capacity regions are equal.  

It is well-known that randomization is not necessary for such an equivalence to hold for single-user channels, as Markov's inequality implies that a fraction of the codewords with the worst probability of error can be purged, while incurring a negligible loss of rate.  The obstacle to utilizing such an approach for classical multiple access channels, and hence for quantum ones as well, is that there is no guarantee that a large enough subset of bad pairs of codewords decomposes as the product of subsets of each sender's codewords.

A particularly attractive feature of the requirements of Section~\ref{section:generalproblem} is that they ensure  \emph{composability}; when combined with other protocols satisfying analogous criteria, the joint protocol will satisfy similar properties.  As an example, recent work on organizing and classifying quantum Shannon-theoretic protocols by means of \emph{resource inequalities} \cite{family}, makes heavy use of such concatenation of quantum information processing protocols.

In the next two subsections, we cast the requirements outlined in Section~\ref{section:generalproblem} into  somewhat simpler forms which are specific to each of the cq and qq cases.  We will use these forms in order to prove the equivalences of entanglement transmission and strong subspace transmission in both the cq and qq cases. 
\subsection{classical-quantum scenario}
Strong subspace transmission can be considered a more ambitious version of entanglement transmission, whereby rather than requiring Bob to transmit half of a maximally entangled state $\ket{\Phi}^{B\tilde{B}},$
it is instead required that he faithfully transmit the $\tilde{B}$ part, presented to him, of \emph{any} bipartite pure state $\ket{\Psi}^{B\tilde{B}},$ where $|B|$ can be any finite number.  The reader should note that this constitutes a generalization of the usual subspace transmission \cite{bkn}, as whenever $\ket{\Psi}^{B\tilde{B}} = \ket{\psi}^B\ket{\varphi}^{\tilde{B}}$, this amounts to requiring that $\ket{\varphi}$ be transmitted faithfully.  We further demand that the maximal error probability for the classical messages be small.

As with entanglement transmission, Alice will send classical information at rate $R$ by preparing one of $2^{nR}$ pure states $\{\ket{\phi_m}^{A'^n}\}_{m\in 2^{nR}}$.  As previously discussed, our more restrictive information transmission constraints can only be met by allowing Alice to employ a stochastic encoding.  We assume that Alice begins by generating some randomness, modeled by the random variable $X$.  To send message $M=m$, she prepares a state $\phi_{f(m)}$, where $f(m)\equiv f_{X}(m)$ is a random encoding function, depending on the randomness in $X$.
In the language of Section~\ref{section:generalproblem}, this amounts to the definition of a c $\rightarrow$ q encoding function $\CE_1\colon M\rightarrow A'^n$.  Observe that our definition there already allows for randomness to be part of the encoding process.

Bob will apply an encoding $\CE\colon \tilde{B}\rightarrow B'^n$ (this is just his encoding $\CE_2$ from Section~\ref{section:generalproblem} without a classical input),  
and Charlie will employ a decoding instrument $\boldsymbol{\CD}\colon C^n\rightarrow \widehat{M}\widehat{B}$.
These maps require a more complicated structure than was required for entanglement generation and transmission.  Indeed, these will be constructed by means of a protocol, to be described below, out of the entanglement transmission codes
which were proved to exist in Section~\ref{section:eteg1}.
The success probability for the protocol, conditioned on $m$ being sent and $\ket{\Psi}^{B\tilde{B}}$ being presented, can be expressed as
\begin{eqnarray*}
P_s(m,\Psi) &=& F\left(\ket{f(m)}^{\widehat{M}}\ket{\Psi}^{B\widehat{B}},
\boldsymbol{\CD}\circ\CN^{\otimes n}
\big(\phi_{f(m)}^{A'^n}\otimes\CE(\Psi^{B\tilde{B}})\big)\right).
\end{eqnarray*}
We will say that
$(f,X,\{\ket{\phi_m}\}_{m\in 2^{nR}},\CE,\boldsymbol{\CD})$ is an $(R,Q,n,\epsilon)$ \emph{cq strong subspace transmission code} for the channel $\CN$ if, for every $m\in 2^{nR}$ and every $\ket{\Psi}^{B\tilde{B}}$,
\begin{eqnarray}
\E_{X}P_s(m,\Psi) \geq 1-\epsilon.
\end{eqnarray}
The rate pair $(R,Q)$ is an \emph{achievable cq rate pair for strong subspace transmission} if there is a sequence of $(R,Q,n,\epsilon_n)$ cq random strong subspace transmission codes with $\epsilon_n\rightarrow 0$, and the capacity region $\CC\CQ(\CN)$ is closure of the collection of all such achievable rates.

\subsection{quantum-quantum scenario}
This scenario is the obvious combination of the relevant concepts from the previous scenario and the qq entanglement transmission scenario.  Alice and Bob are respectively presented with the $\tilde{A}$ and $\tilde{B}$ parts of some pure bipartite states $\ket{\Psi_1}^{A\tilde{A}}$ and $\ket{\Psi_2}^{B\tilde{B}}$.  As before, we place no restriction on $|A|$ and $|B|$, other than that they are finite.
They employ their respective encodings
$\CE_1$ and $\CE_2$ (which are just the encodings from Section~\ref{section:generalproblem} without classical inputs), while Charlie decodes with $\CD$.  As in the above cq case, the structure of these maps will be more complicated than in the previous two scenarios.
$(\CE_1,\CE_2,\CD)$ is then a $(Q_a,Q_b,n,\epsilon)$ \emph{qq strong subspace transmission code} if
\begin{eqnarray}
F\left(\ket{\Psi_1}^{A\widehat{A}}\ket{\Psi_2}^{B\widehat{B}},
\CD\circ\CN^{\otimes n}\circ(\CE_1\otimes\CE_2)(\Psi_1^{A\tilde{A}}\otimes\Psi_2^{B
\tilde{B}})\right)
\geq 1-\epsilon,
\end{eqnarray}
for every pair of pure bipartite states $\ket{\Psi_1}^{A\tilde{A}}$ and $\ket{\Psi_2}^{B\tilde{B}}$.  Achievable rates and the capacity region $\CQ(\CN)$ are defined as in the cq case.  

\section[Equivalence of ET and SST]{Equivalence of entanglement transmission and strong subspace transmission} \label{section:etsst}
Let us first prove the easy directions.  To see that 
$\CC\CQ \subseteq \CC\CQ_{\text{et}}$, 
note that given a strong subspace transmission code, if Alice uses any deterministic value $x$ for her locally generated randomness $X$, the average classical error will be equal to the expected maximal classical error of the randomized code.  Since the ability to transmit any state includes the maximally entangled case, this completes the claim.
The inclusion $\CQ \subseteq \CQ_{\text{et}}$
follows trivially.  As any states can be transmitted, this certainly includes the case of a pair of maximally entangled states.

\subsection{$\CC\CQ_{\text{et}} \subseteq \CC\CQ$}
\begin{proof}
Suppose there exists an $(R,Q,n,\epsilon^2/2)$ entanglement transmission codes with classical message states $\{\ket{\phi_m}^{A'^n}\}_{m\in 2^{nR}},$ quantum encoding $\CE\colon \tilde{B}\rightarrow \widehat{B},$ and decoding instrument $\boldsymbol{\CD}\colon C^n\rightarrow \widehat{M}\widehat{B}$ with trace-reducing components $\{\CD_m:C^n\rightarrow \widehat{B}\}$, which transmits a maximally entangled state $\ket{\Phi}^{A\tilde{A}}$, where $|A|< \infty$ (although $|\tilde{A}| = 2^{nQ}$).

We will initially prove the equivalence by constructing a code which requires  two independent sources of shared common randomness $X$ and $Y$.  $X$ is assumed to be available to Alice and to Charlie, while $Y$ is available to Bob and to Charlie.  Then, we will argue that it is possible to eliminate the dependence on the shared randomness, by using the channel to send a negligibly small ``random seed", which can be recycled to construct a code which asymptotically achieves the same performance as the randomized one.

We begin by demonstrating how shared common randomness between Alice and Charlie allows Alice to send any message with low probability of error.  Setting $\mu=2^{nR}$, let the random variable $X$ be uniformly distributed on the set $\{1,\dotsc,\mu\}$.  To send message $M=m$, Alice computes $m' = m + X$ modulo $\mu$.  She then prepares the state $\ket{\phi_{m'}}$ for transmission through the channel.  Bob encodes the $\tilde{B}$ part of $\ket{\Phi}^{B\tilde{B}}$ with $\CE$, and each sends appropriately through the channel.  Charlie decodes as usual with the instrument $\boldsymbol{\CD}$.  Denoting the classical output as $\widehat{M}'$, his declaration of Alice's message is then
$\widehat{M} = \widehat{M}' - X$ modulo $\mu$.
Defining the trace-reducing maps
$\CM_m\colon \tilde{B}\rightarrow \widehat{B}$ by
\[\CM_m\colon \tau \mapsto
\CD_m\circ\CN^{\otimes n}(\phi_m\otimes\CE(\tau)),\]
and the trace-reducing average map as
\[\CM\colon\tau\rightarrow \frac{1}{\mu}\sum_{m=1}^\mu \CM_m(\tau),\]
we can rewrite the success criterion (\ref{cqIIs}) for entanglement transmission as
\begin{eqnarray*}
F(\ket{\Phi},\CM(\Phi)) \geq 1-\epsilon^2/2,
\end{eqnarray*}
which, together with (\ref{trfide}), implies that for the identity map $\text{id}:\tilde{B}\rightarrow \widehat{B}$,
\begin{eqnarray}
\left|(\CM - \text{id})(\Phi)\right|_1 \leq \epsilon. \label{cqsim}
\end{eqnarray}
The above randomization of the classical part of the protocol can be mathematically expressed by replacing the $\CM_m$ with $\CM_{m+X}$.  As tracing over the common randomness $X$ is equivalent to computing the expectation with respect to $X$, we see that $\E_X\CM_{m+X} = \CM$, or rather
\[\E_X F(\ket{\Phi},\CM_{m+X}(\Phi)) = F(\ket{\Phi},\CM(\Phi)).\]
It is thus clear that the maximal error criterion for the randomized protocol is equal to the average criterion for the original one.

We continue by randomizing the quantum part of the classically randomized protocol.
Setting $d = 2^{nQ} = |\tilde{B}|,$
let $\{U_y\}_{y \in d^2}$ be the collection of Weyl unitaries, or
generalized Pauli operators, on the $d$-dimensional input space.
Observe that for any $\rho$, acting with a uniformly random choice of Weyl unitary has a completely randomizing effect, in the sense that
\[\frac{1}{d^2}\sum_{y=1}^{d^2} U_y\rho U_y^{-1} = \pi_d.\]
Let the random variable $Y$ be uniformly distributed on $\{1,\dotsc,d^2\}$. It will be convenient to define the common randomness state
\[\Upsilon^{Y_BY_C} = \frac{1}{d^2}\sum_{y=1}^{d^2} \proj{y}^{Y_B}\otimes\proj{y}^{Y_C},\]
where the system $Y_B$ is in the possession of Bob, while $Y_C$ is possessed by
Charlie.  Define now the controlled unitaries
$\CU_B\colon Y_B \tilde{B}\rightarrow Y_B\tilde{B}$ and
$\CU_C\colon Y_C \widehat{B}\rightarrow Y_C\widehat{B}$ by
\[\CU_B =
  \sum_{y=1}^{d^2} \proj{y}^{Y_B}\otimes U_y\]
and
\[\CU_C =
  \sum_{y=1}^{d^2} \proj{y}^{Y_C}\otimes U_y^{-1}.\]
Suppose Bob is given the $\tilde{B}$ part of an arbitrary pure state $\ket{\Psi}^{B\tilde{B}}$, and Alice sends the classical message $M=m$.
For encoding, Bob will apply $\CE\circ\CU_B$ to the combined system $\Upsilon\otimes\Psi$.  Charlie decodes with $\CU_C\circ\CD$.  If $\CM$ were equal to the perfect quantum channel $\text{id}\colon\tilde{B}\rightarrow\widehat{B}$, this procedure would result in the state
\[\frac{1}{d^2}\sum_{y=1}^{d^2} \proj{y}^{Y_B}\otimes\proj{y}^{Y_C}
\otimes \Psi.\]
Note that the common randomness is still available for reuse.
Abbreviating $\proj{y}^Y = \proj{y}^{Y_B}\otimes\proj{y}^{Y_C}$,
and  $\ket{\Psi_y}^{B\tilde{B}} = (1^B\otimes U_y)\ket{\Psi}$, we write
\begin{eqnarray}
\sigma^{YB\tilde{B}} &=& \CU_B(\Upsilon\otimes\Psi) \\  &=&\frac{1}{d^2}\sum_{y=1}^{d^2}\proj{y}^Y\otimes\Psi_y.
\end{eqnarray}
Observe that  $\sigma$ is an extension of the maximally mixed state $\pi^{\tilde B}$, and can be seen to arise by storing in $Y$ the result of a von Neumann measurement along the basis $\{\ket{y}^{F}\}_{y\in d^2}$
on the $F$ part of the pure state
\[\ket{\Gamma}^{FB\tilde{B}} =
  \frac{1}{d}\sum_{y=1}^{d^2}\ket{y}^{F}\ket{\Psi_y}^{B\tilde{B}}.\]
Since $\tr_{R'R} \Gamma = \tr_{YR}\sigma = \pi^{\tilde{B}}$, $\ket{\Gamma}$ is maximally entangled between $FB$ and $\tilde{B}$.  So, there exists an isometry $V\colon B\rightarrow FB$ such that
$(V\otimes 1^{\tilde{B}})\ket{\Phi}^{B\tilde{B}} = \ket{\Gamma}.$  This implies that there is a quantum operation $\CO\colon B\rightarrow YB$  such that $(\CO\otimes 1^{\tilde{B}})(\Phi) = \sigma$.
Define the trace-reducing map $\CT\colon\tilde{B}\rightarrow \widehat{B},$ which represents the coded channel with common randomness accounted for, by
\[\CT\colon\tau\mapsto \tr_{Y} \CU_C\circ\CM\circ\CU_B(\Upsilon\otimes\tau).\]
Recalling our denotation of the noiseless quantum channel $\text{id}\colon\tilde{B}\rightarrow \widehat{B}$, 
as well as our convention that id acts as the identity on any system which is not $\tilde{B}$, we now bound
\begin{eqnarray*}
1-F(\ket{\Psi},\CT(\Psi))
&\leq& \big|(\CT - \text{id})
  (\Psi)\big|_1 \\
&\leq& \big|(\CU_C\circ\CM\circ\CU_B - \text{id})
  (\Upsilon\otimes\Psi)\big|_1 \\
&=&\big|(\CM - \text{id})\circ\CU_B(\Upsilon\otimes\Psi)\big|_1 \\
&=&\big|(\CM - \text{id})(\sigma)\big|_1 \\
&\leq& \big|(\CM - \text{id})(\Phi)\big|_1 \\
&\leq& \epsilon,
\end{eqnarray*}
where the first line is by (\ref{trfid}) and the second by monotonicity with respect to $\tr_Y$.  The third follows from unitary invariance of the trace.  The second to last inequality is a consequence of monotonicity with respect to $\CO$, while the last is by (\ref{cqsim}).  Note that by monotonicity, this implies that any density matrix $\Omega^{B\tilde{B}}$ satisfies
\begin{eqnarray}
|\CT(\Omega) - \Omega|_1\leq \epsilon. \label{TOmega}
\end{eqnarray}

We have thus shown that if Alice and Charlie have access to a common randomness source of rate $R$, while Bob and Charlie can access one of rate $2Q$, the conditions for strong subspace transmission can be satisfied.  Next, we will illustrate that, by modifying our protocol, it is possible to reduce the amount of shared randomness required.  Using the previous blocklength-$n$ construction, we will concatenate $N$ such codes, where each utilizes the \emph{same} shared randomness, to construct a new code with blocklength $nN$.
For an arbitrary $\ket{\Psi^{(N)}}^{B\tilde{B}^N}$, further define
the commuting operations $\{\CT_i\}_{i\in N},$ where $\CT_i\colon\tilde{B}_i\rightarrow \widehat{B}_i$ is $\CT$ acting on the $i$'th tensor factor of $\Psi^{(N)}.$  Setting $\xi_0\equiv\Psi^{(N)}$,  we then recursively define the density operators $\xi_i = \CT_i(\xi_{i-1}),$ noting that
$\xi_N = \CT_N\circ\cdots\circ\CT_1(\xi_0) = \CT^{\otimes N}(\Psi^{(N)})$.
Because of (\ref{TOmega}),
$|\xi_{i+1} - \xi_{i}|_1 = |\CT_{i+1}(\xi_i) - \xi_i|_1 \leq \epsilon$,
and we can use the triangle inequality to estimate
\begin{eqnarray*}
\big|\CT^{\otimes N}(\Psi^{(N)}) - \Psi^{(N)}\big|_1
&=& \big|\xi_{N} - \xi_{0}\big|_1 \\
&\leq&
  \sum_{i=1}^N \big|\xi_i - \xi_{i-1} \big|_1 \\
&\leq& N\epsilon.
\end{eqnarray*}
By choosing $N = \frac{1}{\sqrt{\epsilon}}$, it is clear that we have reduced Alice's and Bob's shared randomness rates respectively to $\sqrt{\epsilon}R$ and $2\sqrt{\epsilon}Q$, while the error on the $N$-blocked protocol is now $\sqrt{\epsilon}$.  Next, we argue that by using two more blocks of length $n$, it is possible to simulate
the shared randomness by having Alice send $nR$ random bits $X$ using the first block, while Bob locally prepares two copies of $\Phi$, $\Phi^{B_1\tilde{B}_1}\otimes\Phi^{B_2\tilde{B}_2}$,
and transmits the $\tilde{B}_1\tilde{B}_2$ parts over the channel using both blocks.
Charlie decodes each block separately, obtaining a random variable $\widehat{X}$ and the $\widehat{B}_1$ and $\widehat{B}_2$ parts of the
post-decoded states $\Omega_1^{B_1\widehat{B_1}}$ and $\Omega_2^{B_2\widehat{B_2}}.$
Bob and Charlie then measure their respective parts of $\Omega_1\otimes\Omega_2$ in some previously agreed upon orthogonal bases to obtain a simulation $\widehat{\Upsilon}$ of the perfect shared randomness state which, by monotonicity and telescoping, satisfies
\begin{eqnarray*}
|\Upsilon - \widehat{\Upsilon}|_1 &\leq& |\Phi\otimes\Phi - \Omega_1\otimes\Omega_2|_1 \\
&\leq& \epsilon^2.
\end{eqnarray*}
Further, the noisy shared randomness for the classical messages can be shown to satisfy
\begin{eqnarray*}
\big|\dist(X,X) - \dist(X,\widehat{X})\big|_1 &=& 2\Pr\{X=\widehat{X}\} \\
&\leq& \epsilon^2.
\end{eqnarray*}
By monotonicity of trace distance and the triangle inequality, using the noisy common randomness state $\widehat{\Upsilon}$ increases the estimate for each block by $2\epsilon^2$.  For identical reasons, the same increase is incurred by using the noisy common randomness $(X,\widehat{X})$.  Thus, accounting for both sources of noisy common randomness,
the estimate (\ref{TOmega}) is changed to $2\epsilon$, provided that $\epsilon\leq \frac{1}{4}$.
The noisy common randomness thus increases the bound on the error of the $N$-blocked protocol to $2\sqrt{\epsilon}$, while costing each of Alice and Bob a negligible rate overhead of $\frac{2}{N+2}$ in order to seed the protocol.

The above protocol can be considered as defining an encoding map $\CE'\colon\tilde{B}^N\rightarrow B'^{(N+2)n}$
and decoding instrument
$\boldsymbol{\CD}\colon C^{(N+2)n}\rightarrow \widehat{B}^N \widehat{M}^N$.
Thus, the protocol takes an $(R,Q,n,\epsilon_n)$ cq entanglement transmission code and constructs an $(R',Q',n',\epsilon'_{n'})$ strong subspace transmission code with cq rate pair $(R',Q') = \left(\frac{R}{1+\epsilon'_{n'}},\frac{Q}{1+\epsilon'_{n'}}\right),$
where $n' = \left(2+\frac{1}{\sqrt{\epsilon_n}}\right)n$, and $\epsilon'_{n'} = 2\sqrt{\epsilon_n}$.
Now, if the rates $(R,Q)$ are achievable cq rates for entanglement
transmission, there must exist a sequence of $(R,Q,n,2\epsilon_n^2)$ entanglement transmission codes with $\epsilon_n\rightarrow 0$.  Since this means that $\frac{1}{1+2\sqrt{\epsilon_n}}$ increases to unity, we have shown that for any $\delta > 0$, every rate pair $(R-\delta,Q-\delta)$ is an achievable cq rate pair for strong subspace transmission.  Since the capacity regions for each scenario are defined as the closure of the achievable rates, this completes the proof.
\end{proof}

\subsection{$\CQ_{\text{et}} \subseteq \CQ$}
\begin{proof}
We will employ similar techniques as were used in the previous proof
to obtain this implication.
Suppose there exists a $(Q_a,Q_b,n,\frac{1}{2}\epsilon^2)$ qq entanglement transmission code $(\CE_1,\CE_2,\CD)$, with
$\CE_1\colon\tilde{A}\rightarrow A'^n$, $\CE_2\colon\tilde{B}\rightarrow B'^n$, and $\CD\colon C^n\rightarrow \widehat{A}\widehat{B}.$
Setting $a = |\tilde{A}|= 2^{nQ_a}$ and $b = |\tilde{B}|=2^{nQ_b}$,
define the common randomness states
\[\Upsilon^{X_AX_C}_X = \frac{1}{a^2}\sum_{x=1}^{a^2}\proj{x}^{X_A}\otimes\proj{x}^{X_C}\]
and
\[\Upsilon^{Y_BY_C}_Y = \frac{1}{b^2}\sum_{x=1}^{b^2}\proj{y}^{Y_B}\otimes\proj{y}^{Y_C}\]
These states will be used as partial inputs to the controlled unitaries
\begin{eqnarray*}
\CU_A &=&
 \sum_{x=1}^{a^2}\proj{x}^{X_A}\otimes U_x,\\
\CU_C &=&
 \sum_{x=1}^{a^2}\proj{x}^{X_C}\otimes U_x^{-1},\\
\CV_B &=&
 \sum_{y=1}^{b^2}\proj{y}^{Y_B}\otimes V_x,\\
\CV_C &=&
 \sum_{y=1}^{b^2}\proj{y}^{Y_C}\otimes V_x^{-1}\\
\end{eqnarray*}
where, as before, we have utilized the Weyl unitaries
$\{U_x\}_{x\in a^2}$ and $\{V_y\}_{y\in b^2}$, which respectively completely randomize any states on $a$-dimensional and $b$-dimensional spaces.  Suppose Alice and Bob are respectively presented with the $\tilde{A}$ and $\tilde{B}$ parts of the arbitrary pure states $\ket{\Psi_1}^{A\tilde{A}}$ and $\ket{\Psi_2}^{B\tilde{B}}.$
Writing $\CM = \CD\circ\CN^{\otimes n}\circ(\CE_1\otimes\CE_2)$, and defining the map $\CT\colon\tilde{A}\tilde{B}\rightarrow \widehat{A}\widehat{B}$ by
\[\CT\colon \tau \mapsto
(\CU_C\otimes\CV_C)\circ\CM\circ(\CU_A\otimes\CV_B)
(\tau\otimes\Upsilon_1\otimes\Upsilon_2),\]
the overall joint state of the randomized protocol is given by $\CT(\Psi_1\otimes\Psi_2)$.
Abbreviating
\[\proj{xy}^{XY} = \proj{x}^{X_A}\otimes\proj{x}^{X_C}
\otimes\proj{y}^{Y_B}\otimes\proj{y}^{Y_C}\]
and defining $\ket{\Psi_x}^{A\tilde{A}} = (1^{A}\otimes U_x)\ket{\Psi_1}$,
$\ket{\Psi_y}^{B\tilde{B}} = (1^{B}\otimes V_y)\ket{\Psi_2},$
we write
\[\sigma^{XYAB\tilde{A}\tilde{B}}
= \frac{1}{a^2 b^2}\sum_{xy} \proj{xy}\otimes \Psi_x \otimes \Psi_y.\]
By similar arguments as in the cq case, there exists a map
$\CO\colon AB\rightarrow ABQR$ so that
\[(\CO\otimes 1^{\tilde{A}\tilde{B}})(\Phi_1\otimes\Phi_2) = \sigma.\]
Again, for the same reasons as in the cq case, we have
\begin{eqnarray*}
|(\CT - \text{id})(\Psi_1\otimes\Psi_2)|_1
&\leq& |(\CM - \text{id})(\sigma)|_1 \\
&\leq& |(\CM - \text{id})(\Phi_1\otimes\Phi_2)|_1 \\
&\leq& \epsilon.
\end{eqnarray*}
The rest of the proof is nearly identical to that from the previous section, so we omit these details, so as not to have to repeat our previous arguments here.
\end{proof}

\chapter{Single-letter examples}\label{chapter:singleletter}
Due to the regularized form of our Theorems 1 and 2, the possibility of actually computing the capacity regions seems generally out of reach.  Here we give some examples of channels whose capacity region does in fact admit a single-letter characterization, in the sense that no regularization is necessary.  In the first section below, we show that a certain erasure quantum erasure multiple access channel has an additive cq capacity region.  The next two sections describe classes of channels which have additive single-user capacities.  The contents of these two sections are essentially an elaboration of results which appear elsewhere in \cite{ds}.
The last section demonstrates that the qq capacity region of a certain collective phase-flip channel has an additive capacity region.  
\section[Additivity of $\CC\CQ$ for erasure channel]{Proof of additivity of $\CC\CQ$ for quantum erasure multiple access channel} \label{section:erasure}
Our first example is a multiple access erasure channel $\CN\colon A'B'\rightarrow C$, where $|A'|=2, |B'|=d$ and $|C|=d+1.$  Alice will send classical information while Bob will send quantum.  Fixing bases $\{\ket{0}^{A'},\ket{1}^{A'}\},
 \{\ket{1}^{B'},\dotsc\ket{d}^{B'}\},
\{\ket{0}^C,\dotsc,\ket{d}^C\},$ the channel
has $d+1$ operation elements
\begin{eqnarray*}
N_0 &=& \sum_{j=1}^d\ket{0}^C\bra{0}^{A'}\bra{j}^{B'} \\
N_i &=& \ket{i}^C\bra{1}^{A'}\bra{i}^{B'}, \,\,\,\,\,i=1,\dotsc d.
\end{eqnarray*}
The action of the channel can be interpreted as follows.  First, a projective measurement of Alice's input along $\{\ket{0},\ket{1}\}$ is performed.  If the result is $0$, Charlie's output is prepared in a pure state $\ket{0}$.  Otherwise, Bob's input is transferred perfectly to the remaining degrees of freedom in Charlie's output.  Bob's input is ``erased", or otherwise ejected into the environment, whenever Alice sends $\ket{0}$, and is perfectly preserved when she sends $\ket{1}$.
Indeed, the action of $\CN$ on $\tau^{A'}\otimes\rho^{B'}$ is given by
\[\CN(\tau\otimes\rho) = \tau_{00}\proj{0} + \tau_{11}\rho.\]

We will show that the cq capacity region of this channel, $\CC\CQ(\CN_\text{erasure})$, has a single-letter characterization given by the collection of pairs of nonnegative classical-quantum rates $(R,Q)$ such that 
\begin{eqnarray*}
R &\leq&  H(p) \\ Q &\leq&  (1-2p)\log d
\end{eqnarray*}
for some $0\leq p\leq \frac{1}{2}$, constituting a generalization of results in \cite{bds} on single-user erasure channels to a multiuser setting.  Figure~\ref{erasure} contains a plot of this region for the case where $d=2$.
%\begin{figure}
% \centering
%     \includegraphics[width=.4\textwidth]{erasure.eps}
%    \caption{$\CC\CQ($erasure channel$)$}
%     \label{erasure}
%\end{figure}

In the sense of (\ref{th1arise}), any state $\Omega^{XBC^k}$ which arises from $\CN^{\otimes k}$ can be specified by fixing some pure state ensemble $\{p(x),\ket{\phi_x}^{A'^k}\}$ and a pure bipartite state $\ket{\Psi}^{BB'^k}$.  We thus write
\[\Omega = \sum_x p(x)\proj{x}^X\otimes(1^B\otimes\CN^{\otimes k})(\phi_x\otimes\Psi).\]
For a binary string $y^k$, let $\ket{y^k}^{A'^k}=\ket{y_1}^{A'}\cdots\ket{y_k}^{A'}$ be the associated computational basis state.  Writing
$p(y^k|x) = |\braket{y^k}{\phi_x}|^2$ defines the random variable $Y^k$,
which is correlated with $X$, and can be interpreted as the erasure pattern associated with the state $\Omega$.
We next define another state of the form (\ref{th1arise}),
\[
\Omega'^{XY^kBC^k} = \sum_{x,y^k} p(x)p(y^k|x)\proj{x}^X\otimes\proj{y^k}^{Y^k}\otimes
\CN^{\otimes k}(\proj{y^k}\otimes\Phi),
\]
for
\[\ket{\Phi}^{BB'^k} =  \sum_{j^k}\ket{j^k}^B\ket{j_1}^{B'_1}\cdots\ket{j_k}^{B'_n},\]
where the summation is over $d$-ary strings of length $k$, $j^k = (j_1,\dotsc,j_k).$
Finally, for
\begin{eqnarray*}
q_i &=& \Pr\{Y_i = 0\}, \\
q &=& \frac{1}{k}\sum_{i=1}^kq_i, \\
\ket{\varphi}^{BC} &=&\frac{1}{\sqrt{d}}\sum_{j=1}^d\ket{j}^B\ket{j}^{C},
\end{eqnarray*}
define a third state
\[\omega^{UBC} = q\proj{0}^U\otimes\pi_d^B\otimes\proj{0}^C
+ (1-q) \proj{1}^U\otimes\varphi^{BC}.\]
The above states can easily be seen to satisfy the following chain of inequalities
\begin{eqnarray*}
I(X;C^k)_\Omega &=& I(X;C^k)_{\Omega'} \\
&=& I(X;Y^k)_{\Omega'} \\
&\leq& H(Y^k)_{\Omega'} \\
&\leq& \sum_{i=1}^k H(Y_i)_{\Omega'} \\
&=& \sum_{i=1}^k H(q_i) \\
&\leq& k H(q) \\
&=& k H(U)_\omega \\
&=& k I(U;C)_\omega.
\end{eqnarray*}
The only nontrivial step above is that we have used the concavity of the binary entropy function in the last inequality.
Furthermore, it is not hard to see that
\begin{eqnarray*}
I_c(B\,\rangle C^kX)_\Omega 
&\leq& I_c(B\,\rangle C^kXY^k)_{\Omega'} \\
&=& I_c(B\,\rangle C^kY^k)_{\Omega'}\\
&=& kI_c(B\,\rangle CU)_\omega.
\end{eqnarray*}
Thus, we have shown that for any state $\Omega^{XBC^k}$ arising from
$\CN^{\otimes k}$ in the sense of (\ref{th1arise}), there is a state $\omega^{UBC}$ arising from $\CN$ in the same sense, allowing the multi-letter information quantities to be bounded by single-letter information quantities; i.e. $\CC\CQ(\CN) = \CC\CQ^{(1)}(\CN)$.
\qed

As it is clear that $I(U;C)_\omega = H(q)$, we focus on calculating
\begin{eqnarray*}
I_c(B\,\rangle CU)_\omega
&=& q \left(H(\proj{0}^C) - H(\pi_d^B\otimes\proj{0}^C)\right)
 + (1-q) \Big(H(\pi_d^C) - H(\varphi^{BC})\Big) \\
&=& q (0-\log d) + (1-q)(\log d - 0)\\
&=& (1-2q)\log d.
\end{eqnarray*}
Note that the above quantity is a weighted average of a positive and a  negative coherent information.  It is perhaps tempting to interpret these terms as follows.  The positive term can be considered as resulting from a preservation of quantum information, while the 
negative term can be seen as signifying a complete loss of quantum information to the environment.  The overall coherent information is positive only when $q < \frac{1}{2}$, a result which is in agreement with the result of Bennett et al.\ \cite{bds}
on the quantum capacity of a binary erasure channel.
Varying $0\leq q \leq \frac{1}{2},$
the rate pairs
\begin{eqnarray*}
(R,Q) &=&\big(I(U;C),I_c(B\,\rangle CU)\big)_\omega \\
&=& \big(H(q),(1-2q)\log d\big)
\end{eqnarray*}
can be seen to parameterize the outer boundary of $\CC\CQ(\CN)$,
as is pictured in figure \ref{erasure} for the case $d=2.$

As an aside, we remark that this calculation, together with the quantum channel capacity theorem from \cite{dev}, gives a direct derivation of the quantum capacity of a quantum erasure channel, without relying on the no-cloning and hashing arguments used in \cite{bds}.

\section{Degradable channels} \label{section:degradable}
While for the single-user capacity $Q(\CN)$ of an arbitrary quantum channel 
$\CN\colon A'\rightarrow B$ is known not to be additive in general, there is a certain class of channels for which additivity follows relatively easily.  This is the class of so-called \emph{degradable channels} \cite{ds}.  A channel $\CN$ is degradable if its complement $\CN^c\colon A'\rightarrow E$ is a stochastically degraded version of $\CN$, i.e. if there exists a \emph{degrading channel} $\CN^d\colon B\rightarrow E$ such that 
\[\CN^c = \CN^d\circ\CN.\]
Below, we will give a version of the proof from \cite{ds} of the additivity of the quantum capacity of an arbitrary degradable channel.  Then, we argue that the maximum sum rate bound of the qq capacity region is additive for such channels.

Assume that $\CN_1\colon A'_1\rightarrow C_1$ and $\CN_2\colon A'_2\rightarrow C_2$ are degradable, with isometric extensions
$\CU_i\colon A'_i\rightarrow C_iE_i.$
Fix an input state $\ket{\Psi}^{AA_1'A_2'}$ which gives rise to the 
global state $\ket{\Omega}^{AC^2E^2} = \CU_1\otimes\CU_2(\Psi^{AA'^2})$,
where the $\CU_i$ are isometric extensions of the $\CN_i$.
By degradability, there exist $\CN^d_i$'s so that $\CN_i^c = \CN_i^d\circ\CN_i$, where $\CN^c_i = \tr_{C_i}\CU_i$. 
Letting $\CV_i\colon C_i \rightarrow E_iF_i$ isometrically extend each 
$\CN^d_i$, define $\Theta^{E^2F^2} = \CV_1\otimes\CV_2(\tr_{AE^2}\Omega)$.
Then 
\begin{eqnarray*}
I_c(A\,\rangle C^2)_\Omega &=& H(C^2)_\Omega - H(E^2)_\Omega \\
&=& H(F^2E^2)_{\Theta} - H(E^2)_{\Theta} \\
&=& H(F^2|E^2)_{\Theta} \\
&\leq& H(F_1|E_1)_{\Theta} + H(F_2|E_2)_{\Theta} \\
&=& H(F_1E_1)_{\Theta} - H(E_1)_{\Theta}  
  + H(F_2E_2)_{\Theta} - H(E_2)_{\Theta} \\
&=& H(C_1)_\Omega - H(E_1)_\Omega + H(C_2)_\Omega - H(E_2)_\Omega \\
&=& H(C_1)_\Omega - H(AC^2E_2)_\Omega + H(C_2)_\Omega - H(AC^2E_1)_\Omega \\
&=& I_c(AC_2E_2\,\rangle C_1)_\Omega + I_c(AC_1E_1\,\rangle C_2)_\Omega\\
%&=& H(C_1)_{\omega_1} - H(A_1C_1)_{\omega_1}  
% +  H(C_2)_{\omega_2} - H(A_2C_2)_{\omega_2} \\
&=& I_c(A_1\,\rangle C_1)_{\omega_1}
 +  I_c(A_2\,\rangle C_2)_{\omega_2}
 \end{eqnarray*}
where the inequality is by Lemma~\ref{lemma:CEsubadditivity}.  In the last line, we set $\omega^{A_iC_i}_i = \CN_i(\Psi)$, identifying 
$A_1 \equiv AA'_2$ and $A_2 \equiv AA'_1$.  All other steps are either by the fact that isometries preserve entropy or by other trivial rewritings.

Now, if we are given $k$ identical channels $\CN\colon A'_i\rightarrow C_i$ and we fix an input state $\ket{\Psi}^{AA'^k}$ giving rise to 
$\ket{\Omega}^{AC^nE^n} = \CU^{\otimes k}(\Psi^{AA'^k})$, recursive application of the above yields 
\[I_c(A\,\rangle C^k)_\Omega \leq \sum_i I_c(A_1\,\rangle C_i)_{\omega_i}\]
where $A_i = AA'_1 \cdots A'_{i-1} A'_{i+1} \cdots A'_k$, 
$\omega^{A_iC_i}_i = \CN_i(\Psi)$, and $\CN_i$ is $\CN$ acting on the $i$th tensor factor.  
Choosing \[i^* = \arg\max_i\{I_c(A_i\,\rangle C_i)_{\omega_i}\}\] 
yields 
\[\frac{1}{k}I_c(A\,\rangle C^k)_\Omega \leq 
 I_c(A_{i^*}\,\rangle C_{i^*})_{\omega_{i^*}}
 \leq \max_{\omega^{AC}} I_c(A\,\rangle C)_\omega = Q^{(1)}(\CN),\]
 where the maximization is as over all $\omega = \CN(\phi^{AA'})$.

Let us phrase this conclusion using different notation.
Let $\tau^{A'^kB'^k}$ be arbitrary, and define 
$\tau^{A_iB_i}_i = \tr_{/A_iB_i}\tau,$ where $\tr_{/A_iB_i}$ denotes the partial trace over all systems which are \emph{not} $A_iB_i$.
Then
\[I_c(\tau,\CN^{\otimes k})\leq kI_c(\tau_{i^*},\CN),\] 
where 
\[i^*=\arg\max_i I_c(\tau_{i},\CN).\]
Now, if $\rho^{A'^k}$ and $\sigma^{B'^k}$ are arbitrary, and we define 
$\rho_i = \tr_{/A_i}\rho$ and $\sig_i= \tr_{/B_i}\sig$, 
observe that if $\tau = \rho\otimes\sig$, then 
$\tau_i = \rho_i\otimes\sig_i$.  This immediately implies that 
\[I_c(\rho\otimes\sig,\CN^{\otimes k})\leq kI_c(\rho_{i^*}\otimes\sig_{i^*},\CN),\]   
 where 
\[i^*=\arg\max_i I_c(\rho_{i}\otimes\sig_i,\CN),\]
proving that the maximum sum rate of any degradable channel is additive, even when the inputs are restricted to be product states. This fact will be useful in Section~\ref{section:bitflip}, where we give a channel whose qq capacity region is single-letter.
 
% Now, the more pressing issue is whether or not, for 
% \[\ket{\Omega}^{ABC^2E^2} = \CU_1\circ\CU_2(\Psi^{AA'^2}\otimes\Phi^{BB'^2})\]
% we have    
% \[I_c(A\,\rangle BC^2)_\Omega \leq 
% I_c(A_1\,\rangle B_1C_1)_{\omega_1}
% + I_c(A_2\,\rangle B_2C_2)_{\omega_2}\]
% for appropriate states $\omega_i$.  Let's try this:
% 
% \begin{eqnarray*}
% I_c(AB\,\rangle C^2)_\Omega &=& H(C_1C_2)_\Omega - H(E_1E_2)_\Omega \\
% &=& H(E_1F_1E_2F_2)_\Theta - H(E_1E_2)_\Theta \\
% &=& H(F_1F_2|E_1E_2)_\Theta \\
% &=& H(F_1|E_1)_\Theta + H(F_2|E_2)_\Theta \\
% &=& H(F_1E_1)_\Theta - H(E_1)_\Theta + H(F_2E_2)_\Theta - H(E_2)_\Theta \\
% &=& H(C_1)_\Omega - H(E_1)_\Omega + H(C_2)_\Omega - H(E_2)_\Omega \\
% &=& H(C_1)_\Omega - H(ABC_1C_2E_2)_\Omega 
%  + H(C_2)_\Omega - H(ABC_1C_2E_1)_\Omega \\
% &=& I_c(ABA'_2B'_2\,\rangle C_1)_\Omega 
%  + I_c(ABA'_1B'_1\,\rangle C_2)_\Omega \\
% &=& I_c(A_1B_1\,\rangle C_1)_{\omega_1} 
%  + I_c(A_2B_2\,\rangle C_2)_{\omega_2}
% \end{eqnarray*}
% where $A_1 = AA'_2$, $A_2 = AA'_1$, $B_1 = BB'_2$, and $B_1 = BB'_2$.

\section{Generalized dephasing channels} \label{section:dephasing}
In this section we describe a certain subclass of the class of degradable channels.  These are channels $\CN\colon A'\rightarrow B$ with $|A| = |B| = d$ for which there is a particular orthogonal basis $\{\ket{x}^{A'}\}$ which can be transmitted through the channel without error
\[\CN\big(\proj{x}\big) = \proj{x}\]
although superpositions of these basis vectors are potentially subject to noise.
Here, $\{\ket{x}^B\}$ is a corresponding orthogonal basis for $B$.  
Such a channel has an isometric extension $\CU\colon A'\rightarrow BE$  given by 
\[\CU = \sum_x\ket{x}^B\ket{\phi_x}^E\bra{x}^{A'},\]
where the states $\ket{\phi_x}^E$ are not necessarily orthogonal.
To see that these channels are degradable, observe that for any input state $\rho^{A'}$, 
\begin{eqnarray*}
\CN^c(\rho) &=& \tr_B\CU(\rho) \\
&=& \sum_{x}\bra{x}^B
\left(\sum_{x''x'}\ket{x''}^B\ket{\phi_{x''}}^E\bra{x''}^{A'}
\rho\ket{x'}^{A'}\bra{x'}^B\bra{\phi_{x'}}^E\right)\ket{x}^B  \\
&=& \sum_x\bra{x}\rho\ket{x} \phi_{x}^E.
\end{eqnarray*}
Note that $\CN^c(\rho)$, depends only on the diagonal matrix elements of $\rho$ (when it is expressed in the dephasing basis.  However, these are exactly the matrix elements which are unaffected by the action of $\CN$, making degradability evident.  In fact, the degrading channel is precisely $\CN^c$, i.e. 
\[\CN^c = \CN^c\circ\CN.\]
It is interesting to relate the isometric extension $\CU_\CN$ to the operator sum representation for $\CN$.  To do this, first express $\CU_\CN$ in the flattened representation
\[\CU_\CN = 
\begin{pmatrix}
\ket{\phi_1} &  &  & \\
  & \ket{\phi_2} & & \\
& & \ddots & \\
& & & \ket{\phi_{d}}    
\end{pmatrix}.
\] 
Supposing that $|E|=k$, 
note that the matrix is ``block diagonal", with $d$ $k\times 1$ blocks,
where this is expressed as a map to the system $EB$.
Regrouping the rows into to $k$ groups of size $d$ we rewrite 
\[\CU_\CN = 
\begin{pmatrix}
\braket{1}{\phi_1} &  &  & \\
  & \braket{1}{\phi_2} & & \\
& & \ddots &  \\
& & & \braket{1}{\phi_{d}} \\
\braket{2}{\phi_1} &  &  & \\
  & \braket{2}{\phi_2} & & \\
& & \ddots &  \\
& & & \braket{2}{\phi_{d}} \\  
& & \vdots & \\
\braket{k}{\phi_1} &  &  & \\
  & \braket{k}{\phi_2} & & \\
& & \ddots &  \\
& & & \braket{k}{\phi_{d}}  
\end{pmatrix}
= \begin{pmatrix}
N_1 \\ N_2 \\ \vdots \\N_{k}
  \end{pmatrix}.
\] 

This is just the flattened representation for the map to the system $BE$ (the order of $E$ and $B$ have been reversed).  Note that 
we have identified the $|E|$ blocks with the matrices of the operator sum representation 
\[\CN(\rho) = \sum_{e=1}^k N_e\rho N_e^\dag.\]
So we see that the operator sum matrices are all diagonal in the $\{\ket{i}\}$ basis and are given explicitly as 
\[N_e = \sum_x \bra{e}\ket{\phi_x}\ket{x}^B\bra{x}^{A'}.\]
Reversing the above steps, it is clear that $\CN$ is a generalized dephasing channel if and only if it has an operator sum representation consisting of matrices which commute.

Let us mention that in the special case where the $\{\phi_x\}$ are mutually orthogonal, the channel is \emph{completely dephasing}.  We denote this channel as $\Delta$, and note that it corresponds to a channel which performs a pure state measurement in the dephasing basis while ignoring the result.  This has the effect of setting all of the off-diagonal matrix elements of $\rho$ equal to zero.  $\Delta$ obeys the following equations:
\begin{eqnarray*}
\CN^c &=& \CN^c\circ\Delta \\
H(\Delta(\rho)) &\geq& H(\rho).
\end{eqnarray*}
The first is because $\CN^c$ only depends on the diagonal components of $\rho$, while the second is proved in \cite{cn}.  Observe that the inequality is saturated for diagonal $\rho$.  
Because of this, we may write 
\begin{eqnarray*}
Q(\CN) &=& \max_\rho I_c(\rho,\CN) \\
&=& \max_\rho \Big\{ H\big(\CN(\rho)\big) - H\big(\CN^c(\rho)\big) \Big\}\\
&=& \max_\rho \Big\{H\big(\CN\circ\Delta(\rho)\big) - H\big(\CN^c\circ\Delta(\rho)\big) \Big\} \\
&=& \max_{p(x)} \left\{H(X) - H\Big(\sum_x p(x)\phi_x\Big)\right\}.
\end{eqnarray*}

\section[Additivity of $\CQ$ for phase-flip channel]{Proof of additivity of $\CQ$ for collective phase-flip channel}\label{section:bitflip}

While the description of the capacity region $\CQ$ in Theorem~2
generally requires taking a many-letter limit, we give here an
example of a quantum multiple access channel $\CN_p\colon A'B'\rightarrow C$ for which that description can be single-letterized.
The channel $\CN_p$ takes as input two qubits, one from Alice and the other from Bob.  With probability $p$, the channel causes each qubit to undergo a phase flip, by rotating each by 180$^\circ$ about its z-axis before it is received by the receiver Charlie.  The action of $\CN_p$ on an input density operator $\rho^{A'B'}$ is described in terms of the operator sum representation as 
\[\CN_p(\rho) = (1-p)\rho + p (\sig_z\otimes \sig_z)\rho (\sig_z \otimes \sig_z),\]
where 
\[\sigma_z = \begin{pmatrix} 1 & 0 \\ 0 & -1\end{pmatrix}\]
is the Pauli phase flip matrix.  
We will demonstrate that $\CQ(\CN_p)$ is equal to the collection of all pairs of nonnegative rates $(Q_a,Q_b)$ which satisfy 
\begin{eqnarray*}
 Q_a &\leq& 1 \\
 Q_b &\leq& 1 \\
 Q_a + Q_b &\leq& 2-H(p).
\end{eqnarray*}
\begin{proof}
In order to prove this, we first recall that the maximum of the sum rate bound $I_c(AB\,\rangle C)$ over all inputs of the form (\ref{th2arise}) is additive.  Next, we calculate $Q(\CN_p)$, the single-user capacity of the channel, and observe that it is achieved for inputs of the form (\ref{th2arise}), implying that the maximum sum rate bound equals the capacity.  Then, we show that for the same inputs, the bounds $I_c(A\,\rangle BC)$ and $I_c(B\,\rangle AC)$ on the individual rates are as large as is possible. 
The characterization in terms of a single pentagon will then follow.

We first note that the the operator sum matrices $\sqrt{p}\sig_x\otimes \sig_x$ and $\sqrt{1-p}1_4$ commute.  By results in the previous section, we conclude that $\CN_p$ is an example of a \emph{generalized dephasing channel} and thus, the following two conditions are satisfied:
\begin{itemize}
\item for any state $\Omega^{ABC^k} = \CN^{\otimes k}(\Psi^{ABA'^kB'^k})$ arising from $\CN^{\otimes k}$ (where Alice and Bob can jointly prepare any state at the inputs), there is a state $\omega^{ABC} = \CN(\psi^{ABA'B'})$ for which 
\[I_c(AB\,\rangle C)_\om \geq \frac{1}{k}I_c(AB\,\rangle C^k)_\Omega.\]
Furthermore, the input density operator $\rho^{A'B'} = \tr_{AB}\psi^{ABA'B'}$ is diagonal in the dephasing basis of $\CN_p$.

\item for any state $\Omega'^{ABC^k}= \CN^{\otimes k}(\Psi_1^{AA'^k}\otimes\Psi_2^{BB'^k})$ arising from $\CN^{\otimes k}$ in the sense of (\ref{th2arise}), there is a state $\omega'^{ABC} = \CN(\phi_1^{AA'}\otimes\phi_2^{BB'})$ arising from $\CN$ in the same sense for which 
\[I_c(AB\,\rangle C)_{\om'} \geq \frac{1}{k}I_c(AB\,\rangle C^k)_{\Omega'}.\]
\end{itemize}

The first condition above says that the single-user capacity $Q(\CN_p)$ is additive. 
It also guarantees that the relevant maximization is achieved by an input density operator $\rho^{A'B'}$ which is diagonal in the dephasing basis.
The second condition guarantees that the constrained single-user capacity of $\CN_p$, when the users are constrained to preparing product input states, is additive. 

In order to compute $Q(\CN_p)$, let us first write an isometric extension 
$\CU\colon AB\rightarrow CE$ of $\CN_p$ as 
\[\CU\ket{i}^A\ket{i}^B = \ket{ij}^C\ket{\phi_{ij}}^E,\]
where 
\[\ket{\phi_{00}}^E = \ket{\phi_{11}}^E = \sqrt{1-p}\ket{0}^E +\sqrt{p}\ket{1}^E\equiv \ket{\phi_+}^E\]
and
\[\ket{\phi_{01}}^E =\ket{\phi_{10}}^E = \sqrt{1-p}\ket{0}^E -\sqrt{p}\ket{1}^E\equiv 
\ket{\phi_-}^E.
\]
A complementary channel $\CN^c_p$ is then defined as 
\begin{eqnarray*}
\CN^c_p(\rho) &=& \tr_C\CU(\rho) \\
&=& \sum_{ij} \ket{\phi_{ij}}\bra{i}\bra{j}\rho\ket{i}\ket{j}\bra{\phi_{ij}} \\
&=& \sum_{ij} \rho_{ij}\phi_{ij} \\
&=& (\rho_{00} + \rho_{11})\phi_+ + (\rho_{01}+\rho_{10})\phi_-.
\end{eqnarray*}
Observe that the output of the $\CN^c_p$ depends only on the diagonal elements of $\rho$, when $\rho$ is written in the dephasing basis $\{\ket{00},\ket{01},\ket{10},\ket{11}\}.$
Define $\al = \rho_{00} + \rho_{11}$.
As $Q(\CN_p)$ is achieved when $\rho$ is diagonal in this basis, let us calculate 
\begin{eqnarray*}
H(C) &=& H(A'B') \\
&=& H\big(\{\rho_{00},\rho_{01},\rho_{10},\rho_{11}\}\big) \\
&=& H(\al) + \al H\Big(\frac{\rho_{00}}{\al}\Big) 
           + (1-\al)H\Big(\frac{\rho_{01}}{1-\al}\Big) \\
&\leq& H(\al) + 1,
\end{eqnarray*}
where the inequality is saturated when $\rho_{00} = \rho_{11} = \frac{\al}{2}$
and $\rho_{01} = \rho_{10} = \frac{1-\al}{2}$.
It thus suffices to optimize over the class of states 
\[\rho^{A'B'} = \begin{pmatrix}
\frac{\alpha}{2} & 0 & 0 & 0\\ 0 & \frac{1-\al}{2} & 0 & 0 \\ 0 & 0 & \frac{1-\al}{2} & 0 \\ 0& 0 & 0 & \frac{\al}{2}
              \end{pmatrix}\]
for which 
\[H(C) = H(\rho) = 1 + H(\al),\]
Note that we may express 
\begin{eqnarray*}
\phi_{\pm}^E = \frac{1}{2}\Big(1 \pm \sqrt{p(1-p)}\sig_x - (1-2p)\sig_z\Big),
\end{eqnarray*}
allowing us to write
\begin{eqnarray*}
\CN_p^c(\rho) = \al\phi_+ + (1-\al)\phi_- = 
\frac{1}{2}\Big(1 + (2\al-1)\sqrt{p(1-p)}\sig_x - (1-2p)\sig_z\Big),
\end{eqnarray*}
so that  $H(E) = H\Big(\frac{1}{2}(1 + \sqrt{p(1-p)(2\al-1)^2 + (1-2p)^2})\Big).$
Thus, 
\begin{eqnarray*}
I_c(\rho,\CN) &=& H\big(\CN_p(\rho)\big) - H\big(\CN_p^c(\rho)\big) \\
&=& 1 + H(\al) - H\Big(\frac{1}{2}(1 + \sqrt{p(1-p)(2\al-1)^2 + (1-2p)^2})\Big) \\ &\equiv& h(\al).
\end{eqnarray*}
For fixed $p$, $h(\al)$ is symmetric about $\al = \frac{1}{2}$, and has a   
first derivative which is positive for $0\leq\al<\frac{1}{2}$
(and is thus negative for $\frac{1}{2}<\al\leq 1$).
Because $h(\al)$ is continuous on $0\leq\al\leq 1$, its maximum is attained when $\al= \frac{1}{2}$, so that 
\[\max_\rho I_c(\rho,\CN) = I_c(\pi^{A'B'},\CN) = 1 + H\Big(\frac{1}{2}\Big)  - H(p) = 2-H(p).\]
So we see that the maximum is already achieved for a product state $\pi^{A'B'} = \pi^{A'}\otimes\pi^{B'}$.
Define the Bell states
\begin{eqnarray*}
\ket{\psi_\pm}&=&\frac{1}{\sqrt{2}}\Big(\ket{00} \pm \ket{11}\Big). 
\end{eqnarray*}
As $\ket{\psi_+}$ purifies the maximally mixed state $\pi_2$, let us define the global state 
\[\omega^{ABC} = \CN(\psi_+^{AA'}\otimes\psi_+^{BB'}).\]
Identifying $C=\h{A}\h{B}$ in the obvious way, let us reexpress
\[\omega^{A\h{A}B\h{B}} 
= (1-p)\psi_+^{A\h{A}}\otimes\psi_+^{B\h{B}} + p \psi_-^{A\h{A}}\otimes\psi_-^{B\h{B}}.\]
It is now a simple task to calculate 
\begin{eqnarray*}
H(ABC) &=& H(\omega) = H(p) \\
H(C) &=& H(\pi^{C}) = 2 \\
H(AC) &=& H(A\h{A}) + H(\h{B}) = H(p) + 1 = H(BC). 
\end{eqnarray*}
Combining these gives the relevant coherent informations
\begin{eqnarray*}
I_c(AB\,\rangle C) &=& H(C) - H(ABC) = 2 - H(p) \\
I_c(A\,\rangle BC) &=& H(BC) - H(ABC) = 1 + H(p) -H(p) = 1 \\
I_c(B\,\rangle AC) &=& H(AC) - H(ABC) = 1.
\end{eqnarray*}
As we saw in Section~\ref{section:coherentinformation}, $I_c(A\,\rangle BC)\leq \log|A'| = 1$ and $I_c(B\,\rangle AC)\leq \log|B'| = 1$ for any state arising from $\CN$. The individual rate bounds are thus saturated and the claim follows.
\end{proof}
%\pagebreak

\chapter{Discussion}\label{chapter:discussion}
There have been a number of results analyzing multiterminal coding problems in quantum Shannon theory.
For an i.i.d.\ classical-quantum source $XB$, Devetak and Winter \cite{dw} have proved a Slepian-Wolf-like coding theorem achieving the cq rate pair $(H(X|B),H(B))$ for classical data compression with quantum side information.  Such codes extract classical side information from $B^n$ to aid in compressing $X^n.$  The extraction of side information is done in such a way as to cause a negligible disturbance to $B^n$. Our Theorem~\ref{theo:cq} is somewhat of this flavor.  There, the quantum state of $C^n$ is measured to extract Alice's classical message which, in turn, is used as side information for decoding Bob's quantum information.  
Analogous results to ours were obtained by Winter in his analysis of a multiple access channel with classical inputs and a quantum output, whereby the classical decoded message of one sender can be used as side information to increase the classical capacity of another sender.

We further mention the obvious connection between our coding theorems and the subject of channel codes with side information available to the receiver.  
The more difficult problem of classical and quantum capacities when side information is available at the \emph{encoder} is analyzed by Devetak and Yard in \cite{dy}, constituting quantum generalizations of results obtained by Gelfand and Pinsker \cite{gp} for classical channels with side information.

In an earlier draft of \cite{qmac}, we characterized $\CQ(\CN)$ as the closure of a regularized union of rectangles 
\begin{eqnarray*}
0 \!\!\!&\leq \, R  \,\leq& \!\!\!\frac{1}{k}I_c(A\,\rangle C^k)  \\
0 \!\!\!&\leq \, S  \,\leq&  \!\!\!\frac{1}{k}I_c(B\,\rangle C^k). 
\end{eqnarray*}
This solution had been conjectured on the basis of a duality between classical Slepian-Wolf distributed source coding and classical 
multiple-access channels \cite{ck,coverthomas}, as well as on a purported no-go theorem for distributed data compression of so-called irreducible pure state ensembles that appeared in an early version of \cite{adhw}.  After the earlier preprint was made available, Andreas Winter announced 
\cite{wintertalk} recent progress with Jonathan Oppenheim and Michal Horodecki \cite{how} on the quantum Slepian-Wolf problem, offering a characterization identical in functional form to the classical one, while also supplying an interpretation of negative rates and apparently evading the no-go theorem.  Motivated by the earlier mentioned duality, he informed us that the qq capacity region could also be characterized in direct analogy to the classical case.  Subsequently, we found that we could modify our previous coding theorem to achieve the new region, provided that the rates are nonnegative.  
After those events unfolded, the authors of \cite{adhw} found an error in the proof of their no-go theorem, leading to a revised version consistent with the newer developments. 
Our earlier characterization of $\CQ(\CN)$, while correct, is contained in the rate region of Theorem 2 for any finite $k$, frequently strictly so.  The newer theorem, therefore, gives a more accurate approximation to the rate region for finite $k$.  In fact, for any state arising from the channel which does not saturate the strong subadditivity inequality \cite{hjpw}, the corresponding pentagon and rectangle regions are distinct.  As seen in Section~\ref{section:degradable}, another beneficial feature of the new characterization is
that for any channel which is \emph{degradable}, 
the maximum sum rate bound $R+S \leq \max I_c(AB\,\rangle C)$ is additive, where the maximization is over all states of the form (\ref{th2arise}).  Furthermore, recall that in   Section~\ref{section:bitflip}, the pentagon characterization was single-letterized for the collective phase flip channel. On the other hand, computer calculations have revealed that the rectangle region does \emph{not} lead to a single-letter characterization of that channel.  This seems to indicate that the newer characterization is the ``correct" one, at least for that particular channel.

More recently, we discovered that the same technique used to prove the new characterization of $\CQ(\CN)$ implies a new cq coding theorem, and thus a new characterization of $\CC\CQ(\CN)$.  By techniques nearly identical to those employed in the coding theorem for Theorem~2,
it is possible to achieve the cq rate pair 
\[(R,Q) = \big(I(X;BC),I_c(B\,\rangle C)\big)\]
corresponding to Bob's quantum information being used as side information for decoding Alice's classical message.
This is accomplished by having Charlie isometrically decode Bob's quantum information, then coherently decode to produce an effective channel 
$\CN_1\colon A'\rightarrow BC$ so that Alice can transmit classically at a higher rate.  The new characterization is then a regularized union of pentagons, consisting of pairs of nonnegative rates $(R,Q)$ satisfying
\begin{eqnarray*}
r &\leq& I(X;BC) \\
S &\leq& I_c(B\,\rangle CX)\\
r+S &\leq& I(X;C) + I_c(B\,\rangle CX) = I(X;BC) + I_c(B\,\rangle C).
\end{eqnarray*} 
Surprisingly, it is thus possible to characterize each of $\CC\CQ(\CN)$ and $\CQ(\CN)$ in terms of pentagons, in analogy to the original classical result.  This situation makes apparent the dangers of being satisfied with regularized expressions for capacity regions.  Without being able to prove single-letterization steps in the converses, it is hard to differentiate which characterization is the ``right" one.  While it is intuitively satisfying to see analogous formulae appear in both the classical and quantum theories, the regularized nature of the quantum results blurs the similarity.  Indeed, the problems with single-letterization for single-user channels appear to be amplified when analyzing quantum networks (see e.g. \cite{dch}).  While $\CQ$ is additive for the collective phase flip channel of Section~\ref{section:bitflip}, this behavior does not appear to be generic for the classes of degradable or generalized dephasing channels, as the saturation of the individual rate bounds for that example seem to be the source of additivity.  Perhaps this indicates that the necessity of understanding the capacities of single-user channels at a level beyond regularized optimizations is even more pressing than previously thought.  It should be mentioned that for the erasure channel analyzed in Section~\ref{section:erasure}, the newer description of $\CC\CQ(\CN)$ is not an issue, as the new corner point is contained in the old rectangle for any state arising from any number of parallel instances of the erasure channel.

Consider the full simultaneous classical-quantum region $\CS(\CN)$ defined in Section~\ref{section:generalproblem}. This region can be characterized in a way that generalizes Theorems~1 and 2 as the regularization of the region $\CS^{(1)}(\CN)$,
defined as the vectors of nonnegative rates $(R_a,R_b,Q_a,Q_b)$ satisfying
\begin{eqnarray*}
R_a &\leq& I(X;C|Y) \\
R_b &\leq& I(Y;C|X) \\
R_a + R_b &\leq& I(XY;C) \\
Q_a &\leq& I_c(A\,\rangle BCXY) \\
Q_b &\leq& I_c(B\,\rangle ACXY) \\
Q_a+Q_b &\leq& I_c(AB\,\rangle CXY)
\end{eqnarray*}
for some state of the form
\[\sigma^{XYABC} =
\sum_{x,y}p(x)p(y)\proj{x}^X\otimes\proj{y}^Y
\otimes\CN(\psi_x^{AA'}\otimes\phi_y^{BB'}),\]
arising from the action of $\CN$ on the $A'$ and $B'$ parts of some pure state ensembles
$\{p(x),\ket{\psi_x}^{AA'}\}$, $\{p(y),\ket{\phi_y}^{BB'}\}$.
Briefly, achievability of this region is obtained as follows.  Using techniques introduced in \cite{ds}, each sender ``shapes" their quantum information into HSW codewords.  Decoding is accomplished by first decoding all of the classical information, then using that information as side information for a quantum decoder.  A formal proof of the achievability of this region is found in \cite{future}.  The main result of 
\cite{ds}, the regularized optimization of the cq result from \cite{wintermac} over pairs of input ensembles, and our Theorems 1 and 2 follow as corollaries of the corresponding capacity theorem.  Indeed, the six two-dimensional ``shadows" of the above region, obtained by setting pairs of rates equal to zero, reproduce those aforementioned results.
This characterization, however, only utilizes the rectangle description of $\CC\CQ(\CN)$.  It is indeed possible to write a more accurate regularized description of $\CS(\CN)$ which generalizes the pentagon characterizations of $\CC\CQ(\CN)$ and $\CQ(\CN)$, although we will not pursue that at this time.

%\pagebreak
\chapter{Appendix}\label{section:appendix}
\section{Quantum instruments and coherent information}
\label{section:instrumentcomplement}
For some finite set $\CS$, consider a labelled collection of channels $\{\CN_s\}_{s\in\CS}$, where $\CN_s \colon A'\rightarrow B$.  Define an instrument $\boldsymbol{\CN}\colon A'\rightarrow SB$ to act as
\[\boldsymbol{\CN}\colon \tau \rightarrow 
\sum_s p(s) \proj{s}^S \otimes \CN_s(\tau).\] 
An instrument channel such as $\boldsymbol{\CN}$ may be interpreted as one with classical state information made available to the receiver.  
We will show that every  channel  $\CN^c\colon A'\rightarrow E$ which is complementary to $\boldsymbol{\CN}$ is an instrument as well, as the environment $E$ contains a copy of $S$.  In other words, the classical state information is also available to an eavesdropper with full control of the environment. 

%First, recall the notion of completely dephasing channels.  
%$\Delta\colon \CB(\CH) \rightarrow \CB(\CH)$ is completely dephasing in the orthonormal basis $\{\ket{s}\}$ if it performs a complete measurement in that basis.  Such a channel has operator sum matrices $\{\proj{s}\}$ and has an isometric extension 
%$\CU_\Delta\colon S \rightarrow SE''$, acting as $\CU_\Delta\ket{s} = \ket{s}^{S}\ket{s}^{E''}.$ 

An isometric extension $\CU$ of $\boldsymbol{\CN}$ may be constructed as follows.  First, fix isometric extensions $\CU_s \colon A'\rightarrow E'B$ for the individual $\CN_s$'s.  Then, define $\CU\colon A' \rightarrow SEB$ via 
\[\CU = \sum_s \sqrt{p(s)}\ket{s}^S\ket{s}^{E''}\otimes \CU_s,\]
taking $E=E'E''$.  That this is indeed an isometry is evident, because $\CU^\dag \CU = \sum_s p(s) \CU_s^\dag \CU_s = \sum_s p(s) 1^{A'} = 1^{A'}$.   We may further check that $\CU$ is in fact an extension of $\boldsymbol{\CN}$, by calculating
\begin{eqnarray*}
\tr_E \CU(\tau) &=& \tr_{E'} \tr_{E''}\CU(\tau) \\
&=& \tr_{E'} \sum_s p(s) \proj{s}^S \otimes \CU_s(\tau) \\
&=& \sum_s p(s) \proj{s}^S \otimes \CN_s(\tau) \\
&=& \boldsymbol{\CN}(\tau).
\end{eqnarray*}
Thus, the action of the complementary channel $\CN^c$ can be defined via $\CU$ as
\begin{eqnarray*}
\CN^c(\tau) &=& \tr_{BS} \CU(\tau) \\
&=& \tr_B \sum_s p(s) \proj{s}^{E''}\otimes \CU_s(\tau) \\
&=& \sum_s p(s) \proj{s}^{E''}\otimes \CN^c_{s}(\tau),
\end{eqnarray*}
where the $\CN^c_{s} = \tr_B \CU_s$ are complementary 
channels to the $\CN_s$'s.  
% Let $\rho^{A'}$ be arbitrary, and fix a purification $\ket{\Phi_\rho}^{AA'}$ of $\rho$.   If we then define the global pure state $\ket{\Psi}^{ABSE} = (1^A\otimes\CU)\ket{\Phi_\rho},$ 
% the coherent information can be written as
% \begin{eqnarray*}
% I_c(\rho,\boldsymbol{\CN}) &=& H(\CN(\rho)) - H(\CN^c(\rho)) \\
% &=& H(BS)_\Psi - H(E)_\Psi \\
% &=& H(BS)_\Psi - H(ABS)_\Psi \\
% &=& - H(A|BS)_\Psi \\
% &=& I_c(A\,\rangle BS)_\Psi.
% \end{eqnarray*}
% Thus, our claim in the proof of Theorem~\ref{theo:cq} that the coding theorem from \cite{dev} applies to such channels is valid.

\section{Proof of convexity of $\CC\CQ$ and $\CQ$}
\label{section:convex}
Let $\CN:A'B'\rightarrow C$ be a quantum multiple access channel.  We will prove that
$\CQ(\CN)$ is convex, as the proof for $\CC\CQ$ is identical.
Let $k_0$ and $k_1$ be positive integers, and fix any two states of the form (\ref{th2arise}), $\sigma_0^{A_0B_0C^{k_0}}$ and $\sigma_1^{A_1B_1C^{k_1}}.$
Then $(R_0,S_0),(R_1,S_1)\in \CQ(\CN)$, where for $i\in\{0,1\}$,
\begin{eqnarray*}
R_i &=& \frac{1}{k_i}I_c(A_i\,\rangle C^{k_i})_{\sigma_i} \\
S_i &=& \frac{1}{k_i}I_c(B_i\,\rangle C^{k_i})_{\sigma_i}.
\end{eqnarray*}
We will now show that for any rational $0\leq \lambda \leq 1$, $\lambda(R_0,S_0) + (1-\lambda)(R_1,S_1) \in \CQ(\CN).$  We first write $\lambda = \frac{\alpha}{\beta},$
for integers satisfying $\beta> 0,$ $\beta\geq \alpha \geq 0$. Setting $p_0=\alpha k_1,$ $p_1 = (\beta-\alpha)k_0,$ and  
$k = p_0k_0 + p_1k_1$,
define the composite systems $A = A_0^{p_0}A_1^{p_1}$ and 
$B = B_0^{p_0}B_1^{p_1}$,
as well as the density matrix
$\sigma^{ABC^k} = \sigma_0^{\otimes p_0}\otimes\sigma_1^{\otimes p_1},$ which is also of the form (\ref{th2arise}).
Additivity of coherent information across product states and some simple algebra gives
\begin{eqnarray*}
\frac{1}{k}I_c(A\,\rangle C^k)_{\sigma} &=&
\frac{p_0}{k}I_c(A_0\,\rangle C^{k_0})_{\sigma_0}+
\frac{p_1}{k}I_c(A_1\,\rangle C^{k_1})_{\sigma_1} \\
&=& \frac{p_0k_0 R_0 + p_1k_1R_1}{p_0k_0 + p_1k_1} \\
&=& \lambda R_0 + (1-\lambda) R_1.
\end{eqnarray*}
An identical calculation shows that
$\frac{1}{k}I_c(B\,\rangle C^k)_{\sigma} = \lambda S_0 + (1-\lambda) S_1.$
As $\CQ(\CN)$ was defined as the topological closure of rate pairs corresponding to states which appropriately arise from the channel, the result follows because the set of previously considered $\lambda$'s comprises a dense subset of the unit interval.

\qed

\section{Proof of cardinality bound on $\CX$.}
\label{section:cardinality}
Begin by fixing a finite set $\CX$, a labelled collection of pure states $\{\ket{\phi_x}^{A'}\}_{x\in\CX}$, and a pure bipartite state $\ket{\Psi}^{BB'}.$ For each $x$, these define the states
$\sigma^{BC}_x = \CN(\phi_x\otimes\Psi)$ and $\omega^C_x = \tr_B\sigma_x$.  Assume for now that $|A'| \geq |C|$. Define a mapping $f\colon\CX\rightarrow \mathbb{R}^{|C|^2 + 1}$,
via
\[f \colon x \mapsto f_x\equiv(\omega_x,H(\omega_x),I_c(B\,\rangle C)_{\sigma_x}),\]
where we are considering $\omega_x$ to be synonymous with its $|C|^2-1$ dimensional parameterization.
By linearity, this extends to a map from probability mass functions on $\CX$ to $\mathbb{R}^{|C|^2 + 1},$ where
\[f\colon p(x)\mapsto \sum_x p(x)f_x \equiv (\omega_p, H(C|X)_p,I_c(B\,\rangle CX)_p),\]
Our use of the subscript $p$ should be clear from the context.
The use of Caratheodory's theorem for bounding the support sizes of auxiliary
random variables in information theory (see \cite{ck}) is well-known.  Perhaps less familiar is the observation \cite{wynerziv,salehiaux} that a better bound can often be obtained by use of a related theorem by Fenchel and Eggleston \cite{egg}, which states that if $S\subseteq\mathbb{R}^n$ is the union of at most $n$ connected subsets, and if $y$ is contained in the convex hull of $S$, then $y$ is also contained in the convex hull of at most $n$ points in $S$.  As the map $f$ is linear, it maps the simplex of distributions on $\CX$ into a single connected subset of $\mathbb{R}^{|C|^2+1}$.  Thus, for any distribution $p(x)$, there is another distribution $p'(x)$ which puts positive probability on at most $|C|^2+1$ states, while satisfying $f(p) = f(p').$  If it is instead the case that $|A'|< |C|,$ this bound can be reduced to $|A|^2 + 1$ by replacing the first components of the map $f$ with a parameterization of $\phi_x^{A'}$, as specification of a density matrix on
$A'$ is enough to completely describe the resulting state on $C$.  It is therefore sufficient to consider $|X| \leq \min\{|A'|,|C|\}^2 + 1$ when computing $\CC\CQ^{(1)}(\CN)$.

% \section{Proof that {\small $H(\al) - H\Big(\frac{1}{2}(1 + \sqrt{p(1-p)(2\al-1)^2 + (1-2p)^2})\Big)$} is maximal at $\al=\frac{1}{2}$}
% We will show that for the function
% \[f(\al) = H(\al) - H\Big(\frac{1}{2}(1 + \sqrt{p(1-p)(2\al-1)^2 + (1-2p)^2})\Big),\]
% and for any $0\leq p \leq 1$, we have 
% \[\arg\max_{0\leq \al \leq 1} f(\al) = \frac{1}{2}.\] 
% Taking the derivative with respect to $\al$, we obtain
% \[f'(\al) = \log\left(\frac{1}{\al} + 1\right)^\beta \left(\frac{2}{1 + \beta}\right)^{(2\al-1)p(1-p)},\]
% where 
% \[\beta = \sqrt{1+(2\al - 3)(2\al + 1)p(1-p)}.\]
% Observe that $0\leq \beta\leq 1$.  By exponentiating,  $f'(\al)=0$
% if and only if 
% \[\left(\frac{1}{\al} + 1\right)^\beta \left(\frac{2}{1 + \beta}\right)^{(2\al-1)p(1-p)} = 1\]

\qed


\begin{thebibliography}{50}
\bibitem{ahlswede} R. Ahlswede, ``Multi-way communication channels," Second Intern.\ Sympos.\ on Inf.\ Theory, Thakadsor, 1971, Publ.\ House of the Hungarian Adad.\ of Sciences, pp.\ 23--52, 1973.

\bibitem{ahlswedelober} R. Ahlswede, P. L\"ober,  ``Quantum data processing," \emph{IEEE Trans.\ Inform.\ Theory}, vol.\ 47, no.\ 1, pp.\ 474--478, January 2001.

\bibitem{adhw} C. Ahn, P. Doherty, P. Hayden, A. Winter, ``On the distributed compression of quantum information," to appear in \emph{IEEE Trans. Inform. Theory}, {\tt quant-ph/0403042}.

\bibitem{alickifannes} R. Alicki, M. Fannes, ``Continuity of quantum conditional information," \emph{J. Phys. A,} vol.\ 37, pp.\ L55--L57, January 2004.

\bibitem{bcfjs} H. Barnum, C. M. Caves, C. A. Fuchs, R. Josza, B. Schumacher, ``Noncommuting mixed states cannot be broadcast," \emph{Phys.\ Rev.\ Lett.\ }, vol.\ 76, no.\ 15, pp.\ 2818--2821, 1996.

\bibitem{bkn} H. Barnum, E. Knill, M. Nielsen, ``On quantum fidelities and quantum capacities," \emph{IEEE Trans.\ Inform.\ Theory}, vol.\ 46, no.\ 4, pp.\ 1317--1329, July 2000.

\bibitem{bds} C. Bennett, D. DiVincenzo, J. Smolin, ``Capacities of quantum erasure channels," 
\emph{Phys.\ Rev.\ Lett.,} vol.\ 78, pp.\ 3217–-3220, April 1997.

\bibitem{ces} T. Cover, A. El Gamal, M. Salehi, ``Multiple-access channels with arbitrarily correlated sources," \emph{IEEE Trans.\ Inform.\ Theory,} vol.\ 26, no.\ 6, pp.\ 648--657, November 1980.

\bibitem{coverjulian} T. Cover, D. Julian, ``Concavity of the Second Law of Thermodynamics,"  \emph{IEEE Intern.\ Symp.\ Inform.\ Theory,} Yokohama, Japan, June 2003.

\bibitem{coverthomas} T. Cover, J. A. Thomas, \emph{Elements of Information Iheory,} John--Wiley \& Sons, Inc., 1991.

\bibitem{ck} I. Cs\'iszar, J. K\"orner, \emph{Information Theory: Coding Theorems for Discrete Memoryless Systems,} Akad\'emiai Kiad\'o, Budapest.

\bibitem{davies} E. B. Davies, J. T. Lewis, ``An operational approach to quantum probability," \emph{Comm.\ Math.\ Phys.,} vol.\ 17, no.\ 3, pp.\ 239--260, 1970. 

\bibitem{dev} I. Devetak, ``The private classical information capacity and quantum information capacity of a quantum channel," \emph{IEEE Trans.\ Inform.\ Theory,} vol.\ 55, no. 1, pp.\ 44--55, January 2005.

\bibitem{dw} I. Devetak, A. Winter, ``Classical data compression with quantum side information," \emph{Phys.\ Rev. A}, vol.\ 68, pp.\ 042301-042306, October 2003.

\bibitem{ds} I. Devetak, P. Shor, ``The capacity of a quantum channel for simultaneous transmission of classical and quantum information,"  {\tt quant-ph/0311131}.

\bibitem{dcr} I. Devetak, A. Winter, ``Distilling common randomness from bipartite quantum states," {\tt quant-ph/0304196}, 2003.

\bibitem{dy} I. Devetak, J. Yard, ``Quantum channels with side information," in preparation.

\bibitem{dueck} G. Dueck, ``Maximal error regions are strictly smaller than average error regions for multi-user channels," \emph{Problems of Control and  Information Theory,} vol.\ 7, pp.\ 11--19, 1978.

\bibitem{dch} W. D\"ur, J. I. Cirac, P. Horodecki,
``Non-additivity of quantum capacity for multiparty communication channels," {\tt quant-ph/0403068}.

\bibitem{egg} H. G. Eggleston, \emph{Convexity,} Cambridge University Press, N.Y.\ 1963.

\bibitem{gp} S. I. Gelfand, M. S. Pinsker, ``Coding for a channel with random parameters," \emph{Problems of Control and Information Theory,} vol.\ 9, no.\ 1, pp.\ 19--31, 1980.

\bibitem{family} A. Harrow, I. Devetak, ``A family of quantum protocols," {\tt quant-ph/0308044}.

\bibitem{hjpw} P. Hayden, R. Josza, D. Petz, A. Winter, ``Structure of states which satisfy strong subadditivity of quantum entropy with equality," \emph{Commun.\ Math.\ Phys.,} vol.\ 246, no.\ 2, pp. 359--374, 2004.

\bibitem{holevobound} A. S. Holevo, ``Bounds for the quantity of information transmitted by a quantum channel," \emph{Probl.\ Inf.\ Transm.,} vol.\ 9, pp.\ 177--183, 1973.

\bibitem{holcap} A. S. Holevo, ``The capacity of the quantum channel with general input states," \emph{IEEE Trans.\ Inform.\ Theory,} vol.\ 44, no.\ 1, pp.\ 269-273, January 1998. 

\bibitem{hj} R. Horn, C.R. Johnson, \emph{Matrix Analysis,} Cambridge University Press, Cambridge, 1985.

\bibitem{how} M. Horodecki, J. Oppenheim, A. Winter, ``Quantum information can be negative," preprint available at \\ {\tt http://www.damtp.cam.ac.uk/user/jono/pub/merge.pdf}.

\bibitem{gleb} G. Klimovitch, ``On the classical capacity of a quantum multiple access channel," \emph{IEEE Intern.\ Symp.\ Inform.\ Theory,} Washington D.C.\ June 2001, p.\ 278.

\bibitem{kw} G. Klimovitch, A. Winter, ``Classical capacity of quantum binary adder channels," {\tt quant-ph/0502055}.

\bibitem{monogamy} M. Koashi, A. Winter, ``Monogamy of entanglement and other correlations," \\ \emph{Phys.\ Rev.\ A} vol.\ 69, no.\ 2, pp.\ 022309--022314, February 2004.

\bibitem{liao} Liao, ``Multiple access channels", Ph.D.\ dissertation, Dept.\ of Electrical Engineering, University of Hawaii, 1972.

\bibitem{ssuborig} E. H. Lieb, M. B. Ruskai, ``Proof of the strong subadditivity of quantum-mechanical entropy," \emph{J.\ Math.\ Phys.,}  
vol.\ 14, pp.\ 1938--1941, December 1973.

\bibitem{lloyd} S. Lloyd, ``Capacity of the noisy quantum channel,"
\emph{Phys.\ Rev.\ A,} vol.\ 55, no.\ 3, pp.\ 1613--1622, March 1997.

\bibitem{cn} M. Nielsen, I. Chuang, \emph{Quantum Information and Quantum Computation,} Cambridge University Press, 2001.

\bibitem{ssub} M. Nielsen, D. Petz, ``A simple proof the strong subadditivity inequality," {\tt quant-ph/0408130}.

\bibitem{peres} A. Peres, \emph{Quantum theory: concepts and methods,} Kluwer Academic Publishers, 1995.

\bibitem{peresinforel} A. Peres, D. Terno, ``Quantum information and relativity theory," \emph{Rev.\ Mod.\ Phys.,}
vol.\ 76, pp.\ 93-123, January 2004.

\bibitem{op} Oyha, D. Petz, \emph{Quantum Entropy and Its Use,} Springer, Berlin, 1993.

\bibitem{preskill} J. Preskill, \emph{Lecture Notes for Physics 219, Quantum Computation and Information,} available at \\
{\tt http://www.theory.caltech.edu/\~{}preskill/ph219/index.html\#lecture}.

\bibitem{rustrace} M. B. Ruskai, ``Beyond strong subadditivity: improved bounds on the contraction of generalized relative entropy," \emph{Rev.\ Math.\ Phys.,} vol.\ 6, no.\ 5A, pp.\ 1147--1161, 1994. 

\bibitem{salehiaux} M. Salehi, ``Cardinality bounds on auxiliary random variables in multiple-user theory via the method of Ahlswede and K\"orner," Technical Report No.\ 33, Dept.\ of Statistics, Stanford University, August 1978.

\bibitem{sn96} B. Schumacher, M. A. Nielsen, ``Quantum data processing and error correction," \emph{Phys.\ Rev.\ A}, vol.\ 54, no.\ 4, pp.\ 2629--2635, October 1996.

\bibitem{sw2} B. Schumacher, M. D. Westmoreland, ``Sending classical information via noisy quantum channels," \emph{Phys.\ Rev.\ A,} vol.\ 56, no.\ 1, pp.\ 131--138, July 1997.

\bibitem{shannon} C. Shannon, ``A mathematical theory of communication," \emph{Bell System Technical Journal,}
vol.\ 27, pp.\ 379--423, 623--656, July, October 1948.

\bibitem{shor} P. Shor, ``The quantum channel capacity and coherent information," lecture notes, MSRI Workshop on Quantum Computation, 2002. Available at \\ {\small \tt http://www.msri.org/publications/ln/msri/2002/quantumcrypto/shor/1/}

\bibitem{ss} P. W. Shor, J. A. Smolin, ``Quantum error-correcting codes need not completely reveal the error syndrome," {\tt quant-ph/9604006}.

\bibitem{uhl} A. Uhlmann, ``The `transition probability' in the state space of a *-algebra," \emph{Rep.\ Math.\ Phys.,} vol.\ 9, pp.\ 273-279, 1976.

\bibitem{gentle} A. Winter, ``Coding theorem and strong converse for quantum channels," \emph{IEEE Trans.\ Inform.\ Theory}, vol.\ 45, no.\ 7, pp.\ 2481--2485, November 1999.

\bibitem{wintermac} A. Winter, ``The capacity of the quantum multiple access channel," \emph{IEEE Trans.\ Inform.\ Theory},
vol.\ 47, no.\ 7, pp.\ 3059--3065, November 2001.

\bibitem{wintertalk} A. Winter, F. Verstraete, J. Smolin, J. Oppenheim, M. Horodecki, ``Entanglement of assistance and applications to multiuser quantum information theory, or transmitting partial (quantum) information - which can be negative!," talk given at QIP 2005, Cambridge, Massachusetts, January 2005.

\bibitem{wynerziv} A. Wyner, J. Ziv, ``The rate-distortion function for source decoding with side-information at the decoder," \emph{IEEE Trans.\ Inform.\ Theory,} vol.\ 22, no.\ 1, pp.\ 1--10, January 1976.

\bibitem{future} J. Yard, ``Capacity Theorems for Quantum Multiple Access Channels --- Part II: Simultaneous Classical-Quantum Capacity Region", in preparation.

\bibitem{qmac} J. Yard, I. Devetak, P. Hayden, ``Capacity theorems for quantum multiple access channels --- classical-quantum and quantum-quantum capacity regions,"
{\tt quant-ph/0501045}.

\bibitem{isit} J. Yard, I. Devetak, P. Hayden, ``Capacity theorems for quantum multiple access channels," submitted to \emph{IEEE Intern.\ Symp.\ Inform.\ Theory}, Adelaide, Australia, 2005.


\end{thebibliography}
\end{document}